\documentclass[a4paper]{article}

    \pdfoutput=1
\usepackage{setspace}
\usepackage{times}
\usepackage{graphicx}
\usepackage{amsmath}
\usepackage{bm}
\usepackage{amssymb}
\usepackage{amsfonts}
\usepackage{xfrac}
\usepackage{verbatim}
\usepackage{enumerate}
\usepackage{rotating}
\usepackage{rotfloat}
\usepackage{units}
\usepackage{color}
\usepackage[verbose,a4paper,tmargin=1.0in,bmargin=1.0in,lmargin=1.0in,rmargin=1.0in,nohead]{geometry}
\usepackage[super,sort&compress]{natbib}

\bibpunct{}{}{,}{s}{}{;}

\newcommand{\NPsection}[1]{\subsection*{\textsf{\normalsize\color{black}#1}}}

\usepackage{setspace}
\usepackage{hyperref}
\hypersetup{pdfauthor={Matteo Mariantoni},pdftitle={Implementing the Quantum von Neumann Architecture with Superconducting Circuits}}

\begin{document}

\renewcommand{\topfraction}{1.0}
\renewcommand{\bottomfraction}{1.0}
\renewcommand{\textfraction}{0.0}

% ************************
% *** Manuscript title ***
% ************************
\noindent\parbox{\textwidth}{\flushleft\textsf{\color{black}\Huge
Implementing the Quantum von Neumann Architecture with
Superconducting Circuits}}
    \vspace{3mm}

% **********************************
% *** Author list & affiliations ***
% **********************************
\noindent\parbox{\textwidth}{\flushleft
\textsf{\color{black}\Large
        Matteo~Mariantoni$^{\mathsf{1,4,\S}}$,
        H.~Wang$^{1,*}$,
        T.~Yamamoto$^{1,2}$,
        M.~Neeley$^{1,\dag}$,
        Radoslaw~C.~Bialczak$^{1}$,
        Y.~Chen$^{1}$,
        M.~Lenander$^{1}$,
        Erik~Lucero$^{1}$,
        A.~D.~O'Connell$^{1}$,
        D.~Sank$^{1}$,
        M.~Weides$^{1,\ddag}$,
        J.~Wenner$^{1}$,
        Y.~Yin$^{1}$,
        J.~Zhao$^1$,
        A.~N.~Korotkov$^{3}$,
        A.~N.~Cleland$^{1,4}$,
        and John~M.~Martinis$^{1,4,\S}$
        }
    \vspace{2mm}

\textsf{\textbf{\small $^{1}$Department of Physics, University of California, Santa Barbara, CA 93106-9530, USA\\
$^{2}$Green Innovation Research Laboratories, NEC Corporation, Tsukuba, Ibaraki 305-8501, Japan\\
$^{3}$Department of Electrical Engineering, University of
California, Riverside, CA 92521, USA
$^{4}$California NanoSystems Institute, University of California, Santa Barbara, CA 93106-9530, USA\\
$^{*}$Present address: Department of Physics, Zhejiang University, Hangzhou 310027, China.\\
$^{\dag}$Present address: Lincoln Laboratory, Massachusetts Institute of Technology, 244 Wood Street, Lexington, MA 02420-9108, USA.\\
$^{\ddag}$Present address: National Institute of Standards and Technology, Boulder, CO 80305, USA.\\
$^{\S}$To whom correspondence should be addressed. E-mail: matmar@physics.ucsb.edu (M.~M.); martinis@physics.ucsb.edu (J.~M.~M.)\\
    \vspace{5mm}
    }}}

% ************
% *** Date ***
% ************
\noindent\textsf{\small last updated: \today}\\
{\color{black}\rule[2mm]{\textwidth}{0.1mm}}

    {\noindent\textbf{
The von Neumann architecture for a classical computer comprises
a central processing unit and a memory holding instructions and
data. We demonstrate a quantum central processing unit that
exchanges data with a quantum random-access memory integrated
on a chip, with instructions stored on a classical computer. We
test our quantum machine by executing codes that involve seven
quantum elements: Two superconducting qubits coupled through a
quantum bus, two quantum memories, and two zeroing registers.
Two vital algorithms for quantum computing are demonstrated,
the quantum Fourier transform, with $\mathbf{66}\,\%$ process
fidelity, and the three-qubit Toffoli OR phase gate, with
$\mathbf{98}\,\%$ phase fidelity. Our results, in combination
especially with longer qubit coherence, illustrate a
potentially viable approach to factoring numbers and
implementing simple quantum error correction codes.
    }

% ****************************
% *** General introduction ***
% ****************************
%
Quantum processors~\cite{divincenzo:2000:rules,
nielsen:2000:qcomputing,mermin:2007:qcomputer,wiseman:2010:qmeascontr}
based on nuclear magnetic
resonance~\cite{vandersypen:2001:shornmr,cory:1998:qerrorcorr,weinstein:2001:qft},
trapped
ions~\cite{blatt:2008:tions,chiaverini:2005:sqft,monz:2009:toffoli},
and semiconducting devices~\cite{reilly:2008:spinqubit} were
used to realize Shor's quantum factoring
algorithm~\cite{vandersypen:2001:shornmr} and quantum error
correction~\cite{cory:1998:qerrorcorr,blatt:2008:tions}. The
quantum operations underlying these algorithms include
two-qubit
gates~\cite{nielsen:2000:qcomputing,mermin:2007:qcomputer}, the
quantum Fourier
transform~\cite{weinstein:2001:qft,chiaverini:2005:sqft}, and
three-qubit Toffoli
gates~\cite{lanyon:2009:toffoli,monz:2009:toffoli}. In addition
to a quantum processor, a second critical element for a quantum
machine is a quantum memory, which has been demonstrated, e.g.,
using optical systems to map photonic entanglement into and out
of atomic ensembles~\cite{choi:2008:qmemory}.

Superconducting quantum circuits~\cite{clarke:2008:squibits}
have met a number of milestones, including demonstrations of
two-qubit gates~\cite{plantenberg:2007:cnot,
dicarlo:2009:qprocessor,leek:2009:sideband,yamamoto:2010:czgates,
neeley:2010:ghz,dicarlo:2010:ghz} and the advanced control of
both qubit and photonic quantum
states~\cite{neeley:2010:ghz,dicarlo:2010:ghz,
mariantoni:2011:shell,wang:2011:noon}. We demonstrate a
superconducting integrated circuit that combines a processor,
executing the quantum Fourier transform and a three-qubit
Toffoli-class OR gate, with a memory and a zeroing register in
a single device. This combination of a quantum central
processing unit (quCPU) and a quantum random-access memory
(quRAM), which comprise two key elements of a classical von
Neumann architecture, defines our quantum von Neumann
architecture.

% ********************************************
% *** The quantum von Neumann architecture ***
% ********************************************
%
% ****************
% *** FIGURE 1 ***
% ****************
\begin{figure}[t!]
    \centering
    \includegraphics[width=0.52\columnwidth]{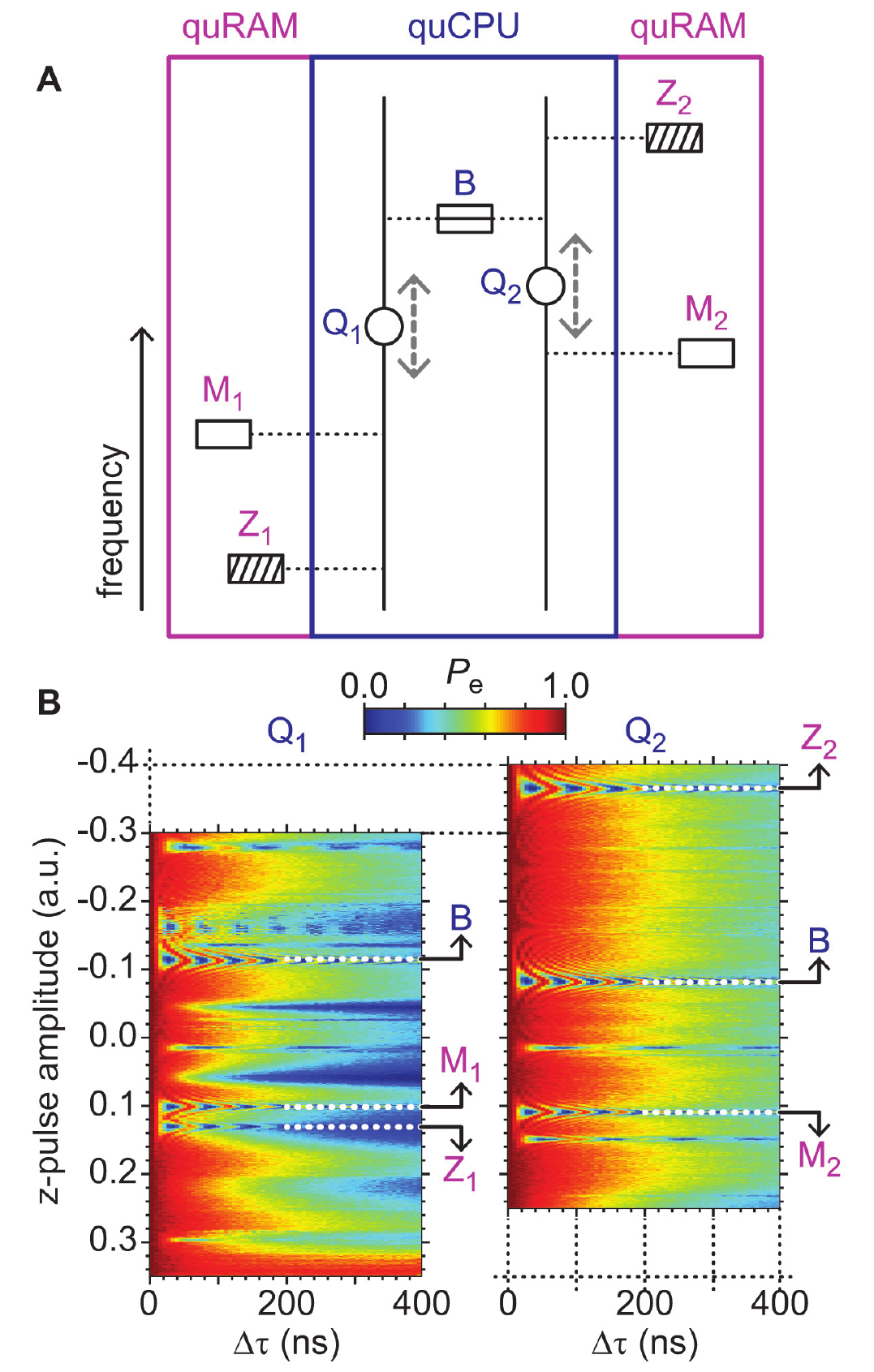}
    \caption{\footnotesize
\textbf{The quantum von Neumann architecture.} (\textbf{A}) The
quCPU (blue box) includes two qubits Q$^{}_1$ and Q$^{}_2$ and
the bus resonator B. The quRAM (magenta boxes) comprises two
memories M$^{}_1$ and M$^{}_2$ and two zeroing registers
Z$^{}_1$ and Z$^{}_2$. The horizontal dotted lines indicate
connections between computational elements. The vertical
direction represents frequency, where the memory and zeroing
registers are fixed in frequency, while the qubit transition
frequencies can be tuned via z-pulses (grey dashed double
arrows). (\textbf{B}) Swap
spectroscopy~\cite{mariantoni:2011:shell} for Q$^{}_1$ (left)
and Q$^{}_2$ (right): Qubit excited state $| \textrm{e}
\rangle$ probability $P^{}_{\textrm{e}}$ (color scale) vs.
z-pulse amplitude (vertical axis) and delay time $\Delta \tau$
(horizontal axis), after exciting the qubit with a $\pi$-pulse.
At zero z-pulse amplitude the qubits are at their idle points,
where they have an energy relaxation time $T^{}_{\textrm{rel}}
{} \simeq {} 400$\,ns. A separate Ramsey experiment yields the
qubits' dephasing time $T^{}_{\textrm{deph}} {} \simeq {}
200$\,ns. By tuning the z-pulse amplitude, the qubit transition
frequencies $f^{}_{\textrm{Q}^{}_1}$ and
$f^{}_{\textrm{Q}^{}_2}$ can be varied between $\simeq 5.5$ and
$8$\,GHz. For z-pulse amplitudes indicated by B and M$^{}_1$
for Q$^{}_1$, and by B and M$^{}_2$ for Q$^{}_2$, the ``chevron
pattern'' of a qubit-resonator interaction is
observed~\cite{mariantoni:2011:shell}. The transition
frequencies of B, M$^{}_1$, and M$^{}_2$ are $f^{}_{\textrm{B}}
{} = {} 6.82$\,GHz, $f^{}_{\textrm{M}^{}_1} {} = {} 6.29$\,GHz,
and $f^{}_{\textrm{M}^{}_2} {} = {} 6.34$\,GHz, respectively.
From the chevron oscillation we obtain the qubit-resonator
coupling strengths, which for both the resonator bus and the
memories are $\simeq {} 20$\,MHz (splitting) for the $|
\textrm{g} \rangle {} \leftrightarrow {} | \textrm{e} \rangle$
qubit transition, and $\approx {} \sqrt{2}$ faster for the $|
\textrm{e} \rangle {} \leftrightarrow {} | \textrm{f} \rangle$
transition ($| \textrm{g} \rangle$, $| \textrm{e} \rangle$, and
$| \textrm{f} \rangle$ are the three lowest qubit
states)~\cite{wang:2011:noon}. For all resonators
$T^{}_{\textrm{rel}} {} \simeq {} 4\,\mu$s. Swap spectroscopy
also reveals that the qubits interact with several modes
associated with spurious two-level systems. Two of them,
Z$^{}_1$ and Z$^{}_2$, are used as zeroing registers. Their
transition frequencies are $f^{}_{\textrm{Z}^{}_1} {} = {}
6.08$\,GHz and $f^{}_{\textrm{Z}^{}_2} {} = {} 7.51$\,GHz,
respectively, with coupling strength to the qubits of $\simeq
{} 17$\,MHz.
    }
\end{figure}

In our architecture (Fig.~1A), the quCPU performs one-, two-,
and three-qubit gates that process quantum information, and the
adjacent quRAM allows quantum information to be written, read
out, and zeroed. The quCPU includes two superconducting phase
qubits~\cite{yamamoto:2010:czgates,neeley:2010:ghz,
mariantoni:2011:shell,wang:2011:noon} Q$^{}_1$ and Q$^{}_2$,
connected through a coupling bus provided by a superconducting
microwave resonator B. The quRAM comprises two superconducting
resonators M$^{}_1$ and M$^{}_2$ that serve as quantum
memories, as well as a pair of zeroing registers Z$^{}_1$ and
Z$^{}_2$, two-level systems that are used to dump quantum
information. The chip geometry is similar to that in
Refs.~\cite{mariantoni:2011:shell,wang:2011:noon}, with the
addition of the two zeroing registers. Figure~1B shows the
characterization of the device by means of swap
spectroscopy~\cite{mariantoni:2011:shell}.

% ********************************************************
% *** Programming the quantum von Neumann architecture ***
% ********************************************************
%
% ****************
% *** FIGURE 2 ***
% ****************
\begin{figure}[t!]
    \centering
    \includegraphics[width=0.99\columnwidth]{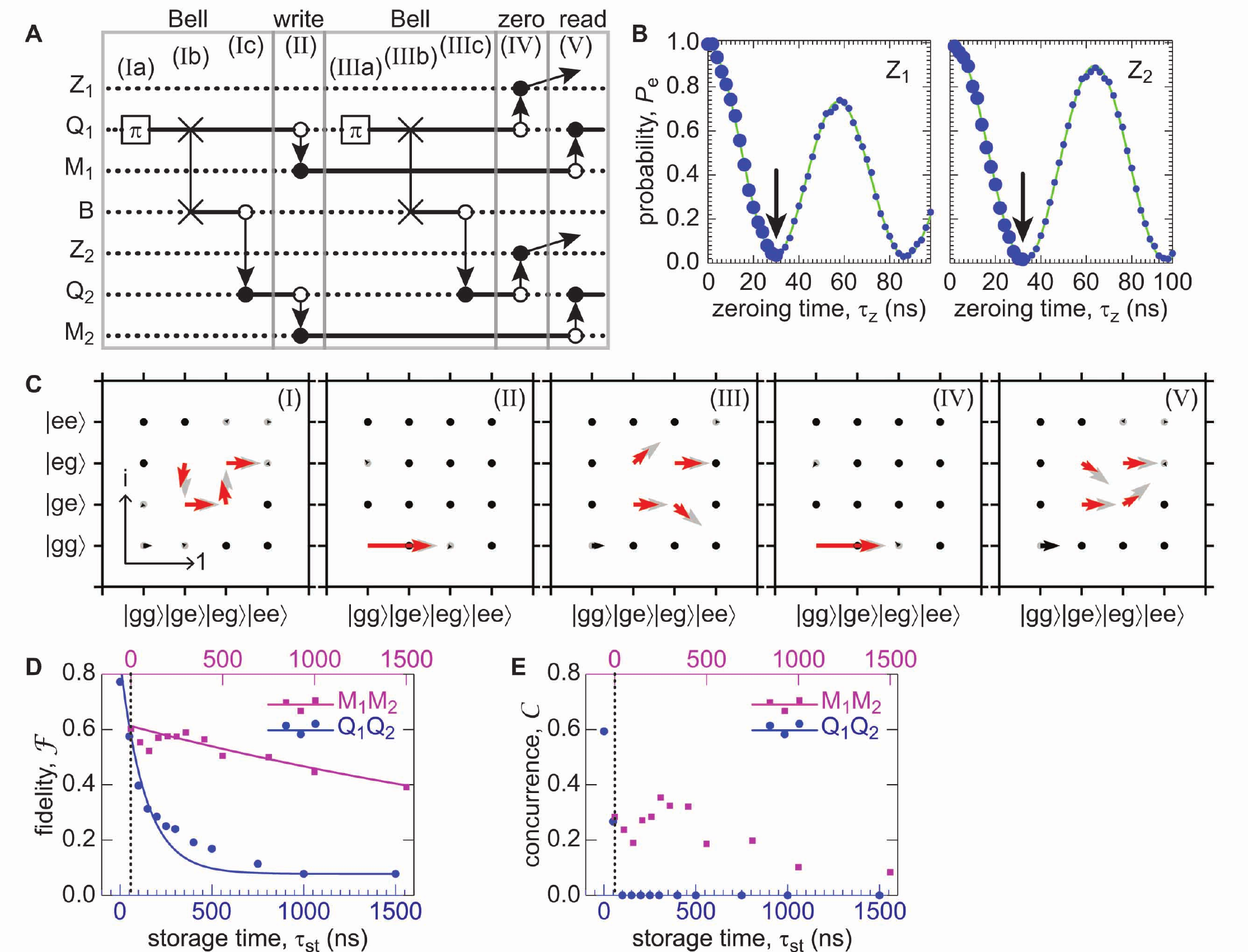}
    \caption{\footnotesize
\textbf{Programming the quantum von Neumann architecture.}
(\textbf{A}) Quantum algorithm comprising $7$ independent
channels interacting through five computational steps. Dotted
and solid lines represent channels in the ground and
excited/superposition states, respectively. A black rectangle
represents a $\pi$-pulse; two crosses connected by a solid line
a $\sqrt{\textrm{iSWAP}}$; an open and a closed circle
connected by a single arrow an iSWAP; oblique arrows indicate
decay from a zeroing register. (\textbf{B}) Calibration of the
zeroing gates. Each qubit is prepared in $| \textrm{e}
\rangle$, interacts on resonance with its zeroing register for
a time $\tau^{}_{\textrm{z}}$, and its probability
$P^{}_{\textrm{e}}$ measured, with $P^{}_{\textrm{e}}$ plotted
vs. $\tau^{}_{\textrm{z}}$ (large and small blue circles). The
solid green line is a decaying cosine fit to the data. The
black arrows indicate the zeroing time for each qubit.
(\textbf{C}) Density matrices $\hat{\rho}^{}_{\textrm{(I)}} ,
\hat{\rho}^{}_{\textrm{(II)}} , \ldots ,
\hat{\rho}^{}_{\textrm{(V)}}$ of the Q$^{}_1$-Q$^{}_2$ state
for each step in A (scale key on bottom left). Grey arrows:
Ideal state. Red and black arrows and black dots: Measured
state (black arrows indicate errors). The off-diagonal elements
of $\hat{\rho}^{}_{\textrm{(I)}}$,
$\hat{\rho}^{}_{\textrm{(III)}}$, and
$\hat{\rho}^{}_{\textrm{(V)}}$ have different angles because of
dynamic phases~\cite{note:methods}. Fidelities:
$\mathcal{F}^{}_{\textrm{(I)}} {} = {} 0.772 \pm 0.003$,
$\mathcal{F}^{}_{\textrm{(II)}} {} = {} 0.916 \pm 0.002$,
$\mathcal{F}^{}_{\textrm{(III)}} {} = {} 0.689 \pm 0.003$,
$\mathcal{F}^{}_{\textrm{(IV)}} {} = {} 0.913 \pm 0.002$, and
$\mathcal{F}^{}_{\textrm{(V)}} {} = {} 0.606 \pm 0.003$.
Concurrences: $\mathcal{C}^{}_{\textrm{(I)}} {} = {} 0.593 \pm
0.006$, $\mathcal{C}^{}_{\textrm{(II)}} {} = {} 0.029 \pm
0.005$, $\mathcal{C}^{}_{\textrm{(III)}} {} = {} 0.436 \pm
0.007$, $\mathcal{C}^{}_{\textrm{(IV)}} {} = {} 0.019 \pm
0.005$, and $\mathcal{C}^{}_{\textrm{(V)}} {} = {} 0.345 \pm
0.008$. (\textbf{D}) Comparison of fidelity $\mathcal{F}$ as a
function of storage time $\tau^{}_{\textrm{st}}$ for a Bell
state stored in Q$^{}_1$ and Q$^{}_2$ (blue circles) vs. that
stored in M$^{}_1$ and M$^{}_2$ (magenta squares; error bars
smaller than symbols). The solid lines are exponential fits to
data. (\textbf{E}) As in D, but for the concurrence
$\mathcal{C}$. In D and E the vertical black dotted line
indicates the time delay ($\simeq {} 59$\,ns) associated with
memory storage, with respect to storage in the qubits, due to
the writing and reading operations (II) and (V) in A.
    }
\end{figure}

The computational capability of our architecture is displayed
in Fig.~2A, where a $7$-channel quantum circuit, yielding a
$128$ dimensional Hilbert space, executes a prototypical
algorithm. First, we create a Bell state between Q$^{}_1$ and
Q$^{}_2$ using a series of $\pi$-pulse,
$\sqrt{\textrm{iSWAP}}$, and iSWAP operations (step I, a to
c)~\cite{wang:2011:noon}. The corresponding density matrix
$\hat{\rho}^{}_{\textrm{(I)}}$ [Fig.~2C (I)] is measured by
quantum state tomography. The Bell state is then written into
the quantum memories M$^{}_1$ and M$^{}_2$ by an iSWAP pulse
(step II)~\cite{wang:2011:noon}, leaving the qubits in their
ground state $| \textrm{g} \rangle$, with density matrix
$\hat{\rho}^{}_{\textrm{(II)}}$ [Fig.~2C (II)]. While storing
the first Bell state in M$^{}_1$ and M$^{}_2$, a second Bell
state with density matrix $\hat{\rho}^{}_{\textrm{(III)}}$
[Fig.~2C (III)] is created between the qubits, using a sequence
similar to the first operation (step III, a to c).

In order to re-use the qubits Q$^{}_1$ and Q$^{}_2$, for
example to read out the quantum information stored in the
memories M$^{}_1$ and M$^{}_2$, the second Bell state has to be
dumped~\cite{reed:2010:reset}. This is accomplished using two
zeroing gates, by bringing Q$^{}_1$ on resonance with Z$^{}_1$
and Q$^{}_2$ with Z$^{}_2$ for a zeroing time
$\tau^{}_{\textrm{z}}$, corresponding to a full iSWAP (step
IV). Figure~2B shows the corresponding dynamics, where each
qubit, initially in the excited state $| \textrm{e} \rangle$,
is measured in the ground state $| \textrm{g} \rangle$ after
$\simeq {} 30$\,ns. The density matrix
$\hat{\rho}^{}_{\textrm{(IV)}}$ of the zeroed two-qubit system
is shown in Fig.~2C (IV). Once zeroed, the qubits can be used
to read the memories (step V), allowing us to verify that, at
the end of the algorithm, the stored state is still entangled.
This is clearly demonstrated by the density matrix shown in
Fig.~2C (V).

The ability to store entanglement in the memories, which are
characterized by much longer coherence times than the qubits,
is key to the quantum von Neumann architecture. We demonstrate
this capability in Fig.~2, D and E, where the fidelity and
concurrence metrics~\cite{horodecki:2009:entanglement} of the
Bell states stored in M$^{}_1$ and M$^{}_2$ are compared to
those for the same states stored in Q$^{}_1$ and Q$^{}_2$. The
experiment is performed as in Fig.~2A, but eliminating steps
(III) and (IV). For the qubits, the storage time
$\tau^{}_{\textrm{st}}$ is defined as the wait time at the end
of step (I), prior to measuring the qubit states, whereas for
the resonators the wait time is that between the write and read
steps. The fidelity of the qubit states decays to below $0.2$
after $400$\,ns, while for the states stored in the memories it
remains above $0.4$ up to $\simeq {} 1.5\,\mu$s. Most
importantly, after only $100$\,ns the state stored in the
qubits does not preserve any entanglement, as indicated by a
zero concurrence, whereas the memories retain their
entanglement for at least $1.5\,\mu$s (Fig.~2E). We expect
taking advantage of our architecture in long computations,
where qubit states can be protected and reused by writing them
into, and reading them out of, the long-lived quRAM.

% *************************************
% *** The quantum Fourier transform ***
% *************************************
%
% ****************
% *** FIGURE 3 ***
% ****************
\begin{figure}[t!]
    \centering
    \includegraphics[width=0.99\columnwidth]{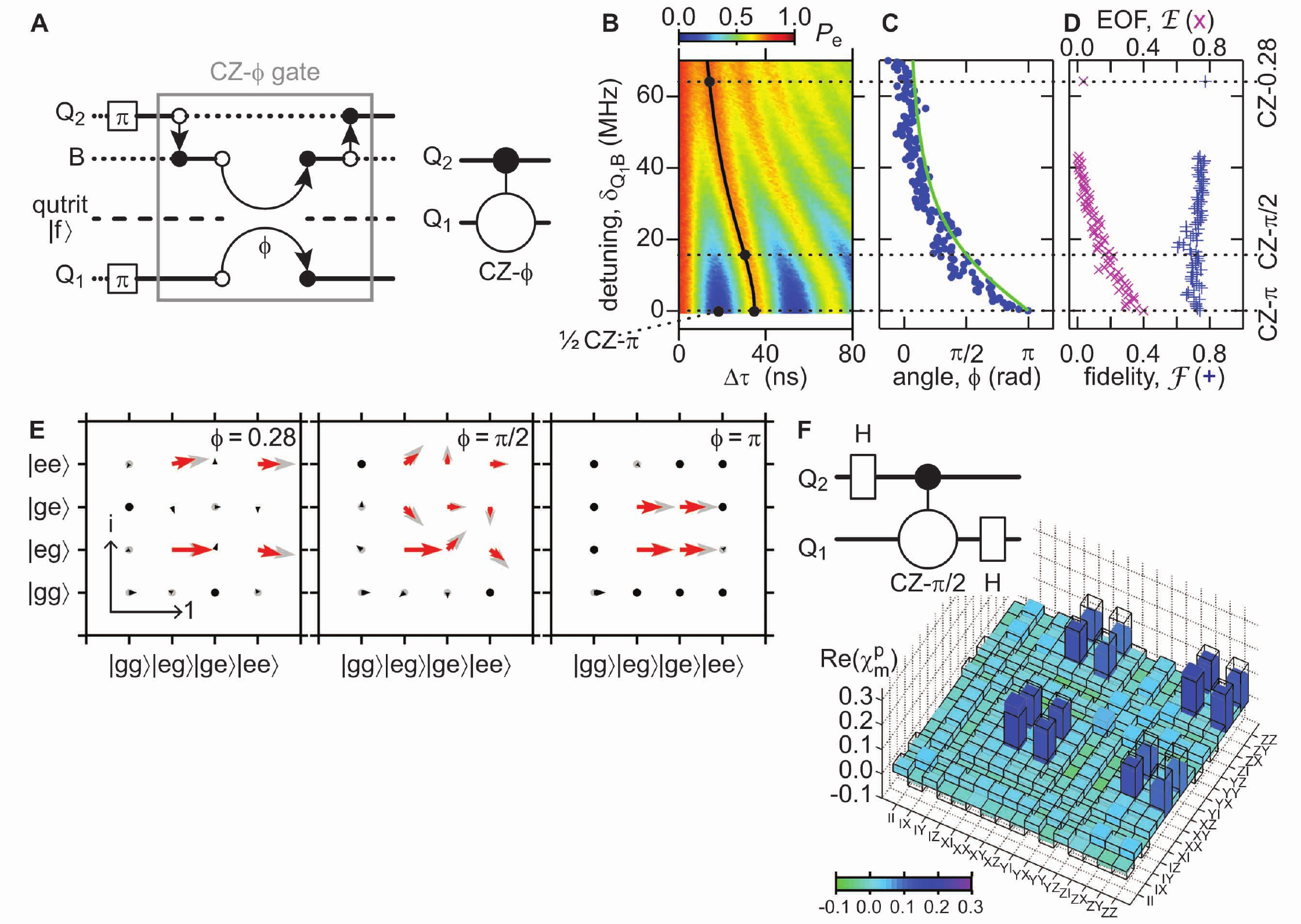}
    \caption{\footnotesize
\textbf{The quantum Fourier transform.} (\textbf{A}) (Left)
Quantum logic circuit of a CZ-$\phi$ gate (enclosed in a grey
box) for $| \textrm{Q}^{}_1 \textrm{Q}^{}_2 \rangle {} = {} |
\textrm{e} \textrm{e} \rangle$. The $| \textrm{f} \rangle$
state of Q$^{}_1$ is indicated by a dashed line. The process
where Q$^{}_1$ acquires the phase $\phi$ is represented by a
pair of open/closed circles, connected by a single arrow in an
arc shape. All other symbols are as in Fig.~2A. (Right)
Shorthand symbol for the CZ-$\phi$ gate. Although the gate
unitary matrix is symmetric, the symbol shows the asymmetric
implementation of the gate. (\textbf{B}) Time-domain swaps
between the states $| \textrm{Q}^{}_1 \textrm{B} \rangle {} =
{} | \textrm{e} 1 \rangle$ and $| \textrm{f} 0 \rangle$, where
we plot the probability $P^{}_{\textrm{e}}$ (color scale) vs.
interaction time $\Delta \tau$ and detuning
$\delta^{}_{\textrm{Q}^{}_1 \textrm{B}}$. The solid black line
indicates combinations of interaction time and detuning that
completely depopulate the non-computational $| \textrm{f}
\rangle$ state. The three black dots on this line correspond to
a CZ-$\pi$, CZ-$\pi / 2$, and CZ-$0.28$ gate (see far right).
The fourth black dot (outside the line) corresponds to a
$\sfrac{1}{2}$\,CZ-$\pi$ gate (see bottom-left), where the $|
\textrm{e} \rangle$ state has been shelved to the
non-computational $| \textrm{f} \rangle$ state. (\textbf{C})
Phase $\phi$ acquired by Q$^{}_1$ as a function of
$\delta^{}_{\textrm{Q}^{}_1 \textrm{B}}$. The blue dots
indicate experimental data and the solid green line the theory
of Eq.~\ref{Equation:1}~\cite{note:methods}. (\textbf{D})
Fidelity $\mathcal{F}$ (blue ``+'' symbols) and EOF (magenta
``$\times$'' symbols) of measured density matrices
$\hat{\rho}^{}_{\phi}$ vs. $\delta^{}_{\textrm{Q}^{}_1
\textrm{B}}$. (\textbf{E}) (Left to Right) Density matrices
$\hat{\rho}^{}_{\phi} {} = {} \hat{\rho}^{}_{0.28}$,
$\hat{\rho}^{}_{\pi / 2}$, and $\hat{\rho}^{}_{\pi}$, obtained
when $\phi {} = {} 0.28$, $\phi {} = {} \pi / 2$, and $\phi {}
= {} \pi$\,rad in Eq.~\ref{Equation:2} (scale key on bottom
left). The arrows are color-coded as in Fig.~2C. The measured
fidelities are $\mathcal{F}^{}_{0.28} {} = {} 0.751 \pm 0.064$,
$\mathcal{F}^{}_{\pi / 2} {} = {} 0.735 \pm 0.017$, and
$\mathcal{F}^{}_{\pi} {} = {} 0.741 \pm 0.030$, and EOF are
$\mathcal{E}^{}_{0.28} {} = {} 0.020 \pm 0.055$ (lower bound
$\mathcal{E}^{}_{0.28} {} = {} 0$), $\mathcal{E}^{}_{\pi / 2}
{} = {} 0.106 \pm 0.031$, and $\mathcal{E}^{}_{\pi} {} = {}
0.401 \pm 0.062$. (\textbf{F}) (Top-Left) Logic circuit for a
two-qubit quantum Fourier transform and, (Bottom), real part of
the corresponding $\chi^{\textrm{p}}_{\textrm{m}}$
matrix~\cite{nielsen:2000:qcomputing,yamamoto:2010:czgates}.
The process fidelity for the real and imaginary (not shown)
part of $\chi^{\textrm{p}}_{\textrm{m}}$ is
$\mathcal{F}^{}_{\chi} {} = {} 0.657 \pm 0.014$. The confidence
intervals are estimated from $10$ measurements for
$\hat{\rho}^{}_{0.28}$, $6$ for $\hat{\rho}^{}_{\pi / 2}$ and
$\hat{\rho}^{}_{\pi}$, and $15$ for
$\chi^{\textrm{p}}_{\textrm{m}}$.
    }
\end{figure}

Two-qubit universal gates are a vital resource for the
operation of the
quCPU~\cite{nielsen:2000:qcomputing,mermin:2007:qcomputer}. A
variety of such gates have been implemented in superconducting
circuits~\cite{plantenberg:2007:cnot,
dicarlo:2009:qprocessor,leek:2009:sideband,yamamoto:2010:czgates,
neeley:2010:ghz,dicarlo:2010:ghz}, with some recent
demonstrations of quantum
algorithms~\cite{dicarlo:2009:qprocessor,yamamoto:2010:czgates}.
Control Z-$\pi$ (CZ-$\pi$) gates are readily realizable with
superconducting qubits, due to easy access to the third energy
state of the qubit, effectively operating the qubit as a
qutrit~\cite{strauch:2003:qlogics,
dicarlo:2009:qprocessor,dicarlo:2010:ghz,yamamoto:2010:czgates}.
However, CZ-$\pi$ gates are just a subset of the more general
class of CZ-$\phi$ gates, obtained for the special case where
the phase $\phi {} = {} \pi$. In our architecture, the full
class of CZ-$\phi$ gates, with $\phi$ from $\simeq {} 0$ to
$\pi$, can be generated by coupling a qutrit close to resonance
with a bus resonator.

Figure~3A shows the quantum logic circuit that generates the
CZ-$\phi$ gate (Left) and a shorthand symbol for the gate
(Right). The logic circuit demonstrates the nontrivial case
where qubits Q$^{}_1$ and Q$^{}_2$ are brought from their
initial ground state to $| \textrm{Q}^{}_1 \textrm{Q}^{}_2
\rangle {} = {} | \textrm{e} \textrm{e} \rangle$ by applying a
$\pi$-pulse to each qubit. The excitation in Q$^{}_2$ is then
transferred into bus resonator B, and Q$^{}_1$'s $| \textrm{e}
\rangle {} \leftrightarrow {} | \textrm{f} \rangle$ transition
brought close to resonance with B for the time required for a
$2 \pi$-rotation, where the states $| \textrm{Q}^{}_1
\textrm{B} \rangle {} = {} | \textrm{e} 1 \rangle$ and $|
\textrm{f} 0 \rangle$ are detuned by a frequency
$\delta^{}_{\textrm{Q}^{}_1 \textrm{B}}$, which we term a
``semi-resonant condition.'' In this process Q$^{}_1$ acquires
the phase~\cite{note:methods}
\begin{equation}
\phi {} = {} \pi - \pi \, \frac{\delta^{}_{\textrm{Q}^{}_1
\textrm{B}}}{\sqrt{\delta^2_{\textrm{Q}^{}_1 \textrm{B}} +
\tilde{g}^2_{\textrm{Q}^{}_1 \textrm{B}}}} \, ,
    \label{Equation:1}
\end{equation}
where $\tilde{g}^{}_{\textrm{Q}^{}_1 \textrm{B}}$ is the
coupling frequency between $| \textrm{e} 1 \rangle$ and $|
\textrm{f} 0 \rangle$. The final step is to move the excitation
from B back into Q$^{}_2$.

The time-domain swaps of $| \textrm{Q}^{}_1 \textrm{B} \rangle$
between the states $| \textrm{e} 1 \rangle$ and $| \textrm{f} 0
\rangle$ are shown in Fig.~3B, where the solid black line
indicates the detunings and corresponding interaction times
used to generate any phase $0 {} \lesssim {} \phi {} \leqslant
{} \pi$ (ideally $\phi {} \rightarrow {} 0$ when
$\delta^{}_{\textrm{Q}^{}_1 \textrm{B}} {} \rightarrow {}
\infty$). These phases are measured by performing two Ramsey
experiments on Q$^{}_1$ for each value of the detuning
$\delta^{}_{\textrm{Q}^{}_1 \textrm{B}}$, one with B in the $|
0 \rangle$ state, and the other with B in the $| 1 \rangle$
state. The relative phase between the Ramsey fringes
corresponds to the value of $\phi$ for the CZ-$\phi$
gate~\cite{note:methods}, as shown in Fig.~3C.

A more sophisticated version of this experiment is performed by
initializing Q$^{}_1$ and Q$^{}_2$ each in the superposition
state $| \textrm{g} \rangle + | \textrm{e} \rangle$. We move
Q$^{}_2$'s state into B, perform a CZ-$\phi$ gate with $0 {}
\lesssim {} \phi {} \leqslant {} \pi$, move the state in B back
into Q$^{}_2$, rotate Q$^{}_1$'s resulting state by $\pi / 2$
about the $y$-axis, and perform a joint measurement of Q$^{}_1$
and Q$^{}_2$. Ideally, this protocol permits to create
two-qubit states ranging from a product state for $\phi {} = {}
0$ to a maximally-entangled state for $\phi {} = {} \pi$. In
the two-qubit basis set $\mathcal{M}^{}_2 {} = {} \{ |
\textrm{g} \textrm{g} \rangle , | \textrm{e} \textrm{g} \rangle
, | \textrm{g} \textrm{e} \rangle , | \textrm{e} \textrm{e}
\rangle \}$, the general density matrix of such two-qubit
states reads
\begin{equation}
 \hat{\rho}^{}_{\phi} {} = {}
 \begin{pmatrix}
  0 & 0 & 0 & 0 \\
  0 & 1 / 2 & ( 1 - e^{- i \phi}_{} ) / 4 & ( 1 + e^{- i \phi}_{} ) / 4 \\
  0 & ( 1 - e^{i \phi}_{} ) / 4 & ( 1 - \cos \phi ) / 4 & ( - i \sin \phi ) / 4 \\
  0 & ( 1 + e^{i \phi}_{} ) / 4 & ( i \sin \phi ) / 4 & ( 1 + \cos \phi ) / 4
 \end{pmatrix} \, .
    \label{Equation:2}
\end{equation}
Figure~3D shows the fidelity and entanglement of formation
(EOF)~\cite{horodecki:2009:entanglement} of two-qubit states
generated using $70$ values of $\phi$. Figure~3E shows three
examples of $\hat{\rho}^{}_{\phi}$ for $\phi {} = {} 0.28$,
$\phi {} = {} \pi / 2$, and $\phi {} = {} \pi$, respectively.

The state generated using $\phi {} = {} \pi / 2$ plays a
central role in the implementation of the two-qubit quantum
Fourier transform. Neglecting bit-order reversal, the quantum
Fourier transform can be realized by applying a Hadamard gate
to Q$^{}_2$, followed by a CZ-$\pi / 2$ gate between Q$^{}_1$
and Q$^{}_2$, and finally a Hadamard on
Q$^{}_1$~\cite{nielsen:2000:qcomputing,weinstein:2001:qft,chiaverini:2005:sqft},
as sketched in Fig.~3F (Top-Left). Representing the input state
of the transform as $| x \rangle$ (position) and the output as
$| p \rangle$ (momentum), assuming $| x \rangle {} \in {}
\mathcal{M}^{}_2$ and the indexes $x$ and $p$ are integers,
with $p {} \in {} \{ 0 , 1 , 2 , 3 \}$, the output state $| p
\rangle {} = {} \sum_{x {} = {} 0}^{3} e^{i \, 2 \pi \, x p /
4}_{} \, | x \rangle / 2$, corresponding to a $4 {} \times {}
4$ unitary operator. This operator can be fully characterized
by means of quantum process
tomography~\cite{nielsen:2000:qcomputing,
yamamoto:2010:czgates}, which allows us to obtain the
$\chi^{\textrm{p}}_{\textrm{m}}$
matrix~\cite{nielsen:2000:qcomputing,yamamoto:2010:czgates}
shown in Fig.~3F (Bottom).

% ***************************************************
% *** The three-qubit Toffoli-class OR phase gate ***
% ***************************************************
%
% ****************
% *** FIGURE 4 ***
% ****************
\begin{figure}[t!]
    \centering
    \includegraphics[width=0.60\columnwidth]{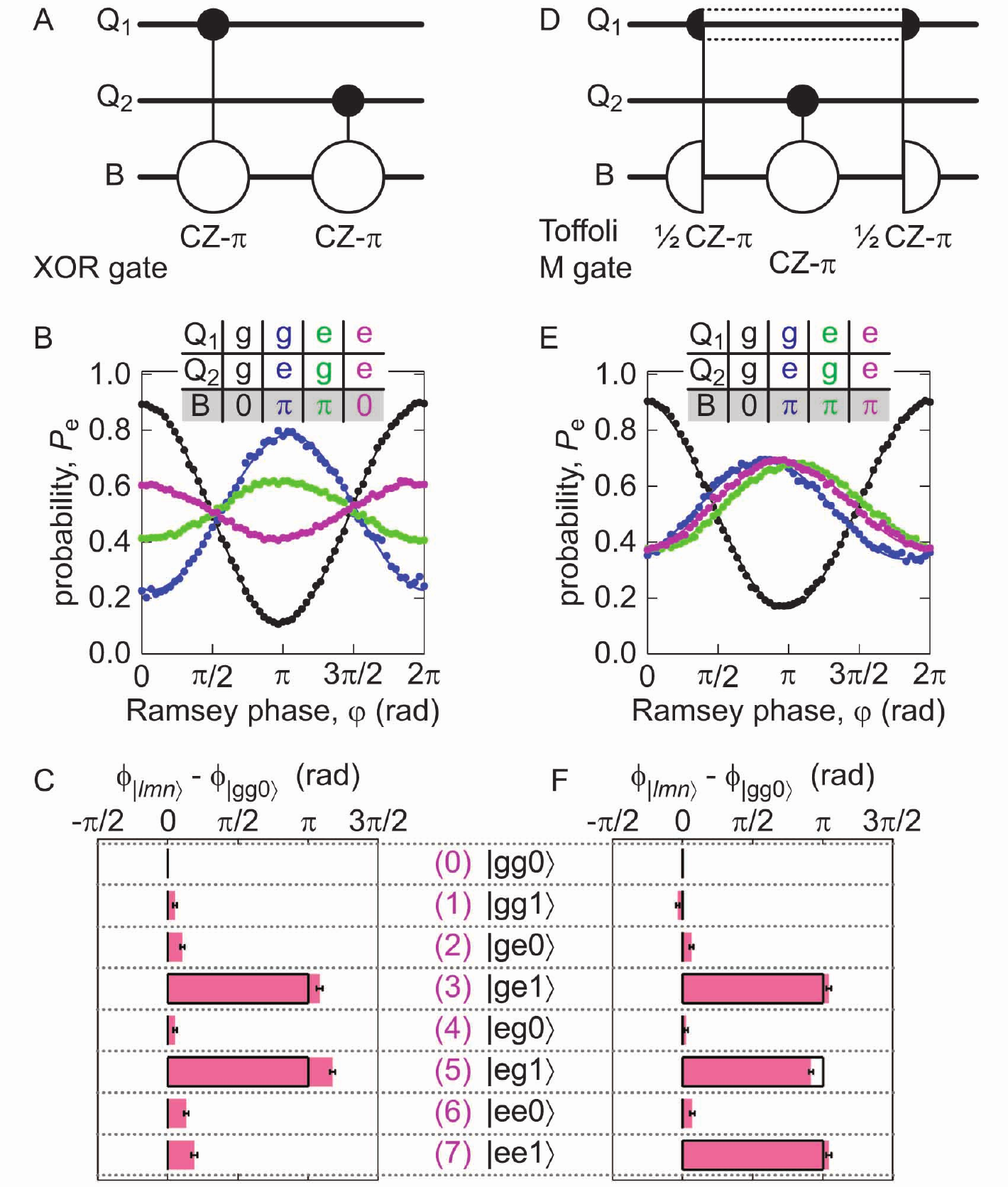}
    \caption{\footnotesize
\textbf{Three-qubit gates: The XOR phase gate and the
Toffoli-class M gate.} (\textbf{A}) Quantum logic circuit for
the XOR phase gate. (\textbf{B}) (Top) XOR-gate truth table.
(Bottom) Ramsey fringes associated with the truth table,
showing the probability $P^{}_{\textrm{e}}$ of measuring
Q$^{}_2$ in $| \textrm{e} \rangle$, vs. the Ramsey phase
$\varphi$, for the control input states in $\mathcal{M}^{}_2$.
Black and magenta dots: $0$ phase. Blue and green dots: $\pi$
phase. The solid lines are least-squares fits to the data used
to extract the truth-table phases. (\textbf{C}) Quantum phase
tomography for the XOR gate: Phase $\phi^{}_{| l m n \rangle} -
\phi^{}_{| \textrm{g} \textrm{g} 0 \rangle}$, for each state $|
l m n \rangle {} \in {} \mathcal{M}^{}_3$. Black open boxes:
Ideal values. Pink areas: Measured values with corresponding
confidence intervals (black lines). (\textbf{D}) Quantum logic
circuit for the M gate, implemented as a
$\sfrac{1}{2}$\,CZ-$\pi$ gate (cf.~Fig.~3B) between Q$^{}_1$
and B (half-dot/half-open circle connected by solid line),
followed by a CZ-$\pi$ gate between Q$^{}_2$ and B, and a
second $\sfrac{1}{2}$\,CZ-$\pi$ gate between Q$^{}_1$ and B.
The dotted black lines connecting the two
$\sfrac{1}{2}$\,CZ-$\pi$ gates indicate qubit shelving to the
$| \textrm{f} \rangle$ state. (\textbf{E}) As in panel B, but
for the M gate. (\textbf{F}) As in C, but for the M gate.
    }
\end{figure}

Finally, by combining the CZ-$\phi$ and zeroing gates, we can
implement a Toffoli-class gate~\cite{barenco:1995:qgates,
lanyon:2009:toffoli,monz:2009:toffoli}, the three-qubit OR
phase gate. This gate, combined with single qubit rotations, is
sufficient for universal computation. A Toffoli gate is a
doubly-controlled quantum operation, where a unitary operation
is applied to a target qubit subject to the state of two
control qubits. The canonical Toffoli is a doubly-controlled
NOT gate; here we consider a doubly-controlled phase gate,
which is equivalent through a change of basis of the target
qubit. In the canonical Toffoli gate, the control gate is
applied if both control qubits, Q$^{}_1$ AND Q$^{}_2$, are in
state $| \textrm{e} \rangle$. In our case, the control gate is
applied conditionally if the controls Q$^{}_1$ OR Q$^{}_2$ are
in $| \textrm{e} \rangle$. Additionally, we have implemented a
three-qubit gate for the logical function XOR, which, even
though not a Toffoli-class gate, helps to understand the more
complex OR gate.

The quantum logic circuits for the XOR and OR gates are drawn
in Fig.~4, A and D. The control qubits are Q$^{}_1$ and
Q$^{}_2$ and the target is the bus resonator B, effectively
acting as the third qubit (as only the states $| 0 \rangle$ and
$| 1 \rangle$ of B are used). The XOR gate is realized as a
series of two CZ-$\pi$ gates between the controls and the
target, and the OR gate as the series $\sfrac{1}{2}$\,CZ-$\pi$,
CZ-$\pi$, and $\sfrac{1}{2}$\,CZ-$\pi$, in an ``M-shape''
configuration.

The truth table for the XOR gate is displayed in Fig.~4B (Top).
The control qubits Q$^{}_1$ and Q$^{}_2$ are assumed to be in
one of the states in $\mathcal{M}^{}_2$, while the target B is
in $| 0 \rangle + | 1 \rangle$. The target acquires a phase
$\pi$, corresponding to a ``true'' result, only when the
controls are in the state $| \textrm{Q}^{}_1 \textrm{Q}^{}_2
\rangle {} = {} | \textrm{g} \textrm{e} \rangle$ or $|
\textrm{e} \textrm{g} \rangle$. For the other non-trivial case
$| \textrm{Q}^{}_1 \textrm{Q}^{}_2 \rangle {} = {} | \textrm{e}
\textrm{e} \rangle$, the target acquires $0$ phase,
corresponding to a ``false'' result. This is due to the action
of the two CZ-$\pi$ gates, giving a global phase $\pi$ when
either of the controls is in $| \textrm{e} \rangle$, and a
phase $2 \pi$ (equivalent to a $0$ phase) when both are in $|
\textrm{e} \rangle$.

The truth table can be experimentally measured by performing
Ramsey experiments on the target, one for each pair of control
states. The experiments are realized by, \textit{(i)},
preparing Q$^{}_2$ in the superposition state $| \textrm{g}
\rangle + | \textrm{e} \rangle$ by means of a $\pi / 2$-pulse;
\textit{(ii)}, moving the state from Q$^{}_2$ into B, thus
creating a $| 0 \rangle + | 1 \rangle$ state in B;
\textit{(iii)}, preparing Q$^{}_1$ and Q$^{}_2$ in each
possible pair of control states in $\mathcal{M}^{}_2$ by means
of $\pi$-pulses; \textit{(iv)}, performing the XOR gate;
\textit{(v)}, zeroing Q$^{}_2$ into Z$^{}_2$ at the end of the
XOR gate; \textit{(vi)}, moving the final target state from B
into the zeroed Q$^{}_2$; \textit{(vii)}, completing the Ramsey
sequence on Q$^{}_2$ with a second $\pi / 2$-pulse with
variable rotation axis relative to the pulse in \textit{(i)}.
The measurement outcomes are displayed in Fig.~4B (Bottom),
together with the least-squares fits used to extract the phase
information associated with each value of the truth table. The
Ramsey fringes for the two control states $| \textrm{g}
\textrm{e} \rangle$ and $| \textrm{e} \textrm{g} \rangle$ are
inverted relative to the reference state $| \textrm{g}
\textrm{g} \rangle$, as expected from the XOR gate truth table.

In general, given the Q$^{}_1$-Q$^{}_2$-B basis set
$\mathcal{M}^{}_3 {} = {} \{ | \textrm{g} \textrm{g} 0 \rangle
, | \textrm{g} \textrm{g} 1 \rangle , | \textrm{g} \textrm{e} 0
\rangle , | \textrm{g} \textrm{e} 1 \rangle , | \textrm{e}
\textrm{g} 0 \rangle , \\ | \textrm{e} \textrm{g} 1 \rangle , |
\textrm{e} \textrm{e} 0 \rangle , | \textrm{e} \textrm{e} 1
\rangle \}$, the vector ${\bm \tau}^{\textrm{XOR}}_{}$ of the
diagonal elements associated with the ideal unitary matrix of
the XOR gate reads
\begin{equation}
 {\bm \tau}^{\textrm{XOR}}_{} {} = {}
 \begin{pmatrix}
  1 , & 1 , & 1 , & -1 , & 1 , & -1 , & 1 , & 1
 \end{pmatrix} \, ,
    \label{Equation:3}
\end{equation}
while all off-diagonal elements of the matrix are zero. Each
element $\tau^{\textrm{XOR}}_k$ can be expressed as a complex
exponential $e^{i \, \phi^{}_{| l m n \rangle}}_{}$, with $| l
m n \rangle {} \in {} \mathcal{M}^{}_3$. The phase $\phi^{}_{|
l m n \rangle}$ can be either $0$, when $\tau^{\textrm{XOR}}_k
{} = {} 1$, or $\pi$, when $\tau^{\textrm{XOR}}_k {} = {} - 1$.
Among the eight values of $\phi^{}_{| l m n \rangle}$, only
seven are physically independent, as the element $e^{i \,
\phi^{}_{| \textrm{g} \textrm{g} 0 \rangle}}_{}$ can be
factored, reducing the set of possible phases to $\phi^{}_{| l
m n \rangle} - \phi^{}_{| \textrm{g} \textrm{g} 0 \rangle}$,
with $| l m n \rangle {} \in {} \mathcal{M}^{}_3 - \{ |
\textrm{g} \textrm{g} 0 \rangle \}$.

In analogy to the truth-table for the target B, a table with
four phase differences can also be obtained for the controls
Q$^{}_1$ and Q$^{}_2$, resulting in a total of twelve phase
differences. These differences can be measured by performing
Ramsey experiments both on the target and the control qubits.
It can be shown that from the twelve phase differences, one can
obtain the seven independent phases associated with the
diagonal elements $\tau^{\textrm{XOR}}_k$~\cite{note:methods},
thus realizing a quantum phase tomography of the Toffoli
gate~\cite{q:phase:tomo}. Figure~4C displays the phase
tomography results for our experimental implementation of the
XOR gate.

The truth table associated with the M gate is reported in
Fig.~4E (Top), where the only difference from the XOR gate is
the phase $\pi$ acquired by the target B when the controls
Q$^{}_1$ and Q$^{}_2$ are loaded in state $| \textrm{Q}^{}_1
\textrm{Q}^{}_2 \rangle {} = {} | \textrm{e} \textrm{e}
\rangle$. In this case, the action of the first
$\sfrac{1}{2}$\,CZ-$\pi$ gate between Q$^{}_1$ and B shelves
the $| 1 \rangle$ state from B to the non-computational state
$| \textrm{f} \rangle$ in Q$^{}_1$, where it remains until the
second $\sfrac{1}{2}$\,CZ-$\pi$ gate. Moving the state of
Q$^{}_1$ outside the computational space during the
intermediate CZ-$\pi$ gate between Q$^{}_2$ and B effectively
turns off the CZ-$\pi$
gate~\cite{ralph:2007:qdittoffoli,lanyon:2009:toffoli}. The
target B thus only acquires a total phase $\pi$ due to the
combined action of the two $\sfrac{1}{2}$\,CZ-$\pi$ gates
(cf.~Fig.~4D). The experimental truth table obtained from
Ramsey fringes is shown in Fig.~4E (Bottom).

The vector ${\bm \tau}^{\textrm{M}}_{}$ of the diagonal
elements associated with the ideal unitary matrix of the M gate
is ${\bm \tau}^{\textrm{M}}_{} {} = {} ( 1 , \, 1 , \, 1 , \,
-1 , \, 1 , \, -1 , \, 1 , \, -1 )$. A similar procedure as for
the XOR gate allows us to obtain the quantum phase tomography
of the M gate (Fig.~4F).

Quantum phase tomography makes it possible to define the phase
fidelity of the XOR and M gate,
\begin{equation}
\mathcal{F}^{}_{\varphi} {} \equiv {} 1 -
\frac{\varepsilon^{}_{\varphi}}{\pi} \, ,
    \label{Equation:4}
\end{equation}
where $\varepsilon^{}_{\varphi}$ is the gate root-mean-square
phase error, with an upper bound of $\pi$. For the XOR gate we
find that $\mathcal{F}^{}_{\varphi} {} = {} 0.954 \pm 0.004$,
and for the M gate $\mathcal{F}^{}_{\varphi} {} = {} 0.979 \pm
0.003$.

% *******************************
% *** Conclusions and outlook ***
% *******************************
%
Our results provide optimism for the near-term implementation
of a larger-scale quantum
processor~\cite{nielsen:2000:qcomputing,mermin:2007:qcomputer,
divincenzo:2000:rules} based on superconducing circuits. Our
architecture shows that proof-of-concept factorization
algorithms~\cite{nielsen:2000:qcomputing,mermin:2007:qcomputer,
vandersypen:2001:shornmr} and simple quantum error correction
codes~\cite{nielsen:2000:qcomputing,mermin:2007:qcomputer,
cory:1998:qerrorcorr,blatt:2008:tions} might be achievable
using this approach.

% ************************
% *** The bibliography ***
% ************************
%

%

% ************************
% *** Acknowledgements ***
% ************************

\NPsection{Acknowledgements}

This work was supported by IARPA under ARO award
W911NF-08-1-0336 and under ARO award W911NF-09-1-0375. M.~M.
acknowledges support from an Elings Postdoctoral Fellowship.
Devices were made at the UC Santa Barbara Nanofabrication
Facility, a part of the NSF-funded National Nanotechnology
Infrastructure Network. The authors thank A.~G.~Fowler for
useful comments on scalability, and M.~H.~Devoret and
R.~J.~Schoelkopf for discussions on Toffoli gates.\\

% ****************************
% *** Author contributions ***
% ****************************

\NPsection{Author Contributions}

M.M. performed the experiments and analyzed the data. M.M. and
H.W. fabricated the sample. T.Y., H.W., and Y.Y. helped with
the Fourier transform and M.N. with three-qubit gates. M.M.,
A.N.C., and J.M.M. conceived the experiment and co-wrote the
manuscript.
    \clearpage

% ******************************
% *** Supplementary Material ***
% ******************************

\setcounter{figure}{0}

\topmargin 0.0cm \oddsidemargin 0.2cm \textwidth 16cm
\textheight 21cm \footskip 1.0cm

\singlespacing

\renewcommand\refname{Supporting References}

\renewcommand{\thefigure}{S\arabic{figure}}
\renewcommand{\thetable}{S\arabic{table}}
\renewcommand{\theequation}{S\arabic{equation}}

\title{
Supplementary Material for
    \\
    \vspace{10.0mm}
\Large{\textbf{Implementing the Quantum von Neumann
Architecture with Superconducting Circuits}} }

    \vspace{10.0mm}

\author{
Matteo Mariantoni,$^{1,4,\S}$ H.~Wang,$^{1}$\footnote{Present
address: Department of Physics, Zhejiang University, Hangzhou
310027, China.}\, T.~Yamamoto,$^{1,2}$
M.~Neeley,$^{1}$\footnote{Present address:
Lincoln Laboratory, Massachusetts Institute of Technology, 244 Wood Street, Lexington, MA 02420-9108, USA.} \\
Radoslaw C.~Bialczak,$^{1}$ Y.~Chen,$^{1}$ M.~Lenander,$^{1}$ Erik Lucero,$^{1}$ A.~D.~O'Connell,$^{1}$ \\
D.~Sank,$^{1}$ M.~Weides,$^{1}$\footnote{Present address:
National Institute of Standards and Technology, Boulder, CO
80305, USA.}\,
J.~Wenner,$^{1}$ Y.~Yin,$^{1}$ J.~Zhao,$^{1}$ \\
A.~N.~Korotkov,$^{3}$ A.~N.~Cleland,$^{1,4}$ John M.~Martinis$^{1,4,\S}$ \\
 \\
\normalsize{$^{1}$Department of Physics, University of California, Santa Barbara, CA 93106-9530, USA} \\
\normalsize{$^{2}$Green Innovation Research Laboratories, NEC Corporation, Tsukuba, Ibaraki 305-8501, Japan} \\
\normalsize{$^{3}$Department of Electrical Engineering, University of California, Riverside, CA 92521, USA} \\
\normalsize{$^{4}$California NanoSystems Institute, University of California,} \\
\normalsize{Santa Barbara, California 93106-9530, USA} \\
    \\
\normalsize{$^\S$To whom correspondence should be addressed. E-mail: matmar@physics.ucsb.edu (M.~M.);} \\
\normalsize{martinis@physics.ucsb.edu (J.~M.~M.)}
    }

\date{}
    \clearpage

\baselineskip24pt

\maketitle

\singlespacing

\noindent \textbf{This PDF file includes:}
    \vspace{1.0mm} \\

    \noindent \hspace{10.0mm} Materials and Methods

    \noindent \hspace{10.0mm} Figs. S1 to S12

    \noindent \hspace{10.0mm} Tables S1 to S3

    \noindent \hspace{10.0mm} References
    \clearpage

\doublespacing

\tableofcontents
    \clearpage

\singlespacing

% *****************************
% *** Materials and Methods ***
% *****************************
%
\section*{Materials and Methods}
    \addcontentsline{toc}{section}{Materials and Methods}

% **************************
% *** Statistical errors ***
% **************************
%
\subsection*{Statistical errors}
    \addcontentsline{toc}{subsection}{Statistical errors}

In this section, we analyze the statistical properties of the
experimental data shown in the main text. First, we explain how
to simulate statistical errors. This procedure was used to
estimate the confidence intervals for the data of Fig.~2 in the
main text. Second, we describe how statistical errors were
obtained from statistical ensembles of independent
measurements. This procedure was used for the data of Fig.~3 in
the main text. Third, we discuss the estimation of statistical
errors due to fits to the data. This procedure was used for the
data of Fig.~4 in the main text.

% ****************************************
% *** Simulation of statistical errors ***
% ****************************************
%
\subsubsection*{Simulation of statistical errors}
    \addcontentsline{toc}{subsubsection}{Simulation of statistical errors}

In this subsection, we discuss two important sources of
statistical errors in our data: Errors associated with qubit's
measurement (binomial-type errors) and errors due to
jitter/fluctuations in the electronics (phase errors). Assuming
binomial-type and phase errors, we describe the procedures used
to simulate the confidence intervals for the elements and
metrics of the density matrices shown in Fig.~2C of the main
text.

% ******************************
% *** Supplementary Figure 1 ***
% ******************************
%
\begin{figure}[t!]
    \centering
    \includegraphics[width=1.14\columnwidth]{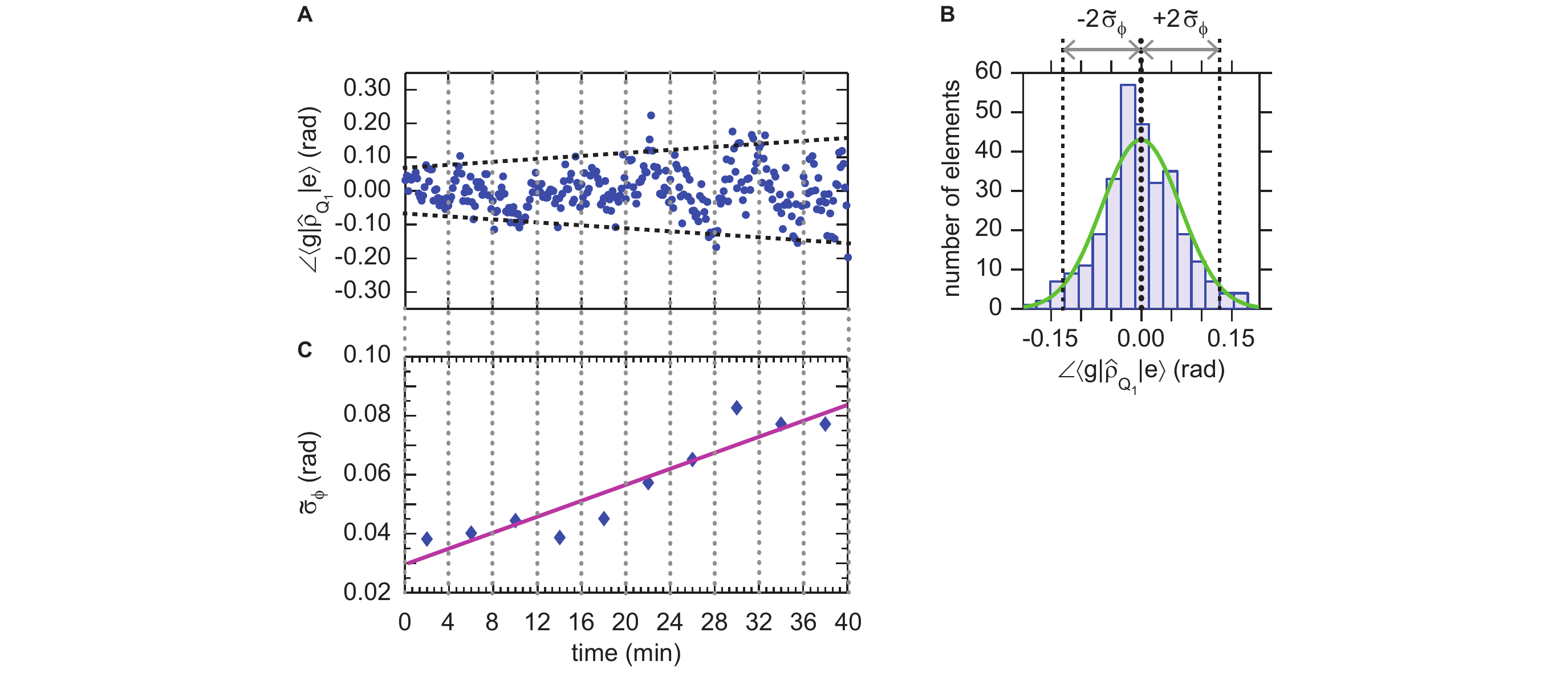}
    \caption{\footnotesize
\textbf{Analysis of phase errors.} (\textbf{A}) Phase angle
$\angle \, \langle \textrm{g} | \hat{\rho}^{}_{\textrm{Q}^{}_1}
| \textrm{e} \rangle$ associated with the off-diagonal elements
of the matrix $\hat{\rho}^{}_{\textrm{Q}^{}_1}$ of
Eq.~\ref{Equation:Supporting:3} plotted vs. time. The time axis
indicates when the QST of each density matrix
$\hat{\rho}^{}_{\textrm{Q}^{}_1}$ was completed. The dashed
black lines are a guide-to-the-eye showing an increase with
time in the data scatter. (\textbf{B}) Histogram associated
with the time-trace data in A, plotting the number of elements
in the time-trace vs. the phase angle $\angle \, \langle
\textrm{g} | \hat{\rho}^{}_{\textrm{Q}^{}_1} | \textrm{e}
\rangle$. The solid green line is a fit to a normal
distribution with mean value of $0$\,rad and standard deviation
$\tilde{\sigma}^{}_{\phi} {} \simeq {} 0.065$\,rad. The $\pm 2
\tilde{\sigma}^{}_{\phi}$ window is indicated. (\textbf{C})
Time-bin average of the data in A, showing the value of
$\tilde{\sigma}^{}_{\phi}$ for each time-bin of $4$\,min, for a
total of $10$ bins (blue diamonds). The bins are indicated by
vertical dotted grey lines, which extend to A for clarity. The
solid magenta line is a linear fit to the data. This fit was
used to estimate the phase errors associated with QST
measurements (tomo and octomo).
    }
    \label{Figure:Supporting:1:Matteo:Mariantoni:201107}
\end{figure}

    \renewcommand{\labelenumi}{\textit{(\roman{enumi})}}
\begin{enumerate}

\item Binomial-type errors are inherent to our qubit
    measurement process, where the measurement is repeated
    a fixed number of times $N$, each measurement trial has
    two possible outcomes, i.e., qubit being in the ground
    state $| \textrm{g} \rangle$ with probability
    $p^{}_{\textrm{g}}$ or in the excited state $|
    \textrm{e} \rangle$ with probability $p^{}_{\textrm{e}}
    {} = {} 1 - p^{}_{\textrm{g}}$, the probability
    $p^{}_{\textrm{e}}$ is to good approximation the same
    for each trial, and the trials can be considered to be
    statistically independent. The measurement outcome
    associated with $| \textrm{g} \rangle$ is counted as
    $0$, and that associated with $| \textrm{e} \rangle$ as
    $1$. Under these assumptions, the qubit measurement
    process can be described by a binomial distribution.

    Given a statistical sample $X^N_{}$ consisting of $N$
    measurement outcomes (i.e., a statistical sample
    $X^N_{}$ from a Bernoulli distribution with parameter
    $p^{}_{\textrm{e}}$), the maximum likelihood estimator
    of $p^{}_{\textrm{e}}$ (i.e., the estimated
    probability) is given by
    \begin{equation}
    P^{}_{\textrm{e}} {} = {} \bar{X}^N_{} {} = {}
    \frac{1}{N} \, \sum_{k {} = {} 1}^N \, X^k_{} \, ,
        \label{Equation:Supporting:1}
    \end{equation}
    where $X^k_{}$ represents the $k$-th outcome among the
    $N$ measured. There are several ways to compute a
    confidence interval for the parameter
    $p^{}_{\textrm{e}}$. The most common result is based on
    the approximation of the binomial distribution with a
    normal distribution. This represents a good
    approximation in our experiments, where the number of
    measurements $N$ is large (typically $N {} \geqslant {}
    1500$). In this case, it can be shown that a confidence
    interval for the parameter $p^{}_{\textrm{e}}$ is given
    by
    \begin{equation}
    P^{}_{\textrm{e}} \pm z^{}_{( 1 - \alpha / 2 )} \,
    \sqrt{\frac{P^{}_{\textrm{e}} ( 1 - P^{}_{\textrm{e}}
    )}{N}} {} = {} P^{}_{\textrm{e}} \pm z^{}_{( 1 - \alpha /
    2 )} \, \tilde{\sigma}^{}_{\textrm{b}} \, ,
        \label{Equation:Supporting:2}
    \end{equation}
    where $z^{}_{( 1 - \alpha / 2 )}$ is the $( 1 - \alpha
    / 2 )$ percentile of a standard normal distribution.
    For example, for a $0.95$ ($95$\%) confidence interval,
    we set $\alpha {} = {} 0.05$, so that $z^{}_{( 1 -
    \alpha / 2 )} {} = {} 1.96$. When analyzing our data we
    approximate the percentile $1.96$ with $2$, thus
    obtaining a slightly wider confidence interval;

\item Phase errors are mostly due to the phase
    jitter/fluctuations in the room-temperature cables and
    electronics used to measure the qubits. In order to
    quantify such errors, the following experiment was
    performed. First, we initialized one of the two qubits,
    e.g., qubit Q$^{}_1$, in the ground state, $|
    \textrm{Q}^{}_1 \rangle {} = {} | \textrm{g} \rangle$;
    second, we applied to Q$^{}_1$ a $\pi / 2$ unitary
    rotation about the $y$-axis, $\hat{R}^{\pi / 2}_y$,
    bringing the qubit into the state $| \textrm{Q}^{}_1
    \rangle {} = {} ( | \textrm{g} \rangle + | \textrm{e}
    \rangle ) / \sqrt{2}$. This state is characterized by
    the density matrix
    \begin{equation}
    \hat{\rho}^{}_{\textrm{Q}^{}_1} {} = {} \frac{1}{2} \,
        \begin{pmatrix}
            1 & 1 \\
            1 & 1 \\
        \end{pmatrix} \, ,
            \label{Equation:Supporting:3}
    \end{equation}
    which represents a ``phase-sensitive'' state due to the
    presence of nonzero off-diagonal elements, thus
    allowing us to measure the phase properties of our
    setup. In fact, if the setup (cables and electronics)
    were ideal, the phase $\angle \, \langle \textrm{g} |
    \hat{\rho}^{}_{\textrm{Q}^{}_1} | \textrm{e} \rangle {}
    = {} - \angle \, \langle \textrm{e} |
    \hat{\rho}^{}_{\textrm{Q}^{}_1} | \textrm{g} \rangle$
    associated with the off-diagonal elements of the matrix
    $\hat{\rho}^{}_{\textrm{Q}^{}_1}$ of
    Eq.~\ref{Equation:Supporting:3} would be zero. We can
    thus assume that any deviation from a zero phase
    corresponds to a phase error; third, we performed a
    single-qubit quantum state tomography (QST) on
    Q$^{}_1$, making possible to measure experimentally
    $\hat{\rho}^{}_{\textrm{Q}^{}_1}$. Using our typical
    settings for a single-qubit
    QST~\cite{steffen:2006:sqqst}, the time needed for each
    QST was approximately $8$\,s; fourth, we repeated a QST
    measurement every $8$\,s for a total time of
    $40$\,minutes, corresponding to $300$ measured density
    matrices; finally, we plotted $\angle \, \langle
    \textrm{g} | \hat{\rho}^{}_{\textrm{Q}^{}_1} |
    \textrm{e} \rangle$ as a function of time. The
    so-obtained time trace is shown in
    Fig.~\ref{Figure:Supporting:1:Matteo:Mariantoni:201107}A.
    Besides negligible slow-varying oscillations in the
    time trace [independent tests have shown that these
    oscillations might be due to temperature changes in the
    room-temperature cables (data not shown)], the overall
    histogram associated with the trace is approximately
    normally distributed about a mean value of $0$\,rad,
    with standard deviation $\tilde{\sigma}^{}_{\phi} {}
    \simeq {} 0.065$\,rad
    (cf.~Fig.~\ref{Figure:Supporting:1:Matteo:Mariantoni:201107}B).
    However, we notice a general increase in the scatter of
    the time-trace data, as indicated by the dashed black
    lines in
    Fig.~\ref{Figure:Supporting:1:Matteo:Mariantoni:201107}A.
    We thus divide the time trace in $10$ sub-traces (time
    bins) with a time length of $4$\,min each, compute the
    standard deviation for each sub-trace, and plot the
    so-obtained $10$ standard deviations as a function of
    time. The result is displayed in
    Fig.~\ref{Figure:Supporting:1:Matteo:Mariantoni:201107}C,
    where the data is overlayed with a linear fit.

\end{enumerate}

The plot of
Fig.~\ref{Figure:Supporting:1:Matteo:Mariantoni:201107}C is
useful in determining the phase errors associated with
different types of two-qubit QST, as well as quantum process
tomography (QPT)~\cite{nielsen:2000:qcomputing,kofman:2009:qpt,
bialczak:2010:qpt,yamamoto:2010:czgates}. In fact, two-qubit
QST can be realized either by applying to each qubit the set of
three unitary operations $\{ \hat{I} , \hat{R}^{\pi / 2}_x ,
\hat{R}^{\pi / 2}_y \}$ ($\hat{I}$ is the $2 \times 2$ identity
matrix, $\hat{R}^{\pi / 2}_x$ a $\pi / 2$ unitary rotation
about the $x$-axis, and $\hat{R}^{\pi / 2}_y$ a $\pi / 2$
unitary rotation about the $y$-axis), which we call ``tomo,''
or the set of six unitary operations $\{ \hat{I} , \hat{R}^{\pi
/ 2}_x , \hat{R}^{\pi / 2}_y , \hat{R}^{- \pi / 2}_x ,
\hat{R}^{- \pi / 2}_y , \hat{R}^{\pi}_x \}$ ($\hat{R}^{- \pi /
2}_x$ is a $- \pi / 2$ unitary rotation about the $x$-axis,
$\hat{R}^{- \pi / 2}_y$ a $- \pi / 2$ unitary rotation about
the $y$-axis, and $\hat{R}^{\pi}_x$ a $\pi$ unitary rotation
about the $x$-axis), which we call ``octomo.''

In the case of two-qubit tomo, the number of operations that
must be applied to the pair of qubits is given by the
permutations of the allowed set of unitary operations, $3^2_{}
{} = {} 9$. This number multiplied by the $4$ possible joint
probabilities for a two-qubit system, $p^{}_{\textrm{gg}} ,
p^{}_{\textrm{ge}} , p^{}_{\textrm{eg}}$, and
$p^{}_{\textrm{ee}}$ (where, e.g., $p^{}_{\textrm{ge}}$ is the
probability to measure the first qubit in the ground state with
the second qubit in the excited state) gives a total of $36$
probabilities. In the case of octomo, the total number of
probabilities is given by the permutations of $6$ unitary
operations for $2$ qubits, $6^2_{} {} = {} 36$, times the $4$
possible joint probabilities for a two-qubit system, for a
total of $144$ probabilities.

In the experiments, the maximum likelihood estimator for each
of the four probabilities $p^{}_{\textrm{gg}} ,
p^{}_{\textrm{ge}} , p^{}_{\textrm{eg}}$, and
$p^{}_{\textrm{ee}}$ is obtained from the outcome of $N$
measurements. We note that, in a joint two-qubit measurement
each outcome consists of $4$ numbers obtained simultaneously,
where each number can be either $0$ or $1$. The statistical
sample consisting of $N$ two-qubit joint measurements will be
hereafter defined as $X^N_{lm}$, with $l , m {} = {} \textrm{g}
, \textrm{e}$. Similarly to Eq.~\ref{Equation:Supporting:1},
the maximum likelihood estimator (i.e., the estimated
probability) for each of the four probabilities
$p^{}_{\textrm{gg}} , p^{}_{\textrm{ge}} , p^{}_{\textrm{eg}}$,
and $p^{}_{\textrm{ee}}$ can thus be obtained from
\begin{equation}
P^{}_{lm} {} = {} \frac{1}{N} \, \sum_{k {} = {} 1}^N \,
X^k_{lm} \, ,
    \label{Equation:Supporting:4}
\end{equation}
where $X^k_{lm}$ represents the $k$-th outcome among the $N$
measured.

For a given $k$, the four possible $X^k_{lm}$, i.e.,
$X^k_{\textrm{gg}}$, $X^k_{\textrm{ge}}$, $X^k_{\textrm{eg}}$,
and $X^k_{\textrm{ee}}$, are measured simultaneously (with
$X^k_{\textrm{gg}} + X^k_{\textrm{ge}} + X^k_{\textrm{eg}} +
X^k_{\textrm{ee}} {} = {} 1$). Hence, the effective number of
events that has to be measured for each tomo is $36 / 4 {} = {}
9$, and for each octomo $144 / 4 {} = {} 36$.

We typically measure $2 500$ events per second, and repeat each
measurement $N {} = {} 15 000$ times. As a consequence, a
two-qubit tomo takes approximately $1$\,min, and a two-qubit
octomo approximately $4$\,min.

All data displayed in Fig.~2C of the main text were obtained
using tomo, while all data in Fig.~3, D and E, were obtained
using octomo. All density matrices used to reconstruct the
$\chi$ matrix of Fig.~3F in the main text were also obtained
with octomo. The standard deviation due to phase errors can be
estimated in each case by looking up the fit in
Fig.~\ref{Figure:Supporting:1:Matteo:Mariantoni:201107}C.

Considering for example a two-qubit octomo with $N {} = {} 15
000$, the statistical properties of the resulting density
matrix $\hat{\rho}$ and of the corresponding metrics [fidelity
$\mathcal{F}$, negativity $\mathcal{N}$, concurrence
$\mathcal{C}$, and entanglement of formation $\mathcal{E}$;
cf.~Ref.~\cite{horodecki:2009:entanglement} and references
therein for an extensive description of these metrics] are
obtained as follows:

    \renewcommand{\labelenumi}{(\arabic{enumi})}
\begin{enumerate}

\item The probabilities $P^{}_{lm}$ associated with
    two-qubit octomo are estimated according to
    Eq.~\ref{Equation:Supporting:4}. As explained above,
    this corresponds to a total of $36 \times 4 {} = {}
    144$ estimated probabilities. To simplify the notation,
    we will hereafter refer to these probabilities as
    $P^{}_i$, with $i {} \in {} \{ 1 , 2 , \ldots , 144
    \}$;

\item The estimated probabilities $P^{}_i$ are corrected
    for measurement errors
    [cf.~Refs.~\cite{mariantoni:2011:shell} and
    \cite{steffen:2006:tqqst} for our standard procedures
    to correct for measurement errors in the case of one
    and two qubits, respectively]. The corrected
    probabilities $P^{}_i$ are stored as a $144 \times 1$
    column vector;

\item For each probability $P^{}_i$, the binomial standard
    deviation $\tilde{\sigma}^{}_{\textrm{b}}$ defined in
    Eq.~\ref{Equation:Supporting:2} is calculated, thus
    obtaining, in the case of octomo, $144$ different
    standard deviations;

% ******************************
% *** Supplementary Figure 2 ***
% ******************************
%
\begin{figure}[t!]
    \centering
    \includegraphics[width=1.14\columnwidth]{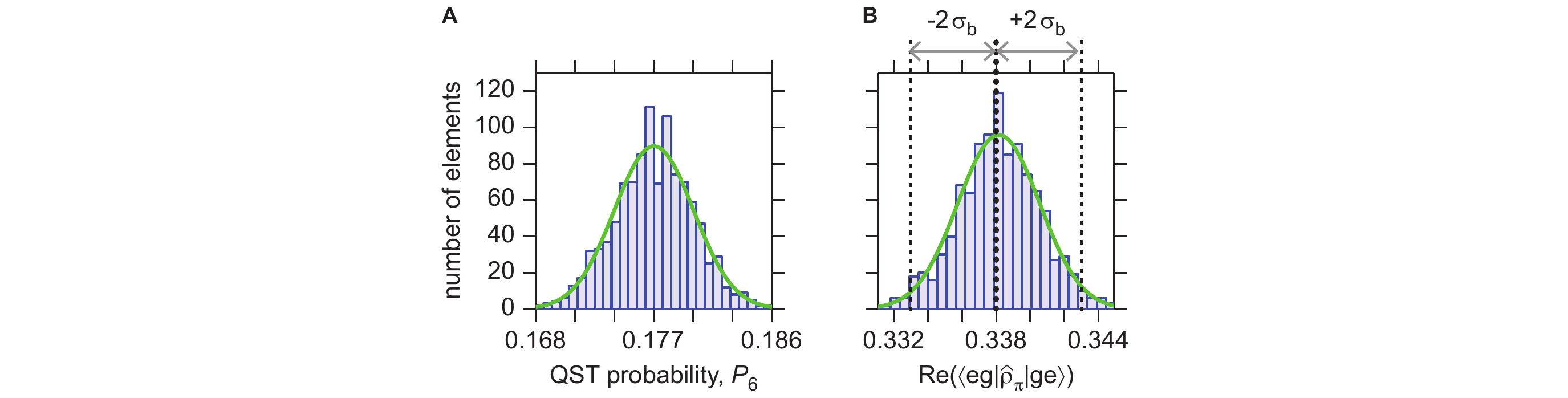}
    \caption{\footnotesize
\textbf{Confidence intervals for a density matrix and its
metrics.} (\textbf{A}) Histogram associated with the $6$-th
probability $P^{}_6$ of the vector of probabilities
$P^{}_i$, plotting the number of elements among the $M$
probabilities obtained in point~($4$) vs. the corresponding
value of the probability $P^{}_6$. The data refers to the
octomo for the state $\hat{\rho}^{}_{\pi}$ of Fig.~3E in
the main text. The solid green line is a fit to a normal
distribution. (\textbf{B}) Histogram for the real part of
the matrix element with mean value $\langle \textrm{eg} |
\hat{\rho}^{}_{\pi} | \textrm{ge} \rangle {} = {} 0.338$
for the state $\hat{\rho}^{}_{\pi}$ of Fig.~3E in the main
text. The solid green line is a fit to a normal
distribution. The $\pm 2 \sigma^{}_{\textrm{b}}$ window is
indicated, where $\sigma^{}_{\textrm{b}}$ is one standard
deviation.
    }
    \label{Figure:Supporting:2:Matteo:Mariantoni:201107}
\end{figure}

\item For each of the $144$ standard deviations
    $\tilde{\sigma}^{}_{\textrm{b}}$ calculated in ($3$), a
    set of $M$ random numbers picked from a normal
    distribution with zero mean value and standard
    deviation $\tilde{\sigma}^{}_{\textrm{b}}$ is
    generated. This results in a matrix of $144 \times M$
    random numbers. Typically, $M {} = {} 1 000$.

    By summing each column of such a matrix to the column
    vector containing the $144$ estimated probabilities
    $P^{}_i$, we obtain a matrix of $144 \times M$
    probabilities, where each column simulates the result
    of a different QST experiment.

    For example,
    Fig.~\ref{Figure:Supporting:2:Matteo:Mariantoni:201107}A
    shows the histogram associated with the $6$-th
    probability $P^{}_6$ of the vector of probabilities
    $P^{}_i$ in the case of the octomo for the state
    $\hat{\rho}^{}_{\pi}$ of Fig.~3E in the main text;

\item Each column of the $144 \times M$ matrix of
    probabilities obtained in point~($4$) is inverted by
    following the usual QST
    rules~\cite{steffen:2006:sqqst,steffen:2006:tqqst}.
    This allows us to find the corresponding density matrix
    $\hat{\rho}^{\textrm{unphys}}_j$, with $j {} \in {} \{
    1 , 2 , \ldots , M \}$, thus obtaining $M$ density
    matrices associated with one state;

\item Physicality constraints are enforced on each,
    generally unphysical, density matrix
    $\hat{\rho}^{\textrm{unphys}}_j$ by means of the MATLAB
    packages SeDuMi~$1.21$ and YALMIP (semidefinite
    programming)~\cite{semidefinite:programming}. The
    physical constraints are such that each final -
    physical - density matrix $\hat{\rho}^{}_j$ should have
    unit trace and be positive semidefinite.

    In order to obtain the mean physical density matrix
    $\hat{\rho}$ associated with the $M$ physical density
    matrices $\hat{\rho}^{}_j$ and the corresponding
    standard deviations, we calculate the mean value and
    standard deviation of the real and imaginary part of
    each matrix element for the $M$ matrices
    $\hat{\rho}^{}_j$. The mean physical matrix
    $\hat{\rho}$ will thus have elements $\langle lm |
    \hat{\rho} | pq \rangle$ (with $| lm \rangle , | pq
    \rangle {} \in {} \mathcal{M}^{}_2$), each of them
    (real and imaginary part) characterized by a given
    standard deviation.
    Figure~\ref{Figure:Supporting:2:Matteo:Mariantoni:201107}B
    shows the histogram for the real part of the matrix
    element with mean value $\langle \textrm{eg} |
    \hat{\rho}^{}_{\pi} | \textrm{ge} \rangle {} = {}
    0.338$ for the state $\hat{\rho}^{}_{\pi}$ of Fig.~3E
    in the main text. As expected, the distribution is
    approximately Gaussian with a $0.95$ confidence
    interval $\pm {} 2 \sigma^{}_{\textrm{b}} {} = {} \pm
    0.005$.

    The knowledge of the $M$ matrices $\hat{\rho}^{}_j$
    also allows us to estimate the confidence intervals for
    the relevant metrics characterizing the state
    $\hat{\rho}$: $\mathcal{F}$, $\mathcal{N}$,
    $\mathcal{C}$, and $\mathcal{E}$. This can easily be
    accomplished by calculating the metrics for each
    $\hat{\rho}^{}_j$, thus obtaining $M$ values for each
    metric, and then computing the mean value and standard
    deviation of the $M$ values associated with each
    metric.

\end{enumerate}

We can follow a similar procedure to account for phase errors.
We now pick two independent sets of $M$ random numbers from a
normal distribution with zero mean value and standard deviation
$\tilde{\sigma}^{}_{\phi}$ (with $\tilde{\sigma}^{}_{\phi}$
opportunely estimated from
Fig.~\ref{Figure:Supporting:1:Matteo:Mariantoni:201107}C
depending on whether a tomo or octomo was used), thus
generating two sets of $M$ phase errors, $\phi^j_1$ and
$\phi^j_2$, with $j {} \in {} \{ 1 , 2 , \ldots , M \}$. In
order to simulate phase errors acting independently on each
qubit, we apply the unitary rotation
\begin{equation}
U^{}_j {} = {}
    \begin{pmatrix}
        1 & 0 & 0 & 0 \\
        0 & e^{i \phi^j_1} & 0 & 0 \\
        0 & 0 & e^{i \phi^j_2} & 0 \\
        0 & 0 & 0 & e^{i ( \phi^j_1 + \phi^j_2 )} \\
    \end{pmatrix}
        \label{Equation:Supporting:5}
\end{equation}
to a $4 {} \times {} 4$ measured density matrix
$\hat{\rho}^{\textrm{meas}}_{}$, thus obtaining the $j$-th
unphysical density matrix
\begin{equation}
\hat{\rho}^{\textrm{unphys}}_j {} = {} U^{}_j \,
\hat{\rho}^{\textrm{meas}}_{} \, U^{\dag}_j \, .
    \label{Equation:Supporting:6}
\end{equation}
We can then proceed as in step~(6) above and obtain a mean
physical density matrix $\hat{\rho}$ and its statistical
properties, as in the case of binomial-type errors. This allows
us also to find the metrics associated with $\hat{\rho}$ and
their statistical properties. Notice that the unitary
transformation of Eq.~\ref{Equation:Supporting:5} simulates
random rotations along the $z$-axis of both qubit $Q^{}_1$ and
qubit $Q^{}_2$.

The total mean physical density matrix is finally obtained by
averaging the mean physical density matrix obtained in the case
of binomial-type errors and the matrix obtained in the case of
phase errors. The same applies to the mean values of all
metrics. The corresponding standard deviations are found by
summing in quadrature the values obtained in the case of
binomial-type and phase errors. For example, the numerical
value with confidence interval of each element of the density
matrices in Fig.~2C of the main text were obtained following
this procedure. These numbers are reported in
Table~\ref{Table:Supporting:1:Matteo:Mariantoni:201107}.

Incidentally, we found that phase errors do not add any
significant contribution to the confidence intervals of the
density matrix elements and of their metrics.

Notice that, the reason why we decided to simulate the
statistical properties of the data in Fig.~2 of the main text
is because we only had $2$ independent measurements of these
data. Such a statistical ensemble is obviously insufficient to
obtain reliable confidence intervals, which, thus, needed to be
simulated.

% *****************************************************
% *** Experimental estimation of statistical errors ***
% *****************************************************
%
\subsubsection*{Experimental estimation of statistical errors}
    \addcontentsline{toc}{subsubsection}{Experimental estimation of statistical errors}

In the case of the density matrices in Fig.~3E and of the
$\chi^{\textrm{p}}_{\textrm{m}}$ matrix of the quantum Fourier
transform in Fig.~3F of the main text we had ensembles of
independent measurements large enough to allow the confidence
intervals estimation directly from the data.

In particular, the density matrix $\hat{\rho}^{}_{0.28}$ in the
left panel of Fig.~3E is the average of a statistical ensemble
of $M {} = {} 10$ independent measurements, and the density
matrices $\hat{\rho}^{}_{\pi / 2}$ and $\hat{\rho}^{}_{\pi}$ in
the center and right panels of Fig.~3E, respectively, are the
average of an ensemble of $M {} = {} 6$ independent
measurements. The standard deviation of each matrix element
(real and imaginary part) as well as the mean value and
standard deviation of all metrics can easily be estimated from
such statistical ensembles.

Finally, the matrix $\chi^{\textrm{p}}_{\textrm{m}}$ of Fig.~3F
is the average of an ensemble of $15$ independent measurements.
This allows us to estimate the mean value and standard
deviation of the process fidelity $\mathcal{F}^{}_{\chi}$
associated with the quantum Fourier transform (cf.~main text).

% ***********************************************
% *** Statistical errors of fitted parameters ***
% ***********************************************
%
\subsubsection*{Statistical errors of fitted parameters}
    \addcontentsline{toc}{subsubsection}{Statistical errors of fitted parameters}

The confidence intervals associated with the quantum phase
tomography data shown in Fig.~4, C and F, of the main text are
dominated by the statistical errors of the coefficients fitted
from the data in Fig.~4, B and E, of the main text. In
particular, the coefficient of interest is the phase of each
curve in Fig.~4, B and E (or, more in general, of each curve in
Fig.~\ref{Figure:Supporting:12:Matteo:Mariantoni:201107}, C and
D).

We remind that the error vector associated with the vector of
coefficients fitted to a curve is given by the square root of
the vector $\vec{S}$ of the diagonal elements from the
estimated covariance matrix of the coefficient estimates, $(
\vec{X}^T_{} \vec{X}^{}_{} )^{- 1}_{} \, \langle s
\rangle^2_{}$. Here, $\vec{X}^{}_{}$ is the Jacobian of the
fitted values with respect to the coefficients, $\vec{X}^T_{}$
is the transpose of $\vec{X}^{}_{}$, and $\langle s
\rangle^2_{}$ is the mean squared error. This procedure allows
us to estimate the errors associated with the fitted phases.
These errors propagate through the quantum tomography process
(cf.~section on ``Quantum phase tomography'' at the end of
these Methods), finally turning into the confidence intervals
reported in Fig.~4, C and F, of the main text.

% ***********************************************
% *** Definition of the qubit reference frame ***
% ***********************************************
%
\subsection*{Definition of the qubit reference frame}
    \addcontentsline{toc}{subsection}{Definition of the qubit reference frame}

In this section, we briefly explain the concepts of reference
frame and reference clock rate associated with a qubit. These
concepts will be useful in understanding the dynamic phases
acquired by the qubits when programming the quantum von Neumann
architecture as well as the sequences used to tune up the
CZ-$\phi$ gates and the XOR and M gate.

In the two-level approximation~\cite{two:level:approximation},
the Hamiltonian of a phase qubit can be written as
    \setlength\arraycolsep{0pt}
\begin{equation}
\widehat{H}^{}_{\textrm{Q}} {} = {} h \frac{f^{}_{\textrm{Q}} (
z )}{2} \hat{\sigma}^{}_z \, ,
    \label{Equation:Supporting:7}
\end{equation}
with ground state $| \textrm{g} \rangle$ and excited state $|
\textrm{e} \rangle$, and eigenenergies $E^{}_{\textrm{g}}$ and
$E^{}_{\textrm{e}}$, respectively. In
Eq.~\ref{Equation:Supporting:7}, $f^{}_{\textrm{Q}} ( z ) {}
\equiv {} \Delta E^{}_{\textrm{eg}} / h {} = {} (
E^{}_{\textrm{e}} - E^{}_{\textrm{g}} ) / h$ represents the
qubit transition frequency, which can be tuned by means of
z-pulses with amplitude $z$, and $\hat{\sigma}^{}_z$ is the
usual spin $1 / 2$ Pauli operator. At the beginning of a
CZ-$\phi$ gate, each qubit is initialized in $| \textrm{g}
\rangle$ at the so-called idle point, which corresponds to a
z-pulse amplitude $z {} = {} 0$. The qubit transition frequency
at the idle point is thus given by $f^{}_{\textrm{Q}} ( z {} =
{} 0 ) {} \equiv {} f^0_{\textrm{Q}}$.

In order to prepare a qubit in the excited state $| \textrm{e}
\rangle$ or in a linear superposition $| \textrm{g} \rangle + |
\textrm{e} \rangle$, the qubit has to be driven by a microwave
pulse. The Hamiltonian governing the interaction between the
qubit and the microwave driving is given
by~\cite{simga:z:driving}
\begin{equation}
\widehat{H}^{}_{\textrm{D}} {} = {} h \Omega^{}_{\textrm{D}} (
\tau ) \hat{\sigma}^{}_y \sin ( 2 \pi f^{}_{\textrm{D}} \tau +
\phi^{}_{\textrm{delay}} ) \, ,
    \label{Equation:Supporting:8}
\end{equation}
where $\Omega^{}_{\textrm{D}} ( \tau )$ is the time-dependent
driving amplitude expressed in unit hertz, $f^{}_{\textrm{D}}$
the driving frequency, $\hat{\sigma}^{}_y$ the usual spin $1 /
2$ Pauli operator, $\tau$ the time, and
$\phi^{}_{\textrm{delay}}$ an arbitrary phase delay. By
calibrating the microwave pulse such that the phase delay
$\phi^{}_{\textrm{delay}} {} = {} \pi / 2$, we can rewrite the
driving Hamiltonian as
\begin{equation}
\widehat{H}^{}_{\textrm{D}} {} = {} h \Omega^{}_{\textrm{D}} (
\tau ) \hat{\sigma}^{}_y \cos ( 2 \pi f^{}_{\textrm{D}} \tau )
\, .
    \label{Equation:Supporting:9}
\end{equation}
By combining the qubit Hamiltonian of
Eq.~\ref{Equation:Supporting:7} and the driving Hamiltonian of
Eq.~\ref{Equation:Supporting:9}, we obtain the total
Hamiltonian of the driven system, $\widehat{H}^{}_{\textrm{QD}}
{} = {} \widehat{H}^{}_{\textrm{Q}} +
\widehat{H}^{}_{\textrm{D}}$.

In our experiments the driving frequency $f^{}_{\textrm{D}}$ is
a fixed parameter that is set to be equal to the qubit
transition frequency at the idle
point~\cite{carrier:frequency},
\begin{displaymath}
f^{}_{\textrm{D}} {} = {} f^0_{\textrm{Q}} \, .
\end{displaymath}
For a given qubit, the microwave driving represents the
\textit{reference frame} associated with that qubit, with
\textit{reference clock rate} given by $f^0_{\textrm{Q}}$.
Defining the detuning between the z-dependent qubit transition
frequency $f^{}_{\textrm{Q}} ( z )$ and the reference clock
rate $f^0_{\textrm{Q}}$ as $\Delta ( z ) {} \equiv {}
f^{}_{\textrm{Q}} ( z ) - f^0_{\textrm{Q}}$, the qubit-driving
Hamiltonian $\widehat{H}^{}_{\textrm{QD}}$ can be expressed in
the uniformly rotating reference frame by applying the unitary
rotation $\widehat{D} {} = {} e^{+ i 2 \pi f^0_{\textrm{Q}}
\tau \hat{\sigma}^{}_z / 2}_{}$~\cite{cohentannoudji:1977:qm}.
The rotated Hamiltonian is thus given by
\begin{eqnarray}
\widehat{\widetilde{H}}^{}_{\textrm{QD}} & {} = {} &
\widehat{D}^{}_{} \, \widehat{H}^{}_{\textrm{QD}}
\widehat{D}^{\dag}_{} - i \, \hbar \widehat{D} \, \frac{d}{d
\tau} \, \widehat{D}^{\dag}_{} {} \nonumber\\[1.5mm]
& {} \approx {} & - h \frac{\Delta ( z )}{2} \hat{\sigma}^{}_z
+ h \frac{\Omega^{}_{\textrm{D}} ( \tau )}{2} \hat{\sigma}^{}_y
\, ,
    \label{Equation:Supporting:10}
\end{eqnarray}
where the counter-rotating terms have been already neglected.
The dynamics associated with the pulse sequences used to tune
up the CZ-$\phi$ gates and the XOR and M gate can be understood
by following the time-evolution of the Hamiltonian of
Eq.~\ref{Equation:Supporting:10}. In particular, the
Hamiltonian $\widehat{\widetilde{H}}^{}_{\textrm{QD}}$
describes the dynamic phases acquired by the qubits when they
are brought outside their reference frame (i.e., qubit
rotations about the $z$-axis). As it will appear clear when
describing the tune-up sequences of the CZ-$\phi$ gates and of
the XOR and M gate, in the experiments we always compensate for
such dynamic phases.

% Table S1
\begin{sidewaystable}[pT!]\footnotesize
    \centering
\caption{\footnotesize \textbf{Numerical values for the density
matrices in Fig.~2C of the main text.} Real and imaginary part
of the elements $\langle lm | \hat{\rho} | pq \rangle$, with
$\hat{\rho} {} = {} \hat{\rho}^{}_{\textrm{(I)}} ,
\hat{\rho}^{}_{\textrm{(II)}} , \ldots ,
\hat{\rho}^{}_{\textrm{(V)}}$ and $| lm \rangle , | pq \rangle
{} \in {} \mathcal{M}^{}_2$. The confidence intervals are given
for the real and imaginary part of each matrix element
separately.}
    \vspace{13.0pt}
    \tabcolsep 4.0pt
    \footnotesize
\begin{tabular}{@{}c|c|c|c|c@{}}
    \hline \hline
$\textrm{(I)}$ & $| \textrm{gg} \rangle$ & $| \textrm{ge} \rangle$ & $| \textrm{eg} \rangle$ & $| \textrm{ee} \rangle$ \\
    \hline
$| \textrm{gg} \rangle$ & $0.122 \pm 0.004$ & $( -0.044 \pm 0.004 ) + ( 0.041 \pm 0.004 ) i$ & $( -0.027 \pm 0.004 ) - ( 0.041 \pm 0.004 ) i$ & $( 0.016 \pm 0.004 ) - ( 0.035 \pm 0.004 ) i$ \\
    \hline
$| \textrm{ge} \rangle$ & $( -0.044 \pm 0.004 ) - ( 0.041 \pm 0.004 ) i$ & $0.419 \pm 0.004$ & $( -0.068 \pm 0.005 ) + ( 0.353 \pm 0.004 ) i$ & $( 0.026 \pm 0.004 ) - ( 0.015 \pm 0.003 ) i$ \\
    \hline
$| \textrm{eg} \rangle$ & $( -0.027 \pm 0.004 ) + ( 0.041 \pm 0.004 ) i$ & $( -0.068 \pm 0.005 ) - ( 0.353 \pm 0.004 ) i$ & $0.406 \pm 0.004$ & $( 0.037 \pm 0.003 ) - ( 0.039 \pm 0.003 ) i$ \\
    \hline
$| \textrm{ee} \rangle$ & $( 0.016 \pm 0.004 ) + ( 0.035 \pm 0.004 ) i$ & $( 0.026 \pm 0.004 ) + ( 0.015 \pm 0.003 ) i$ & $( 0.037 \pm 0.003 ) + ( 0.039 \pm 0.003 ) i$ & $0.053 \pm 0.003$ \\
    \hline \hline
$\textrm{(II)}$ & $| \textrm{gg} \rangle$ & $| \textrm{ge} \rangle$ & $| \textrm{eg} \rangle$ & $| \textrm{ee} \rangle$ \\
    \hline
$| \textrm{gg} \rangle$ & $0.916 \pm 0.004$ & $( 0.021 \pm 0.004 ) - ( 0.015 \pm 0.004 ) i$ & $( -0.042 \pm 0.004 ) - ( 0.048 \pm 0.004 ) i$ & $( -0.010 \pm 0.005 ) - ( 0.040 \pm 0.005 ) i$ \\
    \hline
$| \textrm{ge} \rangle$ & $( 0.021 \pm 0.004 ) + ( 0.015 \pm 0.004 ) i$ & $0.029 \pm 0.002$ & $( 0.025 \pm 0.002 ) - ( 0.004 \pm 0.002 ) i$ & $( 0.017 \pm 0.002 ) - ( 0.023 \pm 0.002 ) i$ \\
    \hline
$| \textrm{eg} \rangle$ & $( -0.042 \pm 0.004 ) + ( 0.048 \pm 0.004 ) i$ & $( 0.025 \pm 0.002 ) + ( 0.004 \pm 0.002 ) i$ & $0.027 \pm 0.002$ & $( 0.019 \pm 0.002 ) - ( 0.017 \pm 0.002 ) i$ \\
    \hline
$| \textrm{ee} \rangle$ & $( -0.010 \pm 0.005 ) + ( 0.040 \pm 0.005 ) i$ & $( 0.017 \pm 0.002 ) + ( 0.023 \pm 0.002 ) i$ & $( 0.019 \pm 0.002 ) + ( 0.017 \pm 0.002 ) i$ & $0.028 \pm 0.002$ \\
    \hline \hline
$\textrm{(III)}$ & $| \textrm{gg} \rangle$ & $| \textrm{ge} \rangle$ & $| \textrm{eg} \rangle$ & $| \textrm{ee} \rangle$ \\
    \hline
$| \textrm{gg} \rangle$ & $0.180 \pm 0.005$ & $( -0.048 \pm 0.004 ) - ( 0.014 \pm 0.004 ) i$ & $( -0.018 \pm 0.004 ) + ( 0.021 \pm 0.004 ) i$ & $( -0.006 \pm 0.005 ) - ( 0.034 \pm 0.005 ) i$ \\
    \hline
$| \textrm{ge} \rangle$ & $( -0.048 \pm 0.004 ) + ( 0.014 \pm 0.004 ) i$ & $0.368 \pm 0.005$ & $( 0.225 \pm 0.005 ) - ( 0.208 \pm 0.005 ) i$ & $( 0.029 \pm 0.004 ) + ( 0.002 \pm 0.004 ) i$ \\
    \hline
$| \textrm{eg} \rangle$ & $( -0.018 \pm 0.004 ) - ( 0.021 \pm 0.004 ) i$ & $( 0.225 \pm 0.005 ) + ( 0.208 \pm 0.005 ) i$ & $0.398 \pm 0.005$ & $( 0.027 \pm 0.004 ) - ( 0.011 \pm 0.004 ) i$ \\
    \hline
$| \textrm{ee} \rangle$ & $( -0.006 \pm 0.005 ) + ( 0.034 \pm 0.005 ) i$ & $( 0.029 \pm 0.004 ) - ( 0.002 \pm 0.004 ) i$ & $( 0.027 \pm 0.004 ) + ( 0.011 \pm 0.004 ) i$ & $0.054 \pm 0.004$ \\
    \hline \hline
$\textrm{(IV)}$ & $| \textrm{gg} \rangle$ & $| \textrm{ge} \rangle$ & $| \textrm{eg} \rangle$ & $| \textrm{ee} \rangle$ \\
    \hline
$| \textrm{gg} \rangle$ & $0.913 \pm 0.004$ & $( 0.012 \pm 0.004 ) - ( 0.019 \pm 0.004 ) i$ & $( -0.050 \pm 0.004 ) + ( 0.057 \pm 0.004 ) i$ & $( 0.002 \pm 0.005 ) - ( 0.033 \pm 0.005 ) i$ \\
    \hline
$| \textrm{ge} \rangle$ & $( 0.012 \pm 0.004 ) + ( 0.019 \pm 0.004 ) i$ & $0.024 \pm 0.002$ & $( 0.023 \pm 0.002 ) - ( 0.000 \pm 0.002 ) i$ & $( 0.019 \pm 0.002 ) - ( 0.018 \pm 0.002 ) i$ \\
    \hline
$| \textrm{eg} \rangle$ & $( -0.050 \pm 0.004 ) - ( 0.057 \pm 0.004 ) i$ & $( 0.023 \pm 0.002 ) + ( 0.000 \pm 0.002 ) i$ & $0.034 \pm 0.002$ & $( 0.017 \pm 0.002 ) - ( 0.018 \pm 0.002 ) i$ \\
    \hline
$| \textrm{ee} \rangle$ & $( 0.002 \pm 0.005 ) + ( 0.033 \pm 0.005 ) i$ & $( 0.019 \pm 0.002 ) + ( 0.018 \pm 0.002 ) i$ & $( 0.017 \pm 0.002 ) + ( 0.018 \pm 0.002 ) i$ & $0.029 \pm 0.002$ \\
    \hline \hline
$\textrm{(V)}$ & $| \textrm{gg} \rangle$ & $| \textrm{ge} \rangle$ & $| \textrm{eg} \rangle$ & $| \textrm{ee} \rangle$ \\
    \hline
$| \textrm{gg} \rangle$ & $0.275 \pm 0.005$ & $( -0.014 \pm 0.004 ) + ( 0.030 \pm 0.004 ) i$ & $( -0.039 \pm 0.004 ) + ( 0.007 \pm 0.004 ) i$ & $( 0.006 \pm 0.005 ) - ( 0.042 \pm 0.005 ) i$ \\
    \hline
$| \textrm{ge} \rangle$ & $( -0.014 \pm 0.004 ) - ( 0.030 \pm 0.004 ) i$ & $0.338 \pm 0.004$ & $( 0.237 \pm 0.005 ) + ( 0.128 \pm 0.005 ) i$ & $( 0.016 \pm 0.004 ) - ( 0.039 \pm 0.004 ) i$ \\
    \hline
$| \textrm{eg} \rangle$ & $( -0.039 \pm 0.004 ) - ( 0.007 \pm 0.004 ) i$ & $( 0.237 \pm 0.005 ) - ( 0.128 \pm 0.005 ) i$ & $0.335 \pm 0.005$ & $( 0.031 \pm 0.004 ) - ( 0.047 \pm 0.004 ) i$ \\
    \hline
$| \textrm{ee} \rangle$ & $( 0.006 \pm 0.005 ) + ( 0.042 \pm 0.005 ) i$ & $( 0.016 \pm 0.004 ) + ( 0.039 \pm 0.004 ) i$ & $( 0.031 \pm 0.004 ) + ( 0.047 \pm 0.004 ) i$ & $0.052 \pm 0.004$ \\
    \hline
\end{tabular}
        \label{Table:Supporting:1:Matteo:Mariantoni:201107}
\end{sidewaystable}
    \clearpage
%

% ********************************************************
% *** Programming the quantum von Neumann architecture ***
% ********************************************************
%
\subsection*{Programming the quantum von Neumann architecture}
    \addcontentsline{toc}{subsection}{Programming the quantum von Neumann architecture}

The phase difference between the off-diagonal elements of the
density matrices shown in Fig.~2C of the main text (red arrows)
are due to the qubits being brought outside their reference
frame during the pulse sequence in Fig.~2A (the qubits acquire
dynamic phases), and to the angle accumulated by the microwave
signal used to excite the qubits. The pulse sequence was
calibrated such that the first density matrix
$\hat{\rho}^{}_{\textrm{(I)}}$ has purely imaginary
off-diagonal elements [cf.~grey and overlayed red arrows in
Fig.~2C (I) of the main text]. We can thus calculate the angles
of the density matrices $\hat{\rho}^{}_{\textrm{(III)}}$ and
$\hat{\rho}^{}_{\textrm{(V)}}$ by knowing the time duration of
the various steps in the sequence and the corresponding qubit
detunings (obtained from independent measurements), as shown by
the grey arrows in the matrices of Fig.~2C, (III) and (V). As
expected, the experimentally measured red arrows overlay the
calculated grey arrows with high accuracy. We will later show a
pulse method that allows us to compensate for dynamic phases
during the experiment, rather than calibrating the phases
\textit{a posteriori} as in Fig.~2C. Such a compensation pulse
method was used to implement the quantum Fourier transform and
the XOR and M gate.

The numerical values of all elements (real and imaginary part)
of each density matrix in Fig.~2C of the main text are reported
in Table~S1. The confidence interval for the real and imaginary
part of each complex number is also indicated. The confidence
intervals correspond to two standard deviations ($95\,\%$
confidence interval), where the standard deviations were
calculated as explained in the section on ``Statistical
errors'' of these Methods.
    \clearpage

% Table S2
\begin{sidewaystable}[pT!]\footnotesize
    \centering
\caption{\footnotesize \textbf{Numerical values for the density
matrices in Fig.~3E of the main text.} Real and imaginary part
of the elements $\langle lm | \hat{\rho}^{}_{\phi} | pq
\rangle$, with $\hat{\rho}^{}_{\phi} {} = {}
\hat{\rho}^{}_{0.28} , \hat{\rho}^{}_{\pi / 2} ,
\hat{\rho}^{}_{\pi}$ and $| lm \rangle , | pq \rangle {} \in {}
\mathcal{M}^{}_2$. The confidence intervals are given for the
real and imaginary part of each matrix element separately.}
    \vspace{13.0pt}
    \tabcolsep 4.0pt
    \footnotesize
\begin{tabular}{@{}c|c|c|c|c@{}}
    \hline \hline
$\phi {} = {} 0.28$ & $| \textrm{gg} \rangle$ & $| \textrm{eg} \rangle$ & $| \textrm{ge} \rangle$ & $| \textrm{ee} \rangle$ \\
    \hline
$| \textrm{gg} \rangle$ & $0.098 \pm 0.016$ & $( -0.049 \pm 0.010 ) + ( 0.025 \pm 0.066 ) i$ & $( 0.032 \pm 0.014 ) - ( 0.020 \pm 0.017 ) i$ & $( -0.031 \pm 0.027 ) + ( 0.051 \pm 0.009 ) i$ \\
    \hline
$| \textrm{eg} \rangle$ & $( -0.049 \pm 0.010 ) - ( 0.025 \pm 0.066 ) i$ & $0.533 \pm 0.074$ & $( 0.032 \pm 0.060 ) + ( 0.087 \pm 0.046 ) i$ & $( 0.311 \pm 0.058 ) - ( 0.069 \pm 0.067 ) i$ \\
    \hline
$| \textrm{ge} \rangle$ & $( 0.032 \pm 0.014 ) + ( 0.020 \pm 0.017 ) i$ & $( 0.032 \pm 0.060 ) - ( 0.087 \pm 0.046 ) i$ & $0.071 \pm 0.043$ & $( -0.002 \pm 0.006 ) - ( 0.078 \pm 0.053 ) i$ \\
    \hline
$| \textrm{ee} \rangle$ & $( -0.031 \pm 0.027 ) - ( 0.051 \pm 0.009 ) i$ & $( 0.311 \pm 0.058 ) + ( 0.069 \pm 0.067 ) i$ & $( -0.002 \pm 0.006 ) + ( 0.078 \pm 0.053 ) i$ & $0.297 \pm 0.093$ \\
    \hline \hline
$\phi {} = {} \pi / 2$ & $| \textrm{gg} \rangle$ & $| \textrm{eg} \rangle$ & $| \textrm{ge} \rangle$ & $| \textrm{ee} \rangle$ \\
    \hline
$| \textrm{gg} \rangle$ & $0.115 \pm 0.027$ & $( -0.067 \pm 0.001 ) - ( 0.056 \pm 0.074 ) i$ & $( 0.004 \pm 0.003 ) - ( 0.074 \pm 0.030 ) i$ & $( -0.027 \pm 0.013 ) + ( 0.035 \pm 0.032 ) i$ \\
    \hline
$| \textrm{eg} \rangle$ & $( -0.067 \pm 0.001 ) + ( 0.056 \pm 0.074 ) i$ & $0.496 \pm 0.045$ & $( 0.145 \pm 0.062 ) + ( 0.159 \pm 0.014 ) i$ & $( 0.190 \pm 0.050 ) - ( 0.142 \pm 0.009 ) i$ \\
    \hline
$| \textrm{ge} \rangle$ & $( 0.004 \pm 0.003 ) + ( 0.074 \pm 0.030 ) i$ & $( 0.145 \pm 0.062 ) - ( 0.159 \pm 0.014 ) i$ & $0.183 \pm 0.064$ & $( 0.002 \pm 0.025 ) - ( 0.144 \pm 0.003 ) i$ \\
    \hline
$| \textrm{ee} \rangle$ & $( -0.027 \pm 0.013 ) - ( 0.035 \pm 0.032 ) i$ & $( 0.190 \pm 0.050 ) + ( 0.142 \pm 0.009 ) i$ & $( 0.002 \pm 0.025 ) + ( 0.144 \pm 0.003 ) i$ & $0.206 \pm 0.082$ \\
    \hline \hline
$\phi {} = {} \pi$ & $| \textrm{gg} \rangle$ & $| \textrm{eg} \rangle$ & $| \textrm{ge} \rangle$ & $| \textrm{ee} \rangle$ \\
    \hline
$| \textrm{gg} \rangle$ & $0.147 \pm 0.032$ & $( -0.016 \pm 0.015 ) - ( 0.041 \pm 0.059 ) i$ & $( -0.026 \pm 0.031 ) - ( 0.034 \pm 0.004 ) i$ & $( 0.040 \pm 0.036 ) + ( 0.009 \pm 0.030 ) i$ \\
    \hline
$| \textrm{eg} \rangle$ & $( -0.016 \pm 0.015 ) + ( 0.041 \pm 0.059 ) i$ & $0.464 \pm 0.025$ & $( 0.338 \pm 0.001 ) + ( 0.001 \pm 0.024 ) i$ & $( 0.041 \pm 0.018 ) + ( 0.036 \pm 0.033 ) i$ \\
    \hline
$| \textrm{ge} \rangle$ & $( -0.026 \pm 0.031 ) + ( 0.034 \pm 0.004 ) i$ & $( 0.338 \pm 0.001 ) - ( 0.001 \pm 0.024 ) i$ & $0.342 \pm 0.030$ & $( 0.045 \pm 0.038 ) + ( 0.015 \pm 0.023 ) i$ \\
    \hline
$| \textrm{ee} \rangle$ & $( 0.040 \pm 0.036 ) - ( 0.009 \pm 0.030 ) i$ & $( 0.041 \pm 0.018 ) - ( 0.036 \pm 0.033 ) i$ & $( 0.045 \pm 0.038 ) - ( 0.015 \pm 0.023 ) i$ & $0.047 \pm 0.017$ \\
    \hline
\end{tabular}
        \label{Table:Supporting:2:Matteo:Mariantoni:201107}
\end{sidewaystable}
    \clearpage

% *************************************
% *** The quantum Fourier transform ***
% *************************************
%
\subsection*{The quantum Fourier transform}
    \addcontentsline{toc}{subsection}{The quantum Fourier transform}

The numerical values of all elements (real and imaginary part)
of each density matrix in Fig.~3E of the main text are reported
in Table~S2. The $95\,\%$ confidence interval for the real and
imaginary part of each complex number is also indicated. The
standard deviations were calculated as explained in the section
on ``Statistical errors'' of these Methods.

As shown in the main text, the CZ-$\phi$ gate is a fundamental
element for the implementation of the quantum Fourier
transform. In the rest of this section, we derive the
analytical expression for the phase $\phi$ of a CZ-$\phi$ gate
by diagonalizing the effective Hamiltonian of the
Q$^{}_1$-B-Q$^{}_2$ system and calculating its time evolution.
We subsequently describe the experimental pulse sequences
required to tune up the CZ-$\phi$ gate and show three examples
of Ramsey experiments used to measure the gate phase $\phi$,
when $\phi {} = {} 0.01$, $\phi {} = {} \pi / 2$, and $\phi {}
= {} \pi$. Finally, we discuss the origin of systematic errors
in the measurement of the phase $\phi$, showing that the global
phase shift in the curve of Fig.~3C of the main text is due to
a drift of the qubit operation point.

% **********************************************************************
% *** Analytical expression for the phase $\phi$ of a CZ-$\phi$ gate ***
% **********************************************************************
%
\subsubsection*{Analytical expression of the phase $\phi$ of a CZ-$\phi$ gate}
    \addcontentsline{toc}{subsubsection}{Analytical expression of the phase $\phi$ of a CZ-$\phi$ gate}

% ******************************
% *** Supplementary Figure 3 ***
% ******************************
%
\begin{figure}[t!]
    \centering
    \includegraphics[width=1.14\columnwidth]{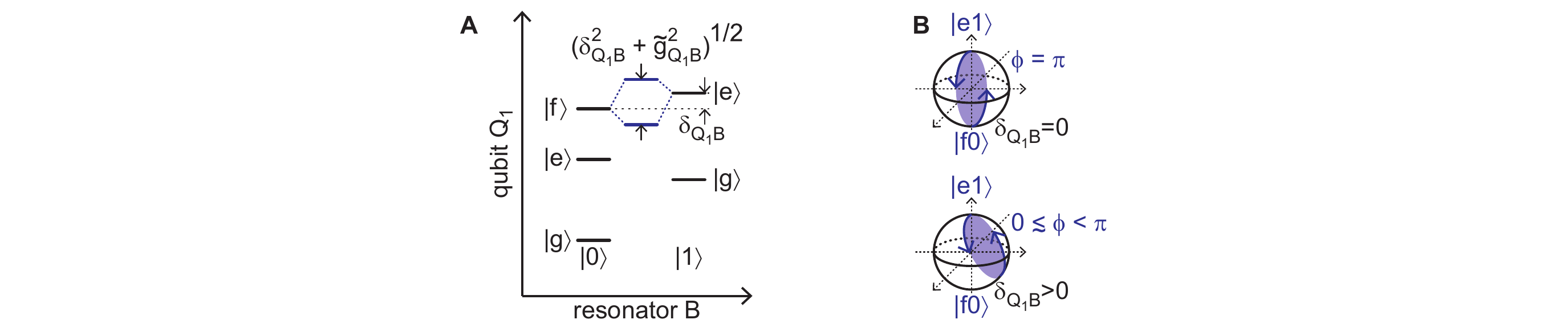}
    \caption{\footnotesize
\textbf{The CZ-$\phi$ gate energy diagram.} (\textbf{A}) Energy
diagram for target qubit Q$^{}_1$ coupled to bus resonator B,
with Q$^{}_1$'s eigenstates indicated by $| \textrm{g}
\rangle$, $| \textrm{e} \rangle$, and $| \textrm{f} \rangle$,
and B's eigenstates indicated by $| 0 \rangle$ and $| 1
\rangle$. In general, states $| \textrm{f} 0 \rangle$ and $|
\textrm{e} 1 \rangle$ are detuned by a quantity
$\delta^{}_{\textrm{Q}^{}_1 \textrm{B}}$. Thus, their effective
Rabi frequency is given by $( \delta^2_{\textrm{Q}^{}_1
\textrm{B}} + \tilde{g}^2_{\textrm{Q}^{}_1 \textrm{B}} )^{1 /
2}_{}$. (\textbf{B}) Bloch sphere interpretation of the $|
\textrm{f} 0 \rangle$-$| \textrm{e} 1 \rangle$ interaction.
When $\delta^{}_{\textrm{Q}^{}_1 \textrm{B}} {} = {} 0$,
Q$^{}_1$ acquires a phase $\phi {} = {} \pi$ (Top). When
$\delta^{}_{\textrm{Q}^{}_1 \textrm{B}} {} > {} 0$, Q$^{}_1$
acquires a phase $0 {} \lesssim {} \phi {} < {} \pi$ (Bottom).
    }
    \label{Figure:Supporting:3:Matteo:Mariantoni:201107}
\end{figure}

The CZ-$\phi$ gate demonstrated in the main text makes use of a
bus resonator B that mediates the interaction between qubit
Q$^{}_1$ and Q$^{}_2$. During the CZ-$\phi$ gate qubit Q$^{}_1$
is used as a qutrit, where the third eigenstate $| \textrm{f}
\rangle$ plays an active role in the implementation of the
gate. Qubit Q$^{}_1$ represents the gate target and qubit
Q$^{}_2$ the gate control. The energy diagram of the Q$^{}_1$-B
coupled system is displayed in
Fig.~\ref{Figure:Supporting:3:Matteo:Mariantoni:201107}A. The
coupled system consists of the states $| \textrm{g} \rangle$,
$| \textrm{e} \rangle$, and $| \textrm{f} \rangle$ of the
target qubit Q$^{}_1$, and of the states $| 0 \rangle$ and $| 1
\rangle$ of the bus resonator B. In the rotating frame of
resonator B and using the rotating-wave approximation, the
system effective Hamiltonian can be written as
\begin{equation}
\widehat{H}^{}_{\textrm{eff}} {} = {} h \Delta | \textrm{e} 1
\rangle \langle \textrm{e} 1 | + h \, ( \Delta +
\delta^{}_{\textrm{Q}^{}_1 \textrm{B}} ) \, | \textrm{f} 0
\rangle \langle \textrm{f} 0 | + h
\frac{\tilde{g}^{}_{\textrm{Q}^{}_1 \textrm{B}}}{2} ( |
\textrm{e} 1 \rangle \langle \textrm{f} 0 | + | \textrm{f} 0
\rangle \langle \textrm{e} 1 | ) \, ,
    \label{Equation:Supporting:11}
\end{equation}
where $\Delta$ represents the frequency detuning of Q$^{}_1$
with respect to the reference frame,
$\delta^{}_{\textrm{Q}^{}_1 \textrm{B}}$ the frequency detuning
between states $| \textrm{f} 0 \rangle$ and $| \textrm{e} 1
\rangle$, and $\tilde{g}^{}_{\textrm{Q}^{}_1 \textrm{B}}$ their
on-resonance coupling. As already mentioned after
Eq.~\ref{Equation:Supporting:10}, in the experiments we
compensate the rotation about the $z$-axis of Q$^{}_1$
associated with the detuning $\Delta$. As a consequence, in
Eq.~\ref{Equation:Supporting:11} we can set $\Delta {} = {} 0$
and rewrite the system effective Hamiltonian as
\begin{equation}
\widehat{H}^{}_{\textrm{eff}} {} = {} h \,
\delta^{}_{\textrm{Q}^{}_1 \textrm{B}} \, | \textrm{f} 0
\rangle \langle \textrm{f} 0 | + h
\frac{\tilde{g}^{}_{\textrm{Q}^{}_1 \textrm{B}}}{2} ( |
\textrm{e} 1 \rangle \langle \textrm{f} 0 | + | \textrm{f} 0
\rangle \langle \textrm{e} 1 | ) \, .
    \label{Equation:Supporting:12}
\end{equation}
The diagonalization of the Hamiltonian of
Eq.~\ref{Equation:Supporting:12} gives the eigenstates
\begin{eqnarray}
| \textrm{Q}^{}_1 \textrm{B} \rangle^{}_- & {} = {} &
\frac{\sqrt{1 + \delta^{}_{\textrm{Q}^{}_1 \textrm{B}} /
\sqrt{\delta^2_{\textrm{Q}^{}_1 \textrm{B}} +
\tilde{g}^2_{\textrm{Q}^{}_1 \textrm{B}}}} \, | \textrm{e} 1
\rangle - \sqrt{1 - \delta^{}_{\textrm{Q}^{}_1 \textrm{B}} /
\sqrt{\delta^2_{\textrm{Q}^{}_1 \textrm{B}} +
\tilde{g}^2_{\textrm{Q}^{}_1 \textrm{B}}}} \, | \textrm{f} 0
\rangle}{\sqrt{2}}
    \label{Equation:Supporting:13:a} \, , \\[2mm]
| \textrm{Q}^{}_1 \textrm{B} \rangle^{}_+ & {} = {} &
\frac{\sqrt{1 - \delta^{}_{\textrm{Q}^{}_1 \textrm{B}} /
\sqrt{\delta^2_{\textrm{Q}^{}_1 \textrm{B}} +
\tilde{g}^2_{\textrm{Q}^{}_1 \textrm{B}}}} \, | \textrm{e} 1
\rangle + \sqrt{1 + \delta^{}_{\textrm{Q}^{}_1 \textrm{B}} /
\sqrt{\delta^2_{\textrm{Q}^{}_1 \textrm{B}} +
\tilde{g}^2_{\textrm{Q}^{}_1 \textrm{B}}}} \, | \textrm{f} 0
\rangle}{\sqrt{2}} \, ,
    \label{Equation:Supporting:13:b}
\end{eqnarray}
with eigenenergies
\begin{eqnarray}
E^{}_- & {} = {} & \frac{\delta^{}_{\textrm{Q}^{}_1 \textrm{B}}
- \sqrt{\delta^2_{\textrm{Q}^{}_1 \textrm{B}} +
\tilde{g}^2_{\textrm{Q}^{}_1 \textrm{B}}}}{2}
    \label{Equation:Supporting:14:a} \, , \\[2mm]
E^{}_+ & {} = {} & \frac{\delta^{}_{\textrm{Q}^{}_1 \textrm{B}}
+ \sqrt{\delta^2_{\textrm{Q}^{}_1 \textrm{B}} +
\tilde{g}^2_{\textrm{Q}^{}_1 \textrm{B}}}}{2} \, ,
    \label{Equation:Supporting:14:b}
\end{eqnarray}
respectively.

Given the initial state $| \textrm{Q}^{}_1 \textrm{B}
\rangle^{}_0 {} = {} | \textrm{e} 1 \rangle$, after a time
$\tau$ the evolution $\widehat{U} ( \tau ) {} = {} \exp ( - i
\widehat{H}^{}_{\textrm{eff}} \tau / \hbar )$ of the effective
Hamiltonian of Eq.~\ref{Equation:Supporting:12} acting on $|
\textrm{Q}^{}_1 \textrm{B} \rangle^{}_0$ results in the state
\begin{eqnarray}
\widehat{U} ( \tau ) \, | \textrm{Q}^{}_1 \textrm{B}
\rangle^{}_0 & {} = {} & \frac{1}{\sqrt{2}} \Big( \sqrt{1 +
\delta^{}_{\textrm{Q}^{}_1 \textrm{B}} /
\sqrt{\delta^2_{\textrm{Q}^{}_1 \textrm{B}} +
\tilde{g}^2_{\textrm{Q}^{}_1 \textrm{B}}}} \, | \textrm{Q}^{}_1
\textrm{B} \rangle^{}_- \, e^{- i E^{}_- \, \tau}_{} {} \nonumber\\[1.5mm]
& & {} + \sqrt{1 - \delta^{}_{\textrm{Q}^{}_1 \textrm{B}} /
\sqrt{\delta^2_{\textrm{Q}^{}_1 \textrm{B}} +
\tilde{g}^2_{\textrm{Q}^{}_1 \textrm{B}}}} \, | \textrm{Q}^{}_1
\textrm{B} \rangle^{}_+ \, e^{- i E^{}_+ \, \tau}_{} \Big) \, .
    \label{Equation:Supporting:15}
\end{eqnarray}
For a time corresponding to a full $2 \pi$-rotation between the
states $| \textrm{f} 0 \rangle$ and $| \textrm{e} 1 \rangle$,
$\tau {} = {} 1 / ( \delta^2_{\textrm{Q}^{}_1 \textrm{B}} +
\tilde{g}^2_{\textrm{Q}^{}_1 \textrm{B}} )^{1 / 2}_{}$, the
Q$^{}_1$-B coupled system is in the state
\begin{eqnarray}
| \textrm{Q}^{}_1 \textrm{B} \rangle & {} = {} &
\frac{1}{\sqrt{2}} \Big[ \sqrt{1 + \delta^{}_{\textrm{Q}^{}_1
\textrm{B}} / \sqrt{\delta^2_{\textrm{Q}^{}_1 \textrm{B}} +
\tilde{g}^2_{\textrm{Q}^{}_1 \textrm{B}}}} \, | \textrm{Q}^{}_1
\textrm{B} \rangle^{}_- \, e^{- i ( \delta^2_{\textrm{Q}^{}_1
\textrm{B}} \, \tau / 2 - \pi )}_{} {} \nonumber\\[1.5mm]
& & {} + \sqrt{1 - \delta^{}_{\textrm{Q}^{}_1 \textrm{B}} /
\sqrt{\delta^2_{\textrm{Q}^{}_1 \textrm{B}} +
\tilde{g}^2_{\textrm{Q}^{}_1 \textrm{B}}}} \, | \textrm{Q}^{}_1
\textrm{B} \rangle^{}_+ \, e^{- i ( \delta^2_{\textrm{Q}^{}_1
\textrm{B}} \, \tau / 2 + \pi )}_{} \Big] \, ,
    \label{Equation:Supporting:16}
\end{eqnarray}
which, for simplicity, can be rewritten as
\begin{equation}
| \textrm{Q}^{}_1 \textrm{B} \rangle {} = {} e^{i \phi}_{} \, |
\textrm{e} 1 \rangle \, ,
    \label{Equation:Supporting:17}
\end{equation}
where the phase $\phi$ is defined as
\begin{equation}
\phi {} \equiv {} \pi - \pi \, \frac{\delta^{}_{\textrm{Q}^{}_1
\textrm{B}}}{\sqrt{\delta^2_{\textrm{Q}^{}_1 \textrm{B}} +
\tilde{g}^2_{\textrm{Q}^{}_1 \textrm{B}}}} \, .
    \label{Equation:Supporting:18}
\end{equation}

Figure~\ref{Figure:Supporting:3:Matteo:Mariantoni:201107}B
depicts the Bloch sphere of the Q$^{}_1$-B coupled system for
the states $| \textrm{e} 1 \rangle$ and $| \textrm{f} 0
\rangle$, showing that the interaction dynamics between the two
states always starts and ends at the same pole of the sphere.
It is during this interaction that the phase $\phi$ of
Eq.~\ref{Equation:Supporting:18} is acquired by the target
qubit Q$^{}_1$. In particular, when $\delta^{}_{\textrm{Q}^{}_1
\textrm{B}} {} = {} 0$ we obtain $\phi {} = {} \pi$, while for
$\delta^{}_{\textrm{Q}^{}_1 \textrm{B}} {} > {} 0$ we obtain
all phases $0 {} \lesssim {} \phi {} < {} \pi$.

An effective Hamiltonian similar to that of
Eq.~\ref{Equation:Supporting:12} governs the interaction
dynamics for the Q$^{}_2$-B system. Hence, the interaction
between the states of both the Q$^{}_1$-B and Q$^{}_2$-B
systems always starts and ends at the same pole of the coupling
Bloch sphere. This has the important consequence that the
CZ-$\phi$ gates used here are insensitive to the relative
phases of qubits Q$^{}_1$ and Q$^{}_2$ when they are brought
into resonance via resonator B. This feature allows us to use
independent reference frames with incommensurate frequencies
(and, hence, no special phase relationship) for each qubit,
thus making possible to tune up each qubit with a separate
calibration sequence.

% ******************************
% *** Supplementary Figure 4 ***
% ******************************
%
\begin{figure}[t!]
    \centering
    \includegraphics[width=1.14\columnwidth]{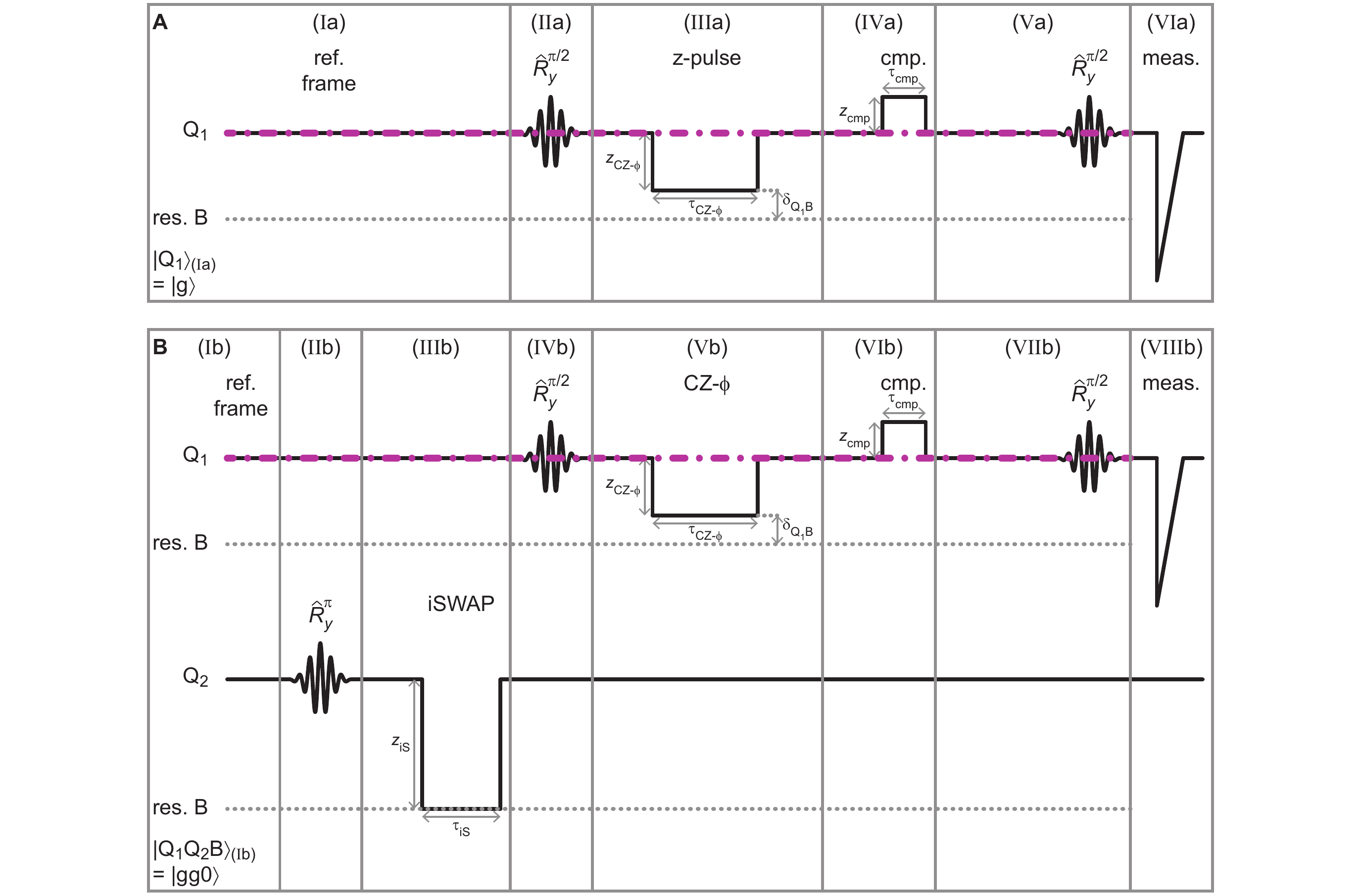}
    \caption{\footnotesize
\textbf{Dynamic phase compensation for qubit Q$^{}_1$.}
(\textbf{A}) Sequence without pulses on qubit Q$^{}_2$.
Sequence steps: $( \textrm{Ia} )$, Q$^{}_1$ is initialized in
state $| \textrm{Q}^{}_1 \rangle^{}_{( \textrm{Ia} )} {} = {} |
\textrm{g} \rangle$ at the idle point. The reference frame of
Q$^{}_1$ is indicated by a dash-dot magenta line. Resonator B,
which is indicated by a dotted grey line, is in the vacuum
state $| 0 \rangle$; $( \textrm{IIa} )$, rotation $\hat{R}^{\pi
/ 2}_y$ on Q$^{}_1$. The Gaussian pulse has a FWHM
$\tau^{}_{\textrm{FWHM}} {} = {} 8$\,ns; $( \textrm{IIIa} )$,
z-pulse on Q$^{}_1$ with amplitude $z^{}_{\textrm{CZ-}\phi}$
and length $\tau^{}_{\textrm{CZ-}\phi}$; $( \textrm{IVa} )$,
compensation pulse on Q$^{}_1$ with amplitude
$z^{}_{\textrm{cmp}}$ and length $\tau^{}_{\textrm{cmp}}$; $(
\textrm{Va} )$, rotation $\hat{R}^{\pi / 2}_y$ on Q$^{}_1$; $(
\textrm{VIa} )$, measurement pulse on Q$^{}_1$. (\textbf{B})
Sequence with pulses on qubit Q$^{}_2$. Sequence steps: $(
\textrm{Ib} )$, the Q$^{}_1$-Q$^{}_2$-B system is initialized
in state $| \textrm{Q}^{}_1 \textrm{Q}^{}_2 \textrm{B}
\rangle^{}_{( \textrm{Ib} )} {} = {} | \textrm{g} \textrm{g} 0
\rangle$, with both qubits at the idle point; $( \textrm{IIb}
)$, rotation $\hat{R}^{\pi}_y$ on Q$^{}_2$. The Gaussian pulse,
in this case, has a FWHM $\tau^{}_{\textrm{FWHM}} {} = {}
7$\,ns; $( \textrm{IIIb} )$, iSWAP between Q$^{}_2$ and B with
amplitude $z^{}_{\textrm{iS}}$ and length
$\tau^{}_{\textrm{iS}} {} = {} 24.97$\,ns; $( \textrm{IVb}
)$-$( \textrm{VIIIb} )$, same as in steps $( \textrm{IIa} )$-$(
\textrm{VIa} )$ of A. Note that, in step $( \textrm{Vb} )$ the
z-pulse on Q$^{}_1$ generates the phase $\phi$ of the CZ-$\phi$
gate.
    }
    \label{Figure:Supporting:4:Matteo:Mariantoni:201107}
\end{figure}
    \clearpage

% *****************************
% *** CZ-$\phi$ gate tuneup ***
% *****************************
%
\subsubsection*{CZ-$\phi$ gate tuneup}
    \addcontentsline{toc}{subsubsection}{CZ-$\phi$ gate tuneup}

Figure~\ref{Figure:Supporting:4:Matteo:Mariantoni:201107}, A
and B, shows the two sequences used to calibrate the pulses
applied to qubit Q$^{}_1$ during the CZ-$\phi$ gate operation.
We note that in the calibration sequence of
Fig.~\ref{Figure:Supporting:4:Matteo:Mariantoni:201107}B a
series of pulses is applied to qubit Q$^{}_1$ as well as to
qubit Q$^{}_2$. We will show that by comparing the results
obtained from the calibration of Q$^{}_1$ without pulsing
Q$^{}_2$
(Fig.~\ref{Figure:Supporting:4:Matteo:Mariantoni:201107}A and
Fig.~\ref{Figure:Supporting:5:Matteo:Mariantoni:201107}A) with
those obtained by pulsing Q$^{}_2$
(Fig.~\ref{Figure:Supporting:4:Matteo:Mariantoni:201107}B and
Fig.~\ref{Figure:Supporting:5:Matteo:Mariantoni:201107}B) it is
possible to measure the phase $\phi$ associated with the
CZ-$\phi$ gate (cf.~also main text and Fig.~3C of the main
text).

Before delving into the analysis of the calibration sequences,
we note that the idle point of qubits Q$^{}_1$ and Q$^{}_2$ was
set at a different position depending on the experiment. For
the experiments of Fig.~2, C to E, in the main text, the idle
point was set in between the memory and bus resonator for both
qubits. This is also the case for the swap spectroscopies shown
in Fig.~1B of the main text. For all the other experiments,
e.g., those described in this section, the idle point was set
above the bus resonator for both qubits.

Consistently with the vertical axis in Fig.~1B of the main
text, a z-pulse in the \textit{upward} direction, which
increases the qubit transition frequency, always corresponds to
a \textit{negative} z-pulse amplitude with respect to the qubit
idle point. The opposite applies to the case of a z-pulse in
the downward direction.

The first calibration sequence for qubit Q$^{}_1$, which is
shown in
Fig.~\ref{Figure:Supporting:4:Matteo:Mariantoni:201107}A,
comprises the following steps:

    \renewcommand{\labelenumi}{(\Roman{enumi}a)}
\begin{enumerate}

\item Qubit Q$^{}_1$ is initialized in the ground state $|
    \textrm{Q}^{}_1 \rangle^{}_{( \textrm{Ia} )} {} = {} |
    \textrm{g} \rangle$ at the idle point, setting the
    qubit reference frame with reference clock rate
    $f^0_{\textrm{Q}^{}_1}$. In
    Fig.~\ref{Figure:Supporting:4:Matteo:Mariantoni:201107}A,
    the reference frame is indicated by the dash-dot
    magenta line. During the entire calibration sequence,
    the bus resonator B is maintained in the vacuum state
    $| 0 \rangle$. Nevertheless, in
    Fig.~\ref{Figure:Supporting:4:Matteo:Mariantoni:201107}A
    we indicate the presence of resonator B by a dotted
    grey line, which helps visualizing the frequency
    detuning between qubit and resonator;

\item Keeping the qubit detuning $\Delta {} = {} 0$, a
    Gaussian microwave pulse with full width at half
    maximum (FWHM) $\tau^{}_{\textrm{FWHM}}$ is applied to
    Q$^{}_1$. The amplitude of the pulse is chosen such
    that $2 \pi \Omega^{}_{\textrm{D}} \tau {} = {} \pi /
    2$. In this case, the time evolution of the Hamiltonian
    of Eq.~\ref{Equation:Supporting:10} yields a $\pi / 2$
    unitary rotation about the
    $y$-axis~\cite{sakurai:1994:mqm}, $\hat{R}^{\pi /
    2}_y$, which brings the qubit into the new state $|
    \textrm{Q}^{}_1 \rangle^{}_{( \textrm{IIa} )} {} = {} (
    | \textrm{g} \rangle + | \textrm{e} \rangle ) /
    \sqrt{2}$;

\item A z-pulse with amplitude $z^{}_{\textrm{CZ-}\phi}$,
    corresponding to a qubit frequency detuning $\Delta (
    z^{}_{\textrm{CZ-}\phi} )$ from the reference frame,
    brings the $| \textrm{e} \rangle {} \leftrightarrow {}
    | \textrm{f} \rangle$ transition of Q$^{}_1$ on or near
    resonance with B for a time
    $\tau^{}_{\textrm{CZ-}\phi}$. In general, the z-pulse
    can be adjusted so that the $| \textrm{e} \rangle {}
    \leftrightarrow {} | \textrm{f} \rangle$ qubit
    transition is detuned by a frequency
    $\delta^{}_{\textrm{Q}^{}_1 \textrm{B}}$ from resonator
    B
    (cf.~Fig.~\ref{Figure:Supporting:4:Matteo:Mariantoni:201107}A,
    and also the section on the ``Analytical expression for
    the phase $\phi$ of a CZ-$\phi$ gate'' in these
    Methods, and the main text). During the z-pulse, since
    the $| \textrm{e} 1 \rangle$-$| \textrm{f} 0 \rangle$
    transition of the Q$^{}_1$-B coupled system is on or
    near resonance, the system remains always in the state
    $| \textrm{Q}^{}_1 \textrm{B} \rangle {} = {} ( |
    \textrm{g} \rangle + | \textrm{e} \rangle ) / \sqrt{2}
    \rangle \otimes | 0 \rangle$. This represents a dark
    state of the time evolution of the coupled system,
    yielding no swaps between Q$^{}_1$ and B, as opposed to
    the bright state $| \textrm{Q}^{}_1 \textrm{B} \rangle
    {} = {} ( | \textrm{g} \rangle + | \textrm{e} \rangle )
    / \sqrt{2} \rangle \otimes | 1 \rangle$. Depending on
    the z-pulse amplitude $z^{}_{\textrm{CZ-}\phi}$ and,
    hence, on the detuning $\delta^{}_{\textrm{Q}^{}_1
    \textrm{B}}$, the z-pulse length is chosen such that
    $\tau^{}_{\textrm{CZ-}\phi} {} = {} 1 / (
    \delta^2_{\textrm{Q}^{}_1 \textrm{B}} +
    \tilde{g}^2_{\textrm{Q}^{}_1 \textrm{B}} )^{1 / 2}_{}$
    (cf.~section on the ``Analytical expression for the
    phase $\phi$ of a CZ-$\phi$ gate'' in these Methods).
    The z-pulse $z^{}_{\textrm{CZ-}\phi}$ moves Q$^{}_1$
    outside its reference frame. As a consequence, the time
    evolution of Eq.~\ref{Equation:Supporting:10} acting on
    the state $| \textrm{Q}^{}_1 \rangle^{}_{( \textrm{IIa}
    )}$ gives the state $| \textrm{Q}^{}_1 \rangle^{}_{(
    \textrm{IIIa} )} {} = {} ( | \textrm{g} \rangle \, e^{-
    i \phi^{}_{\textrm{dyn}}}_{} + | \textrm{e} \rangle \,
    e^{+ i \phi^{}_{\textrm{dyn}}}_{} ) / \sqrt{2}$, where
    $\phi^{}_{\textrm{dyn}} {} \equiv {} - \Delta (
    z^{}_{\textrm{CZ-}\phi} ) \, \tau^{}_{\textrm{CZ-}\phi}
    / 2$. This means that during the z-pulse Q$^{}_1$
    acquires an unwanted dynamic phase
    $\phi^{}_{\textrm{dyn}}$;

\item For the correct operation of the CZ-$\phi$ gate, the
    dynamic phase $\phi^{}_{\textrm{dyn}}$ must be
    compensated. This can be realized by applying a
    compensation z-pulse to Q$^{}_1$, with fixed length
    $\tau^{}_{\textrm{cmp}}$ and variable amplitude
    $z^{}_{\textrm{cmp}}$. In order to avoid crossing the
    resonance with B, the amplitude of the compensation
    pulse is swept in the opposite direction as compared to
    the z-pulse $z^{}_{\textrm{CZ-}\phi}$
    (cf.~Fig.~\ref{Figure:Supporting:4:Matteo:Mariantoni:201107}A).
    In this case, the time evolution of
    $\widehat{\widetilde{H}}^{}_{\textrm{QD}}$ acting on $|
    \textrm{Q}^{}_1 \rangle^{}_{( \textrm{IIIa} )}$ yields
    the state $| \textrm{Q}^{}_1 \rangle^{}_{( \textrm{IVa}
    )} {} = {} ( | \textrm{g} \rangle \, e^{- i
    \phi^{}_{\textrm{a}} / 2}_{} + | \textrm{e} \rangle \,
    e^{+ i \phi^{}_{\textrm{a}} / 2}_{} ) / \sqrt{2}$,
    where $\phi^{}_{\textrm{a}} / 2 {} \equiv {}
    \phi^{}_{\textrm{dyn}} - \Delta ( z^{}_{\textrm{cmp}} )
    \, \tau^{}_{\textrm{cmp}} / 2$;

\item A rotation $\hat{R}^{\pi / 2}_y$ similar to that in
    point~(IIa) is applied to Q$^{}_1$, bringing the qubit
    into the final state $| \textrm{Q}^{}_1 \rangle^{}_{(
    \textrm{Va} )} {} = {} - i \sin ( \phi^{}_{\textrm{a}}
    / 2 ) | \textrm{g} \rangle + \cos (
    \phi^{}_{\textrm{a}} / 2 ) | \textrm{e} \rangle$;

\item Finally, a measurement pulse is applied to Q$^{}_1$
    in order to obtain the probability to find the qubit in
    $| \textrm{e} \rangle$, $P^{}_{\textrm{e}} {} = {} |
    \cos ( \phi^{}_{\textrm{a}} / 2 ) |^2_{} {} = {} ( 1 +
    \cos \phi^{}_{\textrm{a}} ) / 2$. Since
    $\phi^{}_{\textrm{a}}$ depends on the compensation
    pulse amplitude $z^{}_{\textrm{cmp}}$, the probability
    $P^{}_{\textrm{e}}$ is also a function of
    $z^{}_{\textrm{cmp}}$. In order to cancel the effect of
    the dynamic phase $\phi^{}_{\textrm{dyn}}$,
    $z^{}_{\textrm{cmp}}$ has to be chosen such that
    $P^{}_{\textrm{e}}$ reaches a maximum, where the phase
    $\phi^{}_{\textrm{a}} {} = {} 2 K \pi$, with $K {} \in
    {} \mathbb{Z}$.

\end{enumerate}

% ******************************
% *** Supplementary Figure 5 ***
% ******************************
%
\begin{figure}[t!]
    \centering
    \includegraphics[width=1.14\columnwidth]{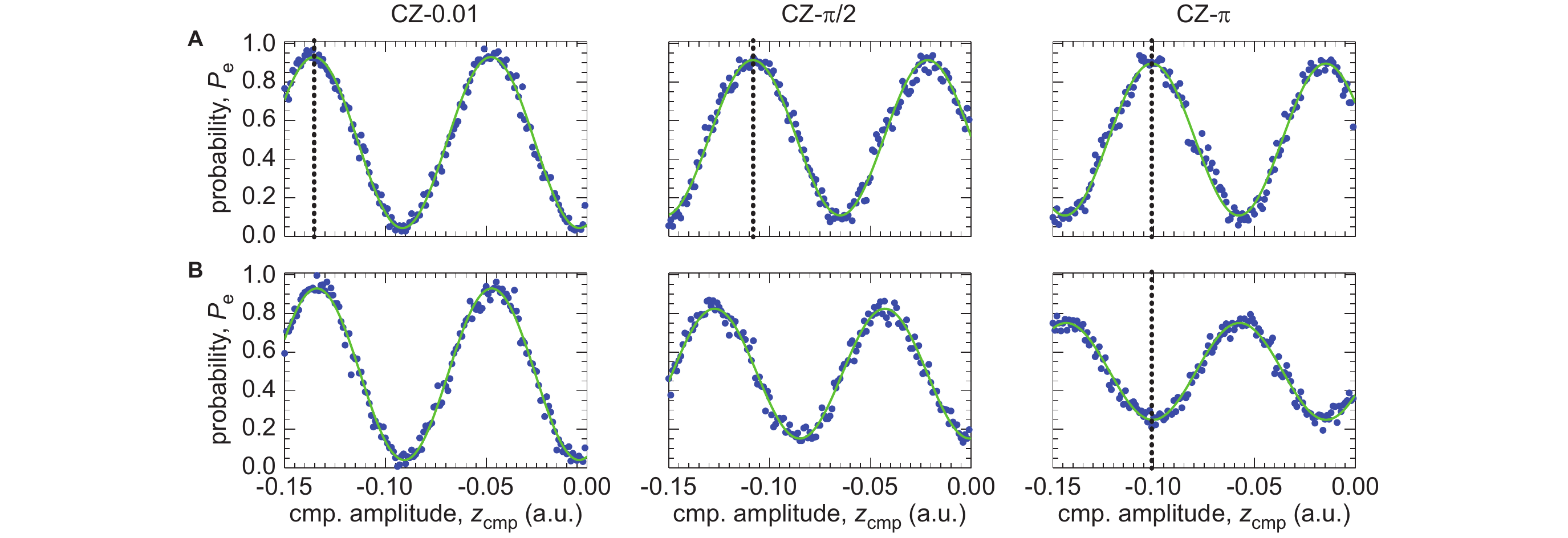}
    \caption{\footnotesize
\textbf{Ramsey experiments for compensating the dynamic phase
of Q$^{}_1$ and measuring $\mathbf{\phi}$.} (\textbf{A})
Probability of measuring Q$^{}_1$ in $| \textrm{e} \rangle$,
$P^{}_{\textrm{e}}$, vs. compensation pulse amplitude
$z^{}_{\textrm{cmp}}$ for the pulse sequence of
Fig.~\ref{Figure:Supporting:4:Matteo:Mariantoni:201107}A. The
compensation pulse is the z-pulse of step $( \textrm{IVa} )$ in
Fig.~\ref{Figure:Supporting:4:Matteo:Mariantoni:201107}A. In
the CZ-$\phi$ gate experiments we always chose a compensation
pulse length $\tau^{}_{\textrm{cmp}} {} = {} 7$\,ns. The blue
dots represent measured data, while the solid green lines are
least-squares fits to a sine function. From left to right, the
panels refer to a CZ-$0.01$ gate, a CZ-$\pi / 2$ gate, and a
CZ-$\pi$ gate, respectively. In each panel, the vertical dotted
black line indicates the amplitude $z^{}_{\textrm{cmp}}$ chosen
to compensate the dynamic phase $\phi^{}_{\textrm{dyn}}$. The
$z^{}_{\textrm{cmp}}$ numerical values expressed in the
arbitrary units of our custom electronics are
$z^{}_{\textrm{cmp}} {} \simeq {} - 0.135$ for the CZ-$0.01$
gate, $z^{}_{\textrm{cmp}} {} \simeq {} - 0.108$ for the
CZ-$\pi / 2$ gate, and $z^{}_{\textrm{cmp}} {} \simeq {} -
0.101$ for the CZ-$\pi$ gate. Notice that the angle of the
CZ-$0.01$ gate is different than that of the CZ-$0.28$ gate in
the main text. This is because the CZ-$0.01$ gate shown here is
the result of a single set of measurements, whereas the
CZ-$0.28$ gate shown in the main text is the average of a set
of $10$ independent measurements. (\textbf{B}) Same as in A,
but for the pulse sequence of
Fig.~\ref{Figure:Supporting:4:Matteo:Mariantoni:201107}B. The
relative phase between the Ramsey fringes of each panel in A
and the corresponding panel in B gives the phase $\phi$ of the
CZ-$\phi$ gate. For the three pairs of Ramsey fringes in this
example, the relative phases are $\phi {} = {} 0.01$\,rad,
$\phi {} = {} \pi / 2$\,rad, and $\phi {} = {} \pi$\,rad. The
vertical dotted black line in the rightmost panel is positioned
at the same value of $z^{}_{\textrm{cmp}}$ as the corresponding
panel in A, but, in this case, it indicates a minimum of the
Ramsey fringe because of the $\pi$ shift introduced by the
gate.
    }
    \label{Figure:Supporting:5:Matteo:Mariantoni:201107}
\end{figure}
    \clearpage

In summary, the two qubit rotations $\hat{R}^{\pi / 2}_y$ at
the beginning and end of the calibration sequence of
Fig.~\ref{Figure:Supporting:4:Matteo:Mariantoni:201107}A
realize a generalized Ramsey experiment, which allows us to
measure the total phase acquired by Q$^{}_1$ during the
z-pulses that bring it outside its reference frame. The
experimental data for the calibration sequence of
Fig.~\ref{Figure:Supporting:4:Matteo:Mariantoni:201107}A are
shown in
Fig.~\ref{Figure:Supporting:5:Matteo:Mariantoni:201107}A, where
the three Ramsey fringes are obtained for three different
values of the detuning $\delta^{}_{\textrm{Q}^{}_1
\textrm{B}}$, corresponding to a CZ-$0.01$, CZ-$\pi / 2$, and
CZ-$\pi$ gate, respectively. For each Ramsey fringe in
Fig.~\ref{Figure:Supporting:5:Matteo:Mariantoni:201107}A, the
z-pulse amplitude $z^{}_{\textrm{cmp}}$ chosen to compensate
the dynamic phase $\phi^{}_{\textrm{dyn}}$ is indicated by a
vertical dotted black line.

The second calibration sequence for qubit Q$^{}_1$ is shown in
Fig.~\ref{Figure:Supporting:4:Matteo:Mariantoni:201107}B. The
sequence, which is the same as the first calibration sequence
with the addition of the pulses applied to qubit Q$^{}_2$,
comprises the following steps:

    \renewcommand{\labelenumi}{(\Roman{enumi}b)}
\begin{enumerate}

\item Qubit Q$^{}_1$, qubit Q$^{}_2$, and resonator B are
    initialized in the ground/vacuum state $|
    \textrm{Q}^{}_1 \rangle^{}_{( \textrm{Ib} )} \otimes |
    \textrm{Q}^{}_2 \rangle^{}_{( \textrm{Ib} )} \otimes |
    \textrm{B} \rangle^{}_{( \textrm{Ib} )} {} = {} |
    \textrm{g} \rangle \otimes | \textrm{g} \rangle \otimes
    | 0 \rangle$, with both qubits biased at the idle
    point;

\item A Gaussian microwave pulse with FWHM
    $\tau^{}_{\textrm{FWHM}}$ is applied to Q$^{}_2$. The
    amplitude of the pulse is chosen such that
    $\Omega^{}_{\textrm{D}} \tau {} = {} \pi$. In this
    case, the time evolution of
    $\widehat{\widetilde{H}}^{}_{\textrm{QD}}$ acting on $|
    \textrm{Q}^{}_2 \rangle^{}_{( \textrm{Ib} )}$ realizes
    a full qubit population transfer, $\hat{R}^{\pi}_y$,
    bringing the qubit into the new state $|
    \textrm{Q}^{}_2 \rangle^{}_{( \textrm{IIb} )} {} = {} |
    \textrm{e} \rangle$;

\item The state $| \textrm{Q}^{}_2 \rangle^{}_{(
    \textrm{IIb} )}$ is moved from Q$^{}_2$ to B by means
    of an iSWAP of length $\tau^{}_{\textrm{iS}}$. At the
    end of the iSWAP, resonator B is in the state $|
    \textrm{B} \rangle^{}_{( \textrm{IIIb} )} {} = {} | 1
    \rangle$;

\item Qubit Q$^{}_1$ is prepared in the state $|
    \textrm{Q}^{}_1 \rangle^{}_{( \textrm{IVb} )} {} = {} (
    | \textrm{g} \rangle + | \textrm{e} \rangle ) /
    \sqrt{2}$ by means of a rotation $\hat{R}^{\pi / 2}_y$;

\item The same z-pulse as in point~(IIIa) is applied to
    Q$^{}_1$. In this case, the Q$^{}_1$-B coupled system
    is in the state $| \textrm{Q}^{}_1 \textrm{B} \rangle
    {} = {} ( | \textrm{g} \rangle + | \textrm{e} \rangle )
    / \sqrt{2} \rangle \otimes | 1 \rangle$, which
    represents a bright state of the time evolution of the
    system, as opposed to the dark state $| \textrm{Q}^{}_1
    \textrm{B} \rangle {} = {} ( | \textrm{g} \rangle + |
    \textrm{e} \rangle ) / \sqrt{2} \rangle \otimes | 0
    \rangle$. Depending on $\tau^{}_{\textrm{CZ-}\phi}$ and
    $z^{}_{\textrm{CZ-}\phi}$, and, thus, on
    $\delta^{}_{\textrm{Q}^{}_1 \textrm{B}}$, at the end of
    the z-pulse the excited state $| \textrm{e} \rangle$ of
    Q$^{}_1$ has acquired a phase $\phi$ [cf.~main text;
    when $\phi {} = {} \pi$, the z-pulse is a $-
    \textrm{SWAP}^2_{}$~\cite{yamamoto:2010:czgates}]. As
    in point~(IIIa), at the end of such a pulse Q$^{}_1$
    has also acquired a dynamic phase
    $\phi^{}_{\textrm{dyn}}$, resulting in the state $|
    \textrm{Q}^{}_1 \rangle^{}_{( \textrm{Vb} )} {} = {} (
    | \textrm{g} \rangle \, e^{- i
    \phi^{}_{\textrm{dyn}}}_{} + | \textrm{e} \rangle \,
    e^{+ i \phi^{}_{\textrm{dyn}}}_{} \, e^{+ i \phi}_{} )
    / \sqrt{2}$;

\item The dynamic phase $\phi^{}_{\textrm{dyn}}$ acquired
    by Q$^{}_1$ is compensated by means of a z-pulse, as in
    point~(IVa). At the end of the compensation pulse
    Q$^{}_1$ is in the state $| \textrm{Q}^{}_1
    \rangle^{}_{( \textrm{VIb} )} {} = {} [ | \textrm{g}
    \rangle \, e^{- i \phi^{}_{\textrm{a}} / 2}_{} + |
    \textrm{e} \rangle \, e^{+ i ( \phi^{}_{\textrm{a}} / 2
    + \phi )}_{} ] / \sqrt{2}$;

\item A rotation $\hat{R}^{\pi / 2}_y$ is applied to
    Q$^{}_1$, bringing the qubit to the final state $|
    \textrm{Q}^{}_1 \rangle^{}_{( \textrm{VIIb} )} {} = {}
    - i e^{i \phi / 2}_{} \sin ( \phi^{}_{\textrm{a}} / 2 +
    \phi / 2 ) | \textrm{g} \rangle + e^{i \phi / 2}_{}
    \cos ( \phi^{}_{\textrm{a}} / 2 + \phi / 2 ) |
    \textrm{e} \rangle$;

\item Finally, a measurement pulse is applied to Q$^{}_1$
    in order to obtain the probability to find the qubit in
    $| \textrm{e} \rangle$, $P^{}_{\textrm{e}} {} = {} |
    e^{i \phi / 2}_{} \cos ( \phi^{}_{\textrm{a}} / 2 +
    \phi / 2 ) |^2_{} {} = {} ( 1 + \cos
    \phi^{}_{\textrm{b}} ) / 2$, where
    $\phi^{}_{\textrm{b}} {} \equiv {} \phi^{}_{\textrm{a}}
    + \phi$.

\end{enumerate}

The phase difference between the probability
$P^{}_{\textrm{e}}$ for the first and second calibration
sequence allows us to measure the CZ-$\phi$ gate phase,
$\phi^{}_{\textrm{b}} - \phi^{}_{\textrm{a}} {} = {} \phi$.
This is illustrated in
Fig.~\ref{Figure:Supporting:5:Matteo:Mariantoni:201107}, where
the phase difference between the Ramsey fringe in each panel of
Fig.~\ref{Figure:Supporting:5:Matteo:Mariantoni:201107}A and
the corresponding fringe in each panel of
Fig.~\ref{Figure:Supporting:5:Matteo:Mariantoni:201107}B gives
the phase of a CZ-$0.01$, CZ-$\pi / 2$, and CZ-$\pi$ gate,
respectively.

The second calibration sequence can also be used to cross check
the amplitude $z^{}_{\textrm{cmp}}$ of the compensation pulse
chosen to cancel the dynamic phase $\phi^{}_{\textrm{dyn}}$.
For example, when $\phi {} = {} \pi$, the Ramsey fringe
obtained from the second calibration sequence should reach a
\textit{minimum} for the same value of $z^{}_{\textrm{cmp}}$
for which it reaches a maximum in the first calibration
sequence. This is confirmed by comparing the experimental data
shown in the rightmost panel of
Fig.~\ref{Figure:Supporting:5:Matteo:Mariantoni:201107}A and
Fig.~\ref{Figure:Supporting:5:Matteo:Mariantoni:201107}B.

% ******************************
% *** Supplementary Figure 6 ***
% ******************************
%
\begin{figure}[t!]
    \centering
    \includegraphics[width=1.14\columnwidth]{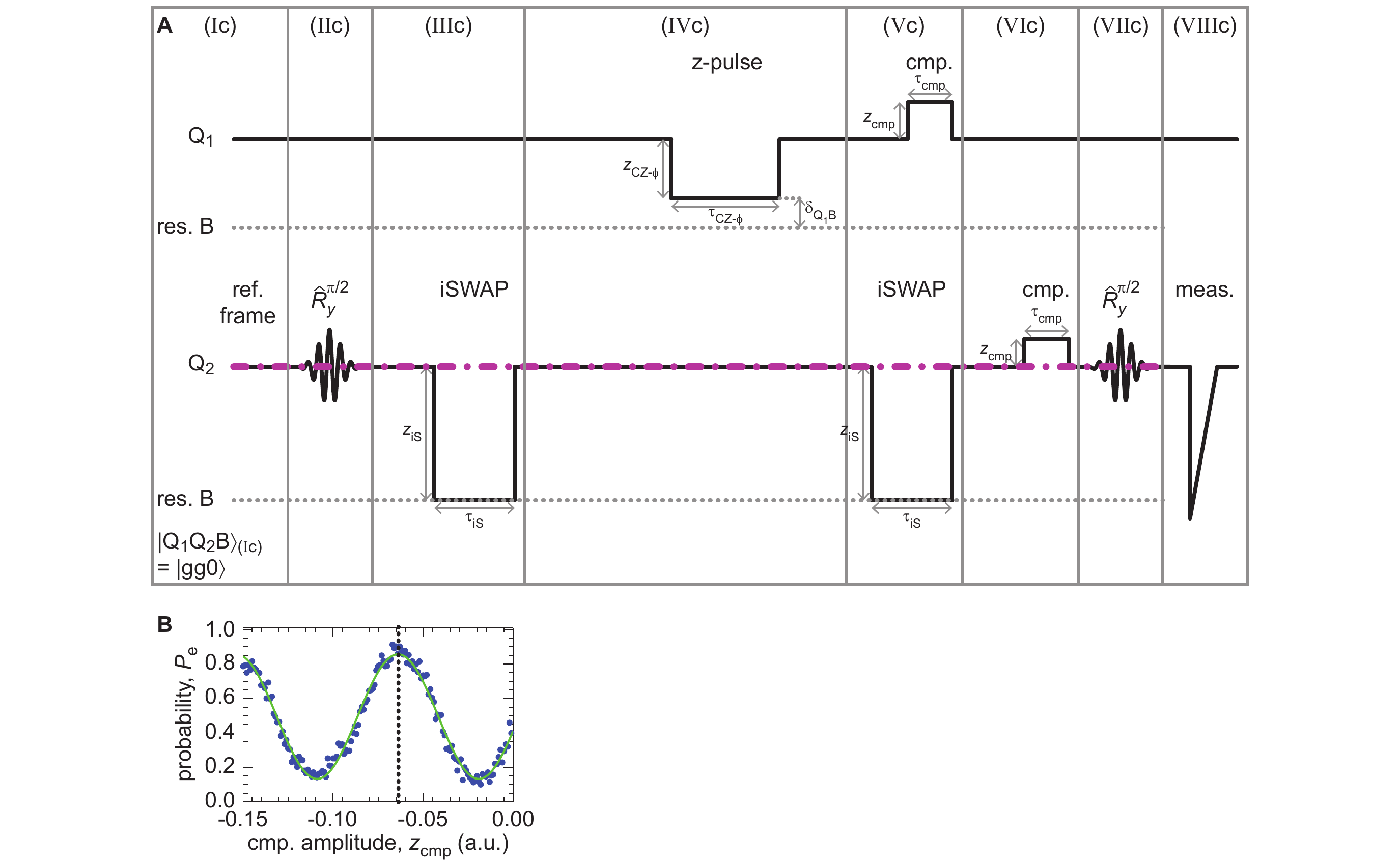}
    \caption{\footnotesize
\textbf{Dynamic phase compensation for qubit Q$^{}_2$.}
(\textbf{A}) Sequence steps: $( \textrm{Ic} )$, the
Q$^{}_1$-Q$^{}_2$-B system is initialized in state $|
\textrm{Q}^{}_1 \textrm{Q}^{}_2 \textrm{B} \rangle^{}_{(
\textrm{Ic} )} {} = {} | \textrm{g} \textrm{g} 0 \rangle$, with
both qubits at the idle point. The reference frame of Q$^{}_2$
is indicated by a dash-dot magenta line. Resonator B is
indicated by a dotted grey line; $( \textrm{IIc} )$, rotation
$\hat{R}^{\pi / 2}_y$ on Q$^{}_2$. The Gaussian pulse has a
FWHM $\tau^{}_{\textrm{FWHM}} {} = {} 7$\,ns; $( \textrm{IIIc}
)$, iSWAP between Q$^{}_2$ and B with amplitude
$z^{}_{\textrm{iS}}$ and length $\tau^{}_{\textrm{iS}} {} = {}
24.97$\,ns; $( \textrm{IVc} )$, z-pulse on Q$^{}_1$ with
amplitude $z^{}_{\textrm{CZ-}\phi}$ and length
$\tau^{}_{\textrm{CZ-}\phi}$; $( \textrm{Vc} )$, iSWAP between
B and Q$^{}_2$ with amplitude $z^{}_{\textrm{iS}}$ and length
$\tau^{}_{\textrm{iS}} {} = {} 24.97$\,ns. At the same time,
compensation pulse on Q$^{}_1$ with amplitude
$z^{}_{\textrm{cmp}}$ and length $\tau^{}_{\textrm{cmp}}$ set
in the sequence of
Fig.~\ref{Figure:Supporting:4:Matteo:Mariantoni:201107}B; $(
\textrm{VIc} )$, compensation pulse on Q$^{}_2$ with amplitude
$z^{}_{\textrm{cmp}}$ and length $\tau^{}_{\textrm{cmp}}$; $(
\textrm{VIIc} )$, rotation $\hat{R}^{\pi / 2}_y$ on Q$^{}_2$;
$( \textrm{VIIIc} )$, measurement pulse on Q$^{}_2$.
(\textbf{B}) Probability of measuring Q$^{}_2$ in $| \textrm{e}
\rangle$, $P^{}_{\textrm{e}}$, vs. compensation pulse amplitude
$z^{}_{\textrm{cmp}}$ for the compensation pulse of step $(
\textrm{VIc} )$ in A. The blue dots represent measured data,
while the solid green line is a least-squares fit to a sine
function. The vertical dotted black line indicates the
amplitude $z^{}_{\textrm{cmp}}$ chosen to compensate the
dynamic phase acquired by Q$^{}_2$ during the sequence in A,
$z^{}_{\textrm{cmp}} {} \simeq {} - 0.064$.
    }
    \label{Figure:Supporting:6:Matteo:Mariantoni:201107}
\end{figure}

Figure~\ref{Figure:Supporting:6:Matteo:Mariantoni:201107}A
shows the sequence used to calibrate the pulses applied to
qubit Q$^{}_2$ during the CZ-$\phi$ gate operation. The
sequence comprises the following steps:

    \renewcommand{\labelenumi}{(\Roman{enumi}c)}
\begin{enumerate}

\item The system is initialized in the state $|
    \textrm{Q}^{}_1 \rangle^{}_{( \textrm{Ic} )} \otimes |
    \textrm{Q}^{}_2 \rangle^{}_{( \textrm{Ic} )} \otimes |
    \textrm{B} \rangle^{}_{( \textrm{Ic} )} {} = {} |
    \textrm{g} \rangle \otimes | \textrm{g} \rangle \otimes
    | 0 \rangle$, with both qubits biased at the idle
    point;

\item A rotation $\hat{R}^{\pi / 2}_y$ with FWHM
    $\tau^{}_{\textrm{FWHM}}$ is applied to Q$^{}_2$;

\item The state of Q$^{}_2$ is moved into B by means of an
    iSWAP;

\item Q$^{}_1$ is moved on resonance or close to resonance
    with B through the same z-pulse as point~(Vb);

\item The state of B is moved back to Q$^{}_2$ via an
    iSWAP. During and between the iSWAPs of point~(IIIc)
    and (Vc), Q$^{}_2$ acquires an unwanted dynamic phase.

    At the same time as the iSWAP in point~(Vc), the
    compensation pulse tuned up in the Q$^{}_1$ calibration
    sequence of
    Fig.~\ref{Figure:Supporting:4:Matteo:Mariantoni:201107},
    A and B [point (IVa) or (VIb)], is applied to Q$^{}_1$
    (this is not strictly necessary due to the independence
    of the calibration sequences for Q$^{}_1$ and
    Q$^{}_2$);

\item A compensation pulse with fixed length
    $\tau^{}_{\textrm{cmp}}$ and variable amplitude
    $z^{}_{\textrm{cmp}}$ is applied to Q$^{}_2$;

\item A rotation $\hat{R}^{\pi / 2}_y$ with FWHM
    $\tau^{}_{\textrm{FWHM}}$ is applied to Q$^{}_2$;

\item The state $| \textrm{e} \rangle$ of Q$^{}_2$ is
    measured, thus obtaining the probability
    $P^{}_{\textrm{e}}$ as a function of
    $z^{}_{\textrm{cmp}}$
    (cf.~Fig.~\ref{Figure:Supporting:6:Matteo:Mariantoni:201107}B).
    Choosing a maximum of the probability
    $P^{}_{\textrm{e}}$ allows us to cancel the effect of
    the unwanted dynamic phase acquired by Q$^{}_2$ during
    and between the two iSWAPs.

\end{enumerate}

% *************************
% *** Systematic errors ***
% *************************
%
\subsubsection*{Systematic errors}
    \addcontentsline{toc}{subsubsection}{Systematic errors}

The Ramsey fringes of
Fig.~\ref{Figure:Supporting:5:Matteo:Mariantoni:201107}, A and
B, which are used to obtain the phase $\phi$ associated with
three particular values of the detuning
$\delta^{}_{\textrm{Q}^{}_1 \textrm{B}}$, can be extended to
any arbitrary value of $\delta^{}_{\textrm{Q}^{}_1 \textrm{B}}$
to obtain all possible values of $\phi$. Figure~3C in the main
text shows, for example, the phase $\phi$ obtained for
$\delta^{}_{\textrm{Q}^{}_1 \textrm{B}} {} \in {} [ 0 , 70
]$\,MHz. In the figure, the theoretical expression for the
phase $\phi$ given by Eq.~\ref{Equation:Supporting:18} (solid
green line) is overlayed to the measured data (blue dots). The
coupling $\tilde{g}^{}_{\textrm{Q}^{}_1 \textrm{B}}$ used to
plot the theoretical curve was estimated from the time-domain
swaps of Fig.~3B in the main text. A qualitative inspection of
the figure shows that theory and measured data are shifted
along the vertical axis by a detuning
$\delta^{}_{\textrm{drift}}$, corresponding to an overall phase
shift along the horizontal axis.

The origin of the detuning $\delta^{}_{\textrm{drift}}$, and of
the corresponding phase shift, could be attributed to two main
causes: \textit{(i) -} Drift of the transition frequency of
qubit Q$^{}_1$ during the experiment; \textit{(ii) -} drift in
the room-temperature electronics. Each pair of Ramsey fringes
used to obtain the phase $\phi$ (e.g., the fringe in the
leftmost panel of
Fig.~\ref{Figure:Supporting:5:Matteo:Mariantoni:201107}A and
the corresponding fringe in the leftmost panel of
Fig.~\ref{Figure:Supporting:5:Matteo:Mariantoni:201107}B) was
measured within a few minutes. This time can be considered
short enough to exclude the electronics drift as a main cause
of the detuning between theory and data. We can thus assume the
drift in the qubit transition frequency as the main reason for
the detuning. This seems a fair assumption since the
measurement of the time-domain swaps of Fig.~3B in the main
text, which were used to calibrate the z-pulse amplitude
$z^{}_{\textrm{CZ-}\phi}$ and swapping time
$\tau^{}_{\textrm{CZ-}\phi}$ necessary to obtain each phase
$\phi$ (cf.~main text and the two previous supporting sections
on the CZ-$\phi$ gate theory and tuneup), took approximately
four hours. Both the swaps and Ramsey fringes were measured
starting from large detuning $\delta^{}_{\textrm{Q}^{}_1
\textrm{B}} {} = {} 70$\,MHz to zero detuning, resulting in a
time delay between swaps and Ramsey fringes for each value of
$\delta^{}_{\textrm{Q}^{}_1 \textrm{B}}$ of approximately four
hours. From independent measurements (not shown), in such a
time interval we expect the qubit transition frequency to drift
by a few mega hertz.

% ******************************
% *** Supplementary Figure 7 ***
% ******************************
%
\begin{figure}[t!]
    \centering
    \includegraphics[width=1.14\columnwidth]{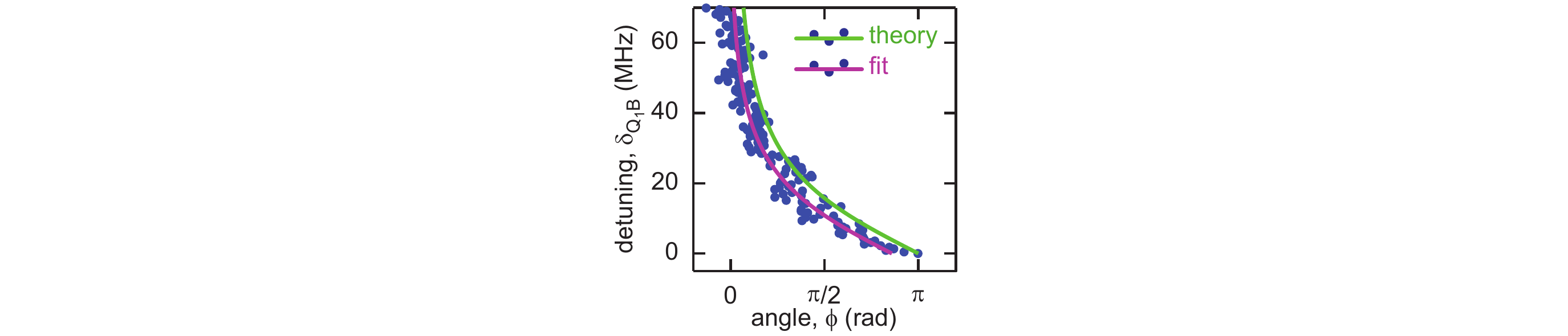}
    \caption{\footnotesize
\textbf{Qubit frequency drift.} Phase $\phi$ acquired by
Q$^{}_1$ as a function of the detuning
$\delta^{}_{\textrm{Q}^{}_1 \textrm{B}}$. The blue dots
indicate the same experimental data as in Fig.~3C of the main
text. The solid green line is the theory given by
Eq.~\ref{Equation:Supporting:18}, and the solid magenta line a
fit to the function given by Eq.~\ref{Equation:Supporting:19},
where $\delta^{}_{\textrm{drift}}$ is the only free fitting
parameter.
    }
    \label{Figure:Supporting:7:Matteo:Mariantoni:201107}
\end{figure}

In order to quantify the detuning $\delta^{}_{\textrm{drift}}$,
we can fit the data of Fig.~3C in the main text with the
function
\begin{equation}
\phi {} \equiv {} \pi - \pi \, \frac{\delta^{}_{\textrm{Q}^{}_1
\textrm{B}}}{\sqrt{\delta^2_{\textrm{Q}^{}_1 \textrm{B}} +
\tilde{g}^2_{\textrm{Q}^{}_1 \textrm{B}}}} + \pi \,
\frac{\delta^{}_{\textrm{drift}}}{\sqrt{\delta^2_{\textrm{Q}^{}_1
\textrm{B}} + \tilde{g}^2_{\textrm{Q}^{}_1 \textrm{B}}}} \, .
    \label{Equation:Supporting:19}
\end{equation}
The function of Eq.~\ref{Equation:Supporting:19} represents the
best fit we found for the data of Fig.~3C in the main text.
Figure~\ref{Figure:Supporting:7:Matteo:Mariantoni:201107} shows
the same data as Fig.~3C of the main text (blue dots), together
with the theory of Eq.~\ref{Equation:Supporting:18} (solid
green line), and the fit of Eq.~\ref{Equation:Supporting:19}
(solid magenta line). The detuning obtained from the fit is $|
\delta^{}_{\textrm{drift}} | {} \simeq {} 4$\,MHz, which is
consistent with our expectation for a qubit frequency drift in
approximately four hours.

We note that the qubit frequency drift as well as the data
scatter in Fig.~3C of the main text (or, equivalently, of
Fig.~\ref{Figure:Supporting:7:Matteo:Mariantoni:201107}) are
peculiar to that measurement, where we intended to show all
phases $\phi {} \in {} ( 0 , \pi ]$ in a single, long scan.
When it will be required to use a specific phase $\phi$ to
perform a quantum Fourier transform during an algorithm, we
will first theoretically estimate the parameters
$z^{}_{\textrm{CZ-}\phi}$ and $\tau^{}_{\textrm{CZ-}\phi}$ and,
then, search for the phase $\phi$ in the close vicinity of
these parameters. This will allow us to measure only a small
portion of the swaps of Fig.~3B in the main text and a few
corresponding Ramsey fringes, which can be realized in a much
shorter time than the long scan of Fig.~3C in the main text,
thus significantly reducing the incidence of systematic errors.

% **********************************
% *** XOR gate and M gate tuneup ***
% **********************************
%
\subsection*{XOR gate and M gate tuneup}
    \addcontentsline{toc}{subsection}{XOR gate and M gate tuneup}

In this section, we describe the experimental pulse sequences
required to tune up the XOR gate and M gate shown in the main
text. We also explain in detail the complete experimental
sequence used to obtain one nontrivial entry of the truth table
associated with the M gate. Finally, we outline the
mathematical procedure at the basis of quantum phase
tomography.

% ******************************
% *** Supplementary Figure 8 ***
% ******************************
%
\begin{figure}[p!]
    \centering
    \includegraphics[width=1.14\columnwidth]{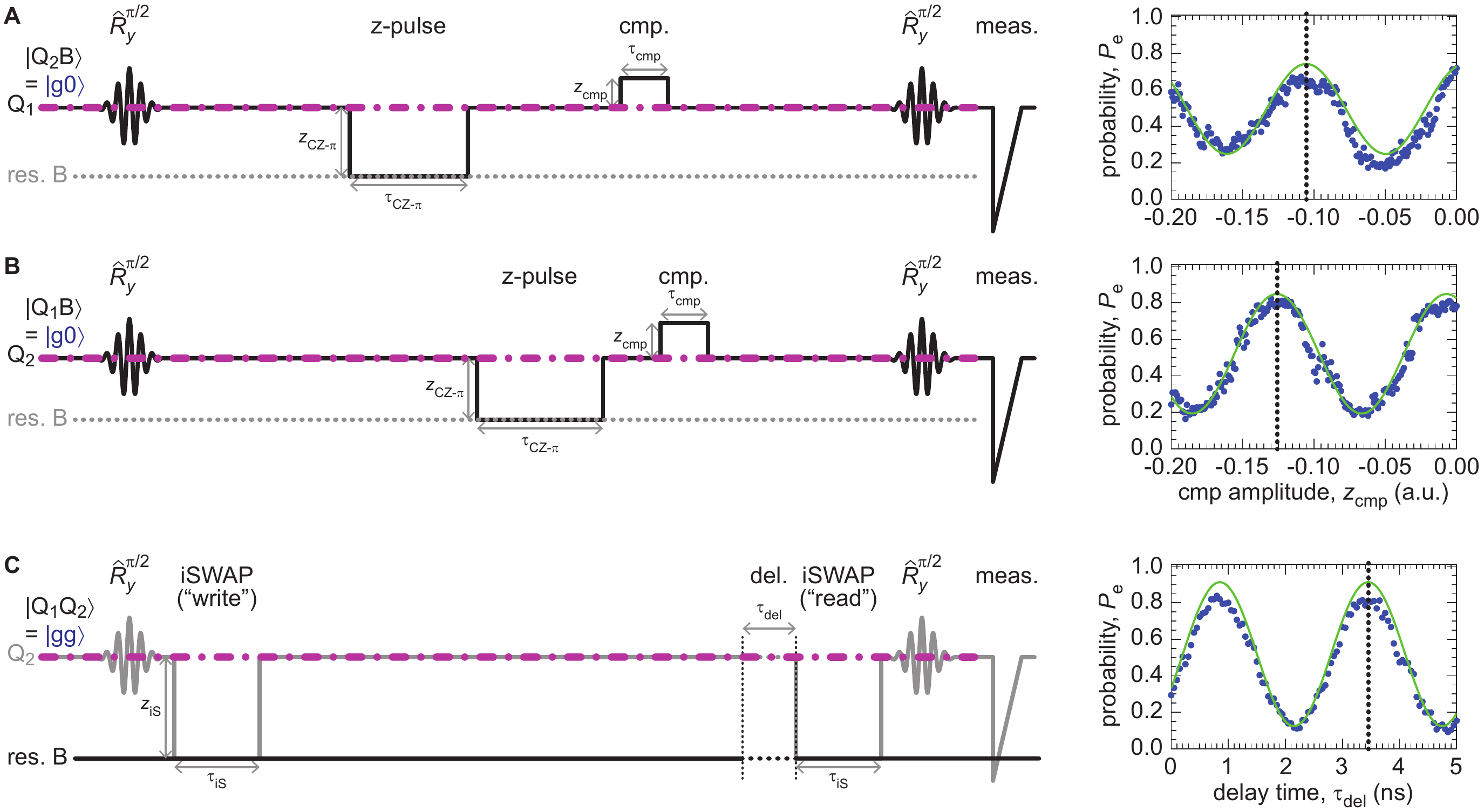}
    \caption{\footnotesize
\textbf{XOR gate tuneup.} (\textbf{A}) (Left) Tuneup
sequence~($1$-XOR). Qubit Q$^{}_1$ is initialized in $|
\textrm{g} \rangle$ at the idle point. Qubit Q$^{}_2$ and
resonator B remain in state $| \textrm{Q}^{}_2 \textrm{B}
\rangle {} = {} | \textrm{g} 0 \rangle$ during the whole
sequence. The reference frame of Q$^{}_1$ is indicated by a
dash-dot magenta line. Resonator B is indicated by a dotted
grey line. (Right) Ramsey fringe corresponding to the sequence
on the left showing the probability of measuring Q$^{}_1$ in $|
\textrm{e} \rangle$, $P^{}_{\textrm{e}}$, plotted vs.
$z^{}_{\textrm{cmp}}$. The blue dots represent measured data,
and the solid green line a least-squares fit to a sine
function. The vertical dotted black line indicates the
amplitude $z^{}_{\textrm{cmp}}$, obtained from the fit, chosen
to compensate the dynamic phase acquired by Q$^{}_1$ during the
first z-pulse in the sequence. (\textbf{B}) Tuneup
sequence~($2$-XOR). As in A, but for Q$^{}_2$. (\textbf{C})
(Left) Tuneup sequence~($3$-XOR). The solid grey line indicates
the ancilla qubit Q$^{}_2$, and the solid black line resonator
B. The reference frame is set by Q$^{}_2$ at the idle point, as
indicated by the dash-dot magenta line. (Right) Ramsey fringe
corresponding to the sequence on the left, where the
probability of measuring Q$^{}_2$ in $| \textrm{e} \rangle$,
$P^{}_{\textrm{e}}$, is plotted vs. $\tau^{}_{\textrm{del}}$.
The blue dots represent measured data, and the solid green line
a least-squares fit to a sine function. The vertical dotted
black line indicates the time $\tau^{}_{\textrm{del}}$,
obtained from the fit, chosen to calibrate away the dynamic
phase acquired by the state in B during and between the
``write'' and ``read'' iSWAP.
    }
    \label{Figure:Supporting:8:Matteo:Mariantoni:201107}
\end{figure}
    \clearpage

\subsubsection*{XOR gate tuneup}
    \addcontentsline{toc}{subsubsection}{XOR gate tuneup}

The quantum logic circuit of the XOR gate considered here is
sketched in Fig.~4A of the main text.
Figure~\ref{Figure:Supporting:8:Matteo:Mariantoni:201107}, A to
C, shows the three sequences and the corresponding Ramsey
fringes required to tune up the XOR gate. The concept behind
each sequence is similar to the compensation of a dynamic phase
acquired during a CZ-$\phi$ gate, which has been elucidated in
the section ``The quantum Fourier transform'' of these Methods.
The third sequence needs special attention as it is applied to
a resonator rather than a qubit state.

    \renewcommand{\labelenumi}{(\arabic{enumi}-XOR)}
\begin{enumerate}

\item The first sequence, which is displayed in
    Fig.~\ref{Figure:Supporting:8:Matteo:Mariantoni:201107}A
    (Left), acts on control qubit Q$^{}_1$. During the
    entire sequence, the control qubit Q$^{}_2$ and the
    target bus resonator B remain in the state $|
    \textrm{Q}^{}_2 \textrm{B} \rangle {} = {} | \textrm{g}
    0 \rangle$, and all pulses acting on Q$^{}_2$ are
    turned off. Qubit Q$^{}_1$ is initialized in the ground
    state $| \textrm{Q}^{}_1 \rangle {} = {} | \textrm{g}
    \rangle$ at the idle point.

    As for the case of the CZ-$\phi$ gate, the sequence
    consists of a Ramsey-type experiment used to determine
    and calibrate away the total dynamic phase acquired by
    Q$^{}_1$ during the XOR gate. Hence, the first step of
    the sequence is an $\hat{R}^{\pi / 2}_y$ unitary
    rotation on Q$^{}_1$, which brings the qubit to the
    equator of the Bloch sphere.

    Then, a z-pulse with amplitude $z^{}_{\textrm{CZ-}\pi}$
    and time $\tau^{}_{\textrm{CZ-}\pi} {} \simeq {}
    39.08$\,ns, brings the $| \textrm{e} \rangle {}
    \leftrightarrow {} | \textrm{f} \rangle$ transition of
    Q$^{}_1$ into resonance with B. The time
    $\tau^{}_{\textrm{CZ-}\pi}$ is inversely proportional
    to the coupling strength between states $|
    \textrm{Q}^{}_1 \textrm{B} \rangle {} = {} | \textrm{e}
    1 \rangle$ and $| \textrm{f} 0 \rangle$. However, we
    note that in the sequence~($1$-XOR), B is always in the
    vacuum state $| 0 \rangle$ and, thus, no dynamics takes
    place between Q$^{}_1$ and B during the z-pulse. The
    only effect of the z-pulse is to detune Q$^{}_1$
    outside its reference frame, which causes the qubit to
    acquire an unwanted dynamic phase
    $\phi^{}_{\textrm{dyn}}$.

    In order to compensate $\phi^{}_{\textrm{dyn}}$, a
    second z-pulse must be applied to Q$^{}_1$. Such a
    pulse is characterized by a fixed time-length
    $\tau^{}_{\textrm{cmp}} {} = {} 5$\,ns and a variable
    amplitude $z^{}_{\textrm{cmp}}$. We note that all
    compensation pulses in the XOR- and M-gate tuneup
    sequences have the same time length of $5$\,ns. By
    continuously varying the amplitude
    $z^{}_{\textrm{cmp}}$, the Ramsey fringe shown in
    Fig.~\ref{Figure:Supporting:8:Matteo:Mariantoni:201107}A
    (Right) is obtained. The dynamic phase
    $\phi^{}_{\textrm{dyn}}$ is totally compensated when
    the probability $P^{}_{\textrm{e}}$ reaches a maximum.
    In the figure, the vertical dotted black line indicates
    the compensation pulse amplitude chosen for this
    purpose, $z^{}_{\textrm{cmp}} {} \simeq {} - 0.106$;

\item The second tuneup sequence is displayed in
    Fig.~\ref{Figure:Supporting:8:Matteo:Mariantoni:201107}B
    (Left). The sequence is analogous to
    sequence~($1$-XOR), but acting on control qubit
    Q$^{}_2$ instead of Q$^{}_1$. In this case,
    $z^{}_{\textrm{cmp}} {} \simeq {} - 0.127$
    [cf.~Fig.~\ref{Figure:Supporting:8:Matteo:Mariantoni:201107}B
    (Right)];

\item The third and last tuneup sequence of the XOR gate,
    which is shown in
    Fig.~\ref{Figure:Supporting:8:Matteo:Mariantoni:201107}C
    (Left), acts on target resonator B. Sequence~($3$-XOR)
    represents a departure from the analogy between the
    tuneup of the XOR gate and the CZ-$\phi$ gate, where
    only two compensation pulses were needed for the gate
    operation.

    As already explained in the main text, resonator B
    plays the role of the third qubit in our implementation
    of three-qubit phase gates. In order to use resonator B
    as an effective qubit, its state must be prepared and
    measured using either qubit Q$^{}_1$ or Q$^{}_2$ as an
    ancilla qubit. In our experiments, we have chosen
    Q$^{}_2$ to perform this function because of slightly
    better coherence times and measurement fidelities
    compared to Q$^{}_1$. It is important to note that
    Q$^{}_2$ is actively used during the XOR gate.
    Consequently, the state of B can only be controlled
    before and/or after the gate operation. This issue does
    not constitute an experimental limitation since B
    represents the target of the gate and, thus, its state
    will not be controlled during the gate. However, once a
    state has been loaded in target B, it has to remain
    stored for a significantly longer time than any state
    stored in the control qubits Q$^{}_1$ and Q$^{}_2$.
    This is not an experimental limitation either, as the
    much longer coherence times of B compared to Q$^{}_1$
    and Q$^{}_2$ (cf.~caption of Fig.~1B in the main text
    for numerical values) largely reduce the effect of a
    longer storing time. This experiment further proves the
    importance of the quantum von Neumann architecture,
    where the ability to store states in a memory makes
    possible to realize longer quantum computations.

    As for the control qubits Q$^{}_1$ and Q$^{}_2$, also
    the state loaded in the target resonator B acquires a
    dynamic phase due to the detuning between the
    transition frequency of B and the reference clock rate
    of Q$^{}_2$. Note that only the frequency detuning with
    respect to Q$^{}_2$ contributes to the dynamic phase of
    the state in B because Q$^{}_2$ is the ancilla qubit
    chosen to manipulate and measure B.

    Following the pulses in
    Fig.~\ref{Figure:Supporting:8:Matteo:Mariantoni:201107}C
    (Left), the Ramsey experiment necessary to compensate
    the dynamic phase acquired by B is indirectly preformed
    through
    Q$^{}_2$~\cite{wang:2008:fockdecay,wang:2009:wignerdecay}.
    Resonator B is initialized in the vacuum state $|
    \textrm{B} \rangle {} = {} | 0 \rangle$, and qubit
    Q$^{}_1$ and Q$^{}_2$ in the state $| \textrm{Q}^{}_1
    \textrm{Q}^{}_2 \rangle {} = {} | \textrm{g} \textrm{g}
    \rangle$, with both qubits at the idle point. While
    Q$^{}_1$ remains in $| \textrm{g} \rangle$ during the
    whole sequence, Q$^{}_2$ is rotated into a linear
    superposition $| \textrm{g} \rangle + | \textrm{e}
    \rangle$ by means of an $\hat{R}^{\pi / 2}_y$ rotation.
    Afterwards, an iSWAP with amplitude
    $z^{}_{\textrm{iS}}$ and time $\tau^{}_{\textrm{iS}} {}
    \simeq {} 25.72$\,ns is used to write the state from
    Q$^{}_2$ into B, which is thus prepared in the state $|
    0 \rangle + | 1 \rangle$. This state remains loaded in
    B until it is read out by Q$^{}_2$ via a second iSWAP
    before the end of the sequence. Between the two iSWAPs,
    the control qubits Q$^{}_1$ and Q$^{}_2$ remain in the
    state $| \textrm{Q}^{}_1 \textrm{Q}^{}_2 \rangle {} =
    {} | \textrm{g} \textrm{g} \rangle$, and all pulses
    acting on both Q$^{}_1$ and Q$^{}_2$ are turned off.
    Due to the excursion outside Q$^{}_2$'s reference
    frame, the state $| 0 \rangle + | 1 \rangle$ in B
    acquires a dynamic phase, which grows until the end of
    the readout iSWAP. Since the resonance frequency of B
    cannot be tuned, in order to calibrate away the effect
    of such a dynamic phase we delay the starting time of
    the readout iSWAP by a variable time
    $\tau^{}_{\textrm{del}}$.

    Finally, after the readout iSWAP, a second
    $\hat{R}^{\pi / 2}_y$ rotation followed by a
    measurement pulse on Q$^{}_2$ completes the Ramsey
    experiment on B. The corresponding Ramsey fringe
    measured as a function of $\tau^{}_{\textrm{del}}$ is
    plotted in
    Fig.~\ref{Figure:Supporting:8:Matteo:Mariantoni:201107}C
    (Right). Similar to sequence~($1$-XOR) and ($2$-XOR),
    choosing $\tau^{}_{\textrm{del}}$ such that the Ramsey
    fringe reaches one maximum allows us to fully
    compensate the dynamic phase acquired by the state in
    B. The vertical dashed black line in the figure
    indicates the delay time chosen in the experiment,
    $\tau^{}_{\textrm{del}} {} \simeq {} 3.46$\,ns. As a
    check, from the fit we also obtained a Ramsey fringe
    frequency of $\simeq {} 385.0$\,MHz, which agrees well
    with the Q$^{}_2$-B detuning $\simeq {} 369.8$\,MHz.

\end{enumerate}

\subsubsection*{M gate tuneup}
    \addcontentsline{toc}{subsubsection}{M gate tuneup}

% ******************************
% *** Supplementary Figure 9 ***
% ******************************
%
\begin{figure}[p!]
    \centering
    \includegraphics[width=1.14\columnwidth]{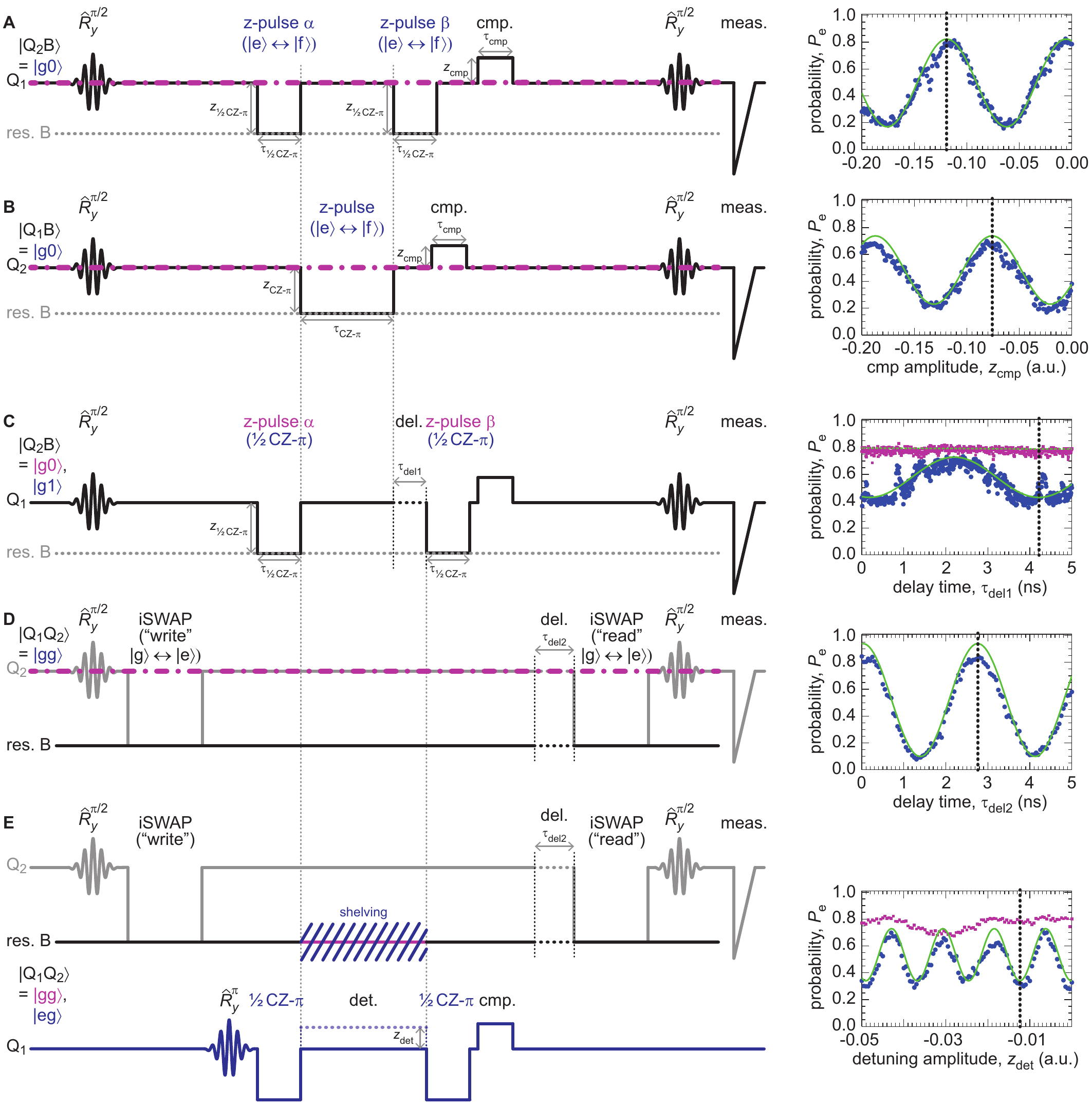}
    \caption{\footnotesize
\textbf{M gate tuneup.} (\textbf{A}) Sequence~($1$-M) for
calibrating the $\nicefrac{1}{2}\textrm{CZ-}\pi$ gates.
(\textbf{B}) Same sequence as in
Fig.~\ref{Figure:Supporting:8:Matteo:Mariantoni:201107}B.
(\textbf{C}) (Left) Sequence~($3$-M) for compensating the
dynamic phase acquired by Q$^{}_1$ during the shelving.
$\tau^{}_{\textrm{del} 1}$: Time delay between the z-pulse
$\alpha$ and z-pulse $\beta$ (i.e., between the two
$\sfrac{1}{2}$\,CZ-$\pi$ gates). (Right) Probability
$P^{}_{\textrm{e}}$ vs. $\tau^{}_{\textrm{del} 1}$ for $|
\textrm{Q}^{}_2 \textrm{B} \rangle {} = {} | \textrm{g} 0
\rangle$ (magenta squares) or $| \textrm{Q}^{}_2 \textrm{B}
\rangle {} = {} | \textrm{g} 1 \rangle$ (blue dots). Dashed and
solid green lines: Least-squares fit to data. (\textbf{D}) Same
sequence as in
Fig.~\ref{Figure:Supporting:8:Matteo:Mariantoni:201107}C.
(\textbf{E}) As in C, but for bus resonator B. In this case, a
detuning $z^{}_{\textrm{det}}$ is applied between the two
$\sfrac{1}{2}$\,CZ-$\pi$ gates. No shelving is indicated by a
solid magenta line, shelving by a hashed blue.
    }
    \label{Figure:Supporting:9:Matteo:Mariantoni:201107}
\end{figure}

The quantum logic circuit of the M gate considered here is
sketched in Fig.~4D of the main text.
Figure~\ref{Figure:Supporting:9:Matteo:Mariantoni:201107}, A to
E, shows the five sequences and the corresponding Ramsey
fringes required to tune up the M gate. The tuneup concept is
similar to that used for the XOR gate, but with a few important
differences due to the $\sfrac{1}{2}$\,CZ-$\pi$ gates.

    \renewcommand{\labelenumi}{(\arabic{enumi}-M)}
\begin{enumerate}

\item The first tuneup sequence for the M gate, which is
    displayed in
    Fig.~\ref{Figure:Supporting:9:Matteo:Mariantoni:201107}A
    (Left), acts on control qubit Q$^{}_1$. The only
    difference between this sequence and sequence~($1$-XOR)
    is that the single z-pulse with amplitude
    $z^{}_{\textrm{CZ-}\pi}$ and length
    $\tau^{}_{\textrm{CZ-}\pi}$ is now split into two
    z-pulses, z-pulse $\alpha$ and z-pulse $\beta$, with
    amplitude $z^{}_{\nicefrac{1}{2} \, \textrm{CZ-}\pi} {}
    = {} z^{}_{\textrm{CZ-}\pi}$ and time
    $\tau^{}_{\nicefrac{1}{2} \, \textrm{CZ-}\pi} {} = {}
    \tau^{}_{\textrm{CZ-}\pi} / 2$. This fact, however,
    does not affect the compensation of the dynamic phase
    acquired by Q$^{}_1$, as the total excursion of
    Q$^{}_1$ outside its reference frame remains unchanged.
    It is worth reminding that the z-pulse $\alpha$ and
    z-pulse $\beta$ bring the qubit $| \textrm{e} \rangle
    {} \leftrightarrow {} | \textrm{f} \rangle$ transition
    on resonance with bus resonator B. As a consequence, in
    sequence~($1$-M) the resonator remains in the vacuum
    state $| 0 \rangle$.

    From the Ramsey fringe in
    Fig.~\ref{Figure:Supporting:9:Matteo:Mariantoni:201107}A
    (Right) we obtain one possible value of the
    compensation pulse amplitude that maximizes the
    probability $P^{}_{\textrm{e}}$ of Q$^{}_1$,
    $z^{}_{\textrm{cmp}} {} \simeq {} - 0.119$;

\item The second tuneup sequence, which is displayed in
    Fig.~\ref{Figure:Supporting:9:Matteo:Mariantoni:201107}B
    (Left), acts on control qubit Q$^{}_2$. The sequence is
    the same as sequence~($2$-XOR) for the XOR gate. As in
    sequence~($1$-M), the z-pulse brings the qubit $|
    \textrm{e} \rangle {} \leftrightarrow {} | \textrm{f}
    \rangle$ transition on resonance with bus resonator B.

    From the Ramsey fringe in
    Fig.~\ref{Figure:Supporting:9:Matteo:Mariantoni:201107}B
    (Right) we obtain one possible value of the
    compensation pulse amplitude that maximizes the
    probability $P^{}_{\textrm{e}}$ of Q$^{}_2$,
    $z^{}_{\textrm{cmp}} {} \simeq {} - 0.077$;

\item The third tuneup sequence, which is displayed in
    Fig.~\ref{Figure:Supporting:9:Matteo:Mariantoni:201107}C
    (Left), acts again on control qubit Q$^{}_1$, with
    control qubit Q$^{}_2$ and target resonator B either in
    state $| \textrm{Q}^{}_2 \textrm{B} \rangle {} = {} |
    \textrm{g} 0 \rangle$ or $| \textrm{Q}^{}_2 \textrm{B}
    \rangle {} = {} | \textrm{g} 1 \rangle$. In addition,
    all pulses acting on Q$^{}_2$ are turned off. Qubit
    Q$^{}_1$ is initialized in the ground state $|
    \textrm{Q}^{}_1 \rangle {} = {} | \textrm{g} \rangle$
    at the idle point.

    In order to understand the dynamics of the interaction
    between Q$^{}_1$ and B, we refer to the energy diagram
    of
    Fig.~\ref{Figure:Supporting:3:Matteo:Mariantoni:201107}A.
    If $| \textrm{B} \rangle {} = {} | 0 \rangle$, after
    the first $\hat{R}^{\pi / 2}_y$ rotation on Q$^{}_1$,
    the Q$^{}_1$-B coupled system is in state $|
    \textrm{Q}^{}_1 \rangle \otimes | \textrm{B} \rangle {}
    = {} ( | \textrm{g} \rangle + | \textrm{e} \rangle )
    \otimes | 0 \rangle$. In this case, during the z-pulse
    $\alpha$ and z-pulse $\beta$ no dynamics takes place.
    As a consequence, in the time interval
    $\tau^{}_{\textrm{sh}}$ \textit{between} the two
    z-pulses, Q$^{}_1$ remains biased at the idle point in
    state $| \textrm{Q}^{}_1 \rangle {} = {} | \textrm{g}
    \rangle + | \textrm{e} \rangle$, without acquiring any
    dynamic phase. The only dynamic phase acquired by
    Q$^{}_1$ is that developed \textit{during} the z-pulse
    $\alpha$ and z-pulse $\beta$, which has already been
    compensated in sequence~($1$-M)
    (cf.~Fig.~\ref{Figure:Supporting:9:Matteo:Mariantoni:201107}A).
    The compensation pulse for such a dynamic phase remains
    turned on during sequence~($3$-M).

    If instead $| \textrm{B} \rangle {} = {} | 1 \rangle$,
    after the first $\hat{R}^{\pi / 2}_y$ rotation on
    Q$^{}_1$, the Q$^{}_1$-B coupled system is in state $|
    \textrm{Q}^{}_1 \rangle \otimes | \textrm{B} \rangle {}
    = {} ( | \textrm{g} \rangle + | \textrm{e} \rangle )
    \otimes | 1 \rangle$. In this case, the reference clock
    rate is given by $f^{}_{\textrm{Q}^{}_1} +
    f^{}_{\textrm{B}}$. After the z-pulse $\alpha$, the
    state $| \textrm{Q}^{}_1 \textrm{B} \rangle {} = {} |
    \textrm{e} 1 \rangle$ gets shelved into the state $|
    \textrm{Q}^{}_1 \textrm{B} \rangle {} = {} | \textrm{f}
    0 \rangle$ for a time $\tau^{}_{\textrm{sh}}$, at the
    end of which the z-pulse $\beta$ is applied. We remind
    that the $| \textrm{g} \rangle {} \leftrightarrow {} |
    \textrm{f} \rangle$ frequency is $( 2
    f^{}_{\textrm{Q}^{}_1} - \delta^{}_{\textrm{nl}} )$,
    where $\delta^{}_{\textrm{nl}}$ is the qubit
    nonlinearity defined as the frequency difference
    between the $| \textrm{e} \rangle {} \leftrightarrow {}
    | \textrm{f} \rangle$ and the $| \textrm{g} \rangle {}
    \leftrightarrow {} | \textrm{e} \rangle$ qubit
    transitions. In this experiment,
    $f^{}_{\textrm{Q}^{}_1} {} \simeq {} 7.2161$\,GHz,
    $\delta^{}_{\textrm{nl}} {} \simeq {} 140.6$\,MHz, and
    $f^{}_{\textrm{B}} {} \simeq {} 6.8150$\,GHz. During
    the time $\tau^{}_{\textrm{sh}}$, the coupled system is
    in state $| \textrm{Q}^{}_1 \textrm{B} \rangle {} = {}
    ( | \textrm{g} 1 \rangle + | \textrm{f} 0 \rangle )$
    and Q$^{}_1$ acquires a dynamic phase
    $\phi^{}_{\textrm{sh}} {} = {} ( f^{}_{\textrm{Q}^{}_1}
    - \delta^{}_{\textrm{nl}} - f^{}_{\textrm{B}} ) \,
    \tau^{}_{\textrm{sh}}$. This dynamic phase is
    independent from the dynamic phase acquired during the
    z-pulse $\alpha$ and z-pulse $\beta$.

    In order to compensate the phase
    $\phi^{}_{\textrm{sh}}$, we delay the starting point of
    the z-pulse $\beta$ by a time $\tau^{}_{\textrm{del}
    1}$. By continuously varying $\tau^{}_{\textrm{del}
    1}$, the two Ramsey fringes plotted in
    Fig.~\ref{Figure:Supporting:9:Matteo:Mariantoni:201107}C
    (Right) are obtained. The magenta squares correspond to
    the case $| \textrm{Q}^{}_2 \textrm{B} \rangle {} = {}
    | \textrm{g} 0 \rangle$. As expected, in this case
    nothing happens as the dynamic phases acquired during
    the z-pulse $\alpha$ and z-pulse $\beta$ were already
    corrected by the compensation pulse tuned up in
    sequence~($1$-M). The Ramsey fringe thus remains at the
    maximum probability chosen in that sequence. This
    fringe indicates that no resonator state has been
    shelved to the qutrit state $| \textrm{Q}^{}_1 \rangle
    {} = {} | \textrm{f} \rangle$. The blue dots, instead,
    correspond to the case $| \textrm{Q}^{}_2 \textrm{B}
    \rangle {} = {} | \textrm{g} 1 \rangle$. In this case,
    the sinusoidal dependence of the fringe clearly shows
    an excursion outside the Q$^{}_1$-B reference frame,
    which causes the dynamic phase $\phi^{}_{\textrm{sh}}$
    to be acquired by the state in Q$^{}_1$. This phase is
    totally compensated when the probability
    $P^{}_{\textrm{e}}$ of Q$^{}_1$ reaches a minimum. The
    reason why a minimum has to be chosen is because the
    Ramsey fringe is obtained with one excitation in the
    system, instead of no excitation as in
    sequences~($1$-M) and ($2$-M). This is analogous to the
    CZ-$\phi$ gate tuneup sequence displayed in
    Fig.~\ref{Figure:Supporting:4:Matteo:Mariantoni:201107}B.
    This can also be understood from the phase-gate cube of
    Fig.~\ref{Figure:Supporting:11:Matteo:Mariantoni:201107}E
    (or the second row in
    Table~\ref{Table:Supporting:3:Matteo:Mariantoni:201107}),
    as the Ramsey fringe in sequence~($3$-M) measures the
    phase difference between vertex ($5$) and vertex ($1$)
    of the cube, which is $\pi$\,rad instead of $0$\,rad.
    The vertical dotted black line in
    Fig.~\ref{Figure:Supporting:9:Matteo:Mariantoni:201107}C
    (Right) indicates the delay time chosen in the
    experiment, $\tau^{}_{\textrm{del} 1} {} \simeq {}
    4.2$\,ns. As a check, a least-squares fit to the data
    (solid green line) allows us to extract the frequency
    of the Ramsey fringe, which is $\simeq {} 248.2$\,MHz.
    This number is close to the expected value $(
    f^{}_{\textrm{Q}^{}_1} - \delta^{}_{\textrm{nl}} -
    f^{}_{\textrm{B}} ) {} \simeq {} 260.5$\,MHz (note that
    the data consists of one Ramsey oscillation period
    only. Hence, the $\simeq {} 10$\,MHz difference between
    the theoretically expected value and that obtained from
    the fit is within the fit confidence interval);

\item The fourth tuneup sequence, which is displayed in
    Fig.~\ref{Figure:Supporting:9:Matteo:Mariantoni:201107}D
    (Left), acts on target resonator B. The sequence is the
    same as sequence~($3$-XOR) for the XOR gate. In this
    sequence, the two iSWAPs bring the qubit $| \textrm{g}
    \rangle {} \leftrightarrow {} | \textrm{e} \rangle$
    transition on resonance with bus resonator B.

    From the Ramsey fringe in
    Fig.~\ref{Figure:Supporting:9:Matteo:Mariantoni:201107}D
    (Right) we obtain one possible value of the delay time
    that maximizes the probability $P^{}_{\textrm{e}}$ of
    Q$^{}_2$, $\tau^{}_{\textrm{del} 2} {} \simeq {}
    2.77$\,ns (vertical dotted black line). Since the XOR
    gate and M gate experiments were performed several
    hours apart, the delay time for the M gate differs
    slightly from that for the XOR gate because of drifts
    in the qubit transition frequency (cf.~section on
    ``Systematic errors'' in these Supporting Online
    Material). In addition, from a least-squares fit to the
    data (solid green line) we extracted a Ramsey fringe
    frequency $\simeq {} 365.8$\,MHz, which agrees well
    with the detuning between the reference clock rate of
    Q$^{}_2$ and the transition frequency of B, $\simeq {}
    369.8$\,MHz;

\item The fifth tuneup sequence, which is displayed in
    Fig.~\ref{Figure:Supporting:9:Matteo:Mariantoni:201107}E
    (Left), acts again on target resonator B.

    Sequence~($3$-M) served to compensate the dynamic phase
    acquired by Q$^{}_1$ during the shelving dynamics.
    Because B takes also part in the shelving, its state
    acquires a similar dynamic phase that must be
    compensated. Sequence~($5$-M) is analogous to
    sequence~($3$-M), with two differences. First, the
    Ramsey experiment is now performed on B via Q$^{}_2$
    [as for sequence~($4$-M)]. Second, instead of adding a
    time delay, qubit Q$^{}_1$ is detuned in the z
    direction by a quantity $z^{}_{\textrm{det}}$ for the
    entire interval between the z-pulse $\alpha$ and
    z-pulse $\beta$. In fact, we are not allowed to use two
    times the same degree of freedom, i.e., the delay time
    $\tau^{}_{\textrm{del} 1}$ of sequence~($3$-M), for the
    compensation of two independent dynamic phases. By
    continuously varying $z^{}_{\textrm{det}}$, the two
    Ramsey fringes plotted in
    Fig.~\ref{Figure:Supporting:9:Matteo:Mariantoni:201107}E
    (Right) are obtained. The interpretation of the fringes
    is the same as for sequence~($3$-M). The detuning value
    chosen in the experiment to compensate the dynamic
    phase acquired by the state in B during the shelving is
    $z^{}_{\textrm{det}} {} \simeq {} - 0.012$ (vertical
    dotted black line);

\item The sixth and last tuneup sequence for the M gate,
    which is not shown in
    Fig.~\ref{Figure:Supporting:9:Matteo:Mariantoni:201107},
    consists in repeating sequence~($1$-M). The
    compensation pulse for Q$^{}_1$ needs to be
    recalibrated due to the detuning $z^{}_{\textrm{det}}$
    set in sequence~($5$-M). The final value of the
    compensation pulse amplitude for Q$^{}_1$ chosen in the
    experiment is $z^{}_{\textrm{cmp}} {} \simeq {} -
    0.125$.

\end{enumerate}

\subsubsection*{M gate pulse sequence}
    \addcontentsline{toc}{subsubsection}{M gate pulse sequence}

% *******************************
% *** Supplementary Figure 10 ***
% *******************************
%
\begin{figure}[t!]
    \centering
    \includegraphics[width=1.14\columnwidth]{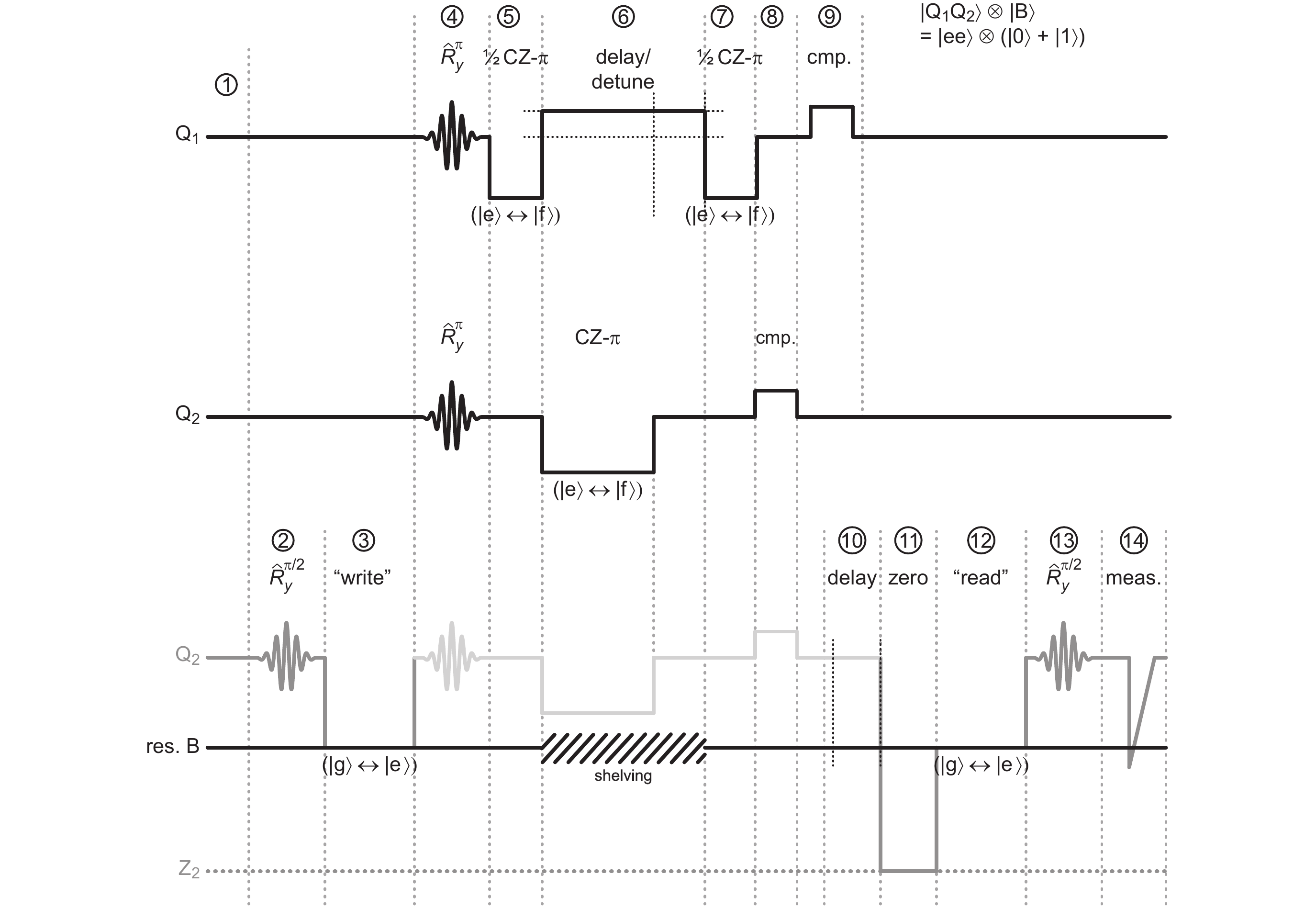}
    \caption{\footnotesize
\textbf{M gate pulse sequence.} Pulse sequence used to measure
the Ramsey fringe in Fig.~4E of the main text (magenta dots).
The vertical dotted grey lines separate the sequence in $14$
time frames. Each frame is described in the text.
    }
    \label{Figure:Supporting:10:Matteo:Mariantoni:201107}
\end{figure}

Figure~\ref{Figure:Supporting:10:Matteo:Mariantoni:201107}
shows the complete pulse sequence utilized to measure the entry
of the M gate truth table associated with state $|
\textrm{Q}^{}_1 \textrm{Q}^{}_2 \rangle \otimes | \textrm{B}
\rangle {} = {} | \textrm{e} \textrm{e} \rangle \otimes ( | 0
\rangle + | 1 \rangle )$. Step~($1$): Both control qubits
Q$^{}_1$ and Q$^{}_2$ and target resonator B are initialized in
the ground state, $| \textrm{Q}^{}_1 \textrm{Q}^{}_2 \textrm{B}
\rangle {} = {} | \textrm{g} \textrm{g} 0 \rangle$. The qubits
are biased at the idle point. Step~($2$): Q$^{}_2$ is prepared
in state $| \textrm{g} \rangle + | \textrm{e} \rangle$ by means
of an $\hat{R}^{\pi / 2}_y$ rotation. Step~($3$): The state in
qubit Q$^{}_2$ is written into B via an iSWAP. Until this step,
qubit Q$^{}_2$ serves as ancilla qubit to load resonator B. The
iSWAP effectively zeros Q$^{}_2$, which can be now used as a
control qubit in the M gate. Step~($4$): Both control qubits
Q$^{}_1$ and Q$^{}_2$ are loaded in state $| \textrm{e}
\rangle$ by means of an $\hat{R}^{\pi}_y$ rotation. Step~($5$):
First $\sfrac{1}{2}$\,CZ-$\pi$ gate between Q$^{}_1$ and B.
Step~($6$): CZ-$\pi$ gate between Q$^{}_2$ and B. In the same
time frame, the delay and detune necessary to compensate the
dynamic phase on Q$^{}_1$ and B due to the shelving are
applied. Step~($7$): Second $\sfrac{1}{2}$\,CZ-$\pi$ gate
between Q$^{}_1$ and B. Step~($8$): Compensation pulse on
Q$^{}_2$. Step~($9$): Compensation pulse on Q$^{}_1$.
Step~($10$): Compensation delay for the dynamic phase on B.
Step~($11$): Zeroing gate applied to Q$^{}_2$. This step is
necessary to re-use Q$^{}_2$ as ancilla qubit for controlling
B. The zeroing is performed through an iSWAP between Q$^{}_2$
and Z$^{}_2$. Step~($12$): The state of B is read out by the
zeroed Q$^{}_2$ via an iSWAP. Steps~($13$) and ($14$): A second
$\hat{R}^{\pi / 2}_y$ rotation on Q$^{}_2$ followed by a
measurement pulse completes the Ramsey experiment on B. The
Ramsey fringe obtained from this sequence is plotted in Fig.~4E
of the main text (magenta dots).

\subsection*{Quantum phase tomography}
    \addcontentsline{toc}{subsection}{Quantum phase tomography}

The most general unitary operation describing a three-qubit
controlled-phase quantum gate can be written as
\begin{equation}
 \widetilde{U}^{\phi}_{} {} = {}
 \begin{pmatrix}
  e^{i \, \phi^{}_{| \textrm{g} \textrm{g} 0 \rangle}}_{} & 0 & 0 & 0 & 0 & 0 & 0 & 0 \\
  0 & e^{i \, \phi^{}_{| \textrm{g} \textrm{g} 1 \rangle}}_{} & 0 & 0 & 0 & 0 & 0 & 0 \\
  0 & 0 & e^{i \, \phi^{}_{| \textrm{g} \textrm{e} 0 \rangle}}_{} & 0 & 0 & 0 & 0 & 0 \\
  0 & 0 & 0 & e^{i \, \phi^{}_{| \textrm{g} \textrm{e} 1 \rangle}}_{} & 0 & 0 & 0 & 0 \\
  0 & 0 & 0 & 0 & e^{i \, \phi^{}_{| \textrm{e} \textrm{g} 0 \rangle}}_{} & 0 & 0 & 0 \\
  0 & 0 & 0 & 0 & 0 & e^{i \, \phi^{}_{| \textrm{e} \textrm{g} 1 \rangle}}_{} & 0 & 0 \\
  0 & 0 & 0 & 0 & 0 & 0 & e^{i \, \phi^{}_{| \textrm{e} \textrm{e} 0 \rangle}}_{} & 0 \\
  0 & 0 & 0 & 0 & 0 & 0 & 0 & e^{i \, \phi^{}_{| \textrm{e} \textrm{e} 1 \rangle}}_{} \\
 \end{pmatrix} \, .
    \label{Equation:Supporting:20}
\end{equation}
In the ideal case, the amplitude of each diagonal element of
Eq.~\ref{Equation:Supporting:20} is unity, while the phase
$\phi^{}_{| l m n \rangle}$ depends on which state $| l m n
\rangle {} \in {} \mathcal{M}^{}_3$ is considered (cf.~main
text). All off-diagonal elements are zero.

It is straightforward to show that only seven of the eight
phases of Eq.~\ref{Equation:Supporting:20} are physically
independent. In fact, the first complex exponential $e^{i \,
\phi^{}_{| \textrm{g} \textrm{g} 0 \rangle}}_{}$ can be
factored out from the equation, allowing us to write the matrix
\begin{equation}
{\footnotesize
 U^{\phi}_{} {} = {}
 \begin{pmatrix}
  e^{i \, 0}_{} & 0 & 0 & 0 & 0 & 0 & 0 & 0 \\
  0 & e^{i \, ( \phi^{}_{| \textrm{g} \textrm{g} 1 \rangle} - \phi^{}_{| \textrm{g} \textrm{g} 0 \rangle} )}_{} & 0 & 0 & 0 & 0 & 0 & 0 \\
  0 & 0 & e^{i \, ( \phi^{}_{| \textrm{g} \textrm{e} 0 \rangle} - \phi^{}_{| \textrm{g} \textrm{g} 0 \rangle} )}_{} & 0 & 0 & 0 & 0 & 0 \\
  0 & 0 & 0 & e^{i \, ( \phi^{}_{| \textrm{g} \textrm{e} 1 \rangle} - \phi^{}_{| \textrm{g} \textrm{g} 0 \rangle} )}_{} & 0 & 0 & 0 & 0 \\
  0 & 0 & 0 & 0 & e^{i \, ( \phi^{}_{| \textrm{e} \textrm{g} 0 \rangle} - \phi^{}_{| \textrm{g} \textrm{g} 0 \rangle} )}_{} & 0 & 0 & 0 \\
  0 & 0 & 0 & 0 & 0 & e^{i \, ( \phi^{}_{| \textrm{e} \textrm{g} 1 \rangle} - \phi^{}_{| \textrm{g} \textrm{g} 0 \rangle} )}_{} & 0 & 0 \\
  0 & 0 & 0 & 0 & 0 & 0 & e^{i \, ( \phi^{}_{| \textrm{e} \textrm{e} 0 \rangle} - \phi^{}_{| \textrm{g} \textrm{g} 0 \rangle} )}_{} & 0 \\
  0 & 0 & 0 & 0 & 0 & 0 & 0 & e^{i \, ( \phi^{}_{| \textrm{e} \textrm{e} 1 \rangle} - \phi^{}_{| \textrm{g} \textrm{g} 0 \rangle} )}_{} \\
 \end{pmatrix} \, ,
    \label{Equation:Supporting:21}
}
\end{equation}
which is equivalent to $\widetilde{U}^{\phi}_{}$ up to a global
phase $\phi^{}_{| \textrm{g} \textrm{g} 0 \rangle}$.

The phases associated with each diagonal element in
Eq.~\ref{Equation:Supporting:21} can be grouped in a column
vector ${\bm \tau}^{}_{}$ defined as
\begin{equation}
 {\bm \tau}^{}_{} {} = {}
 \begin{pmatrix}
  0 \\
  \phi^{}_{| \textrm{g} \textrm{g} 1 \rangle} - \phi^{}_{| \textrm{g} \textrm{g} 0 \rangle} \\
  \phi^{}_{| \textrm{g} \textrm{e} 0 \rangle} - \phi^{}_{| \textrm{g} \textrm{g} 0 \rangle} \\
  \phi^{}_{| \textrm{g} \textrm{e} 1 \rangle} - \phi^{}_{| \textrm{g} \textrm{g} 0 \rangle} \\
  \phi^{}_{| \textrm{e} \textrm{g} 0 \rangle} - \phi^{}_{| \textrm{g} \textrm{g} 0 \rangle} \\
  \phi^{}_{| \textrm{e} \textrm{g} 1 \rangle} - \phi^{}_{| \textrm{g} \textrm{g} 0 \rangle} \\
  \phi^{}_{| \textrm{e} \textrm{e} 0 \rangle} - \phi^{}_{| \textrm{g} \textrm{g} 0 \rangle} \\
  \phi^{}_{| \textrm{e} \textrm{e} 1 \rangle} - \phi^{}_{| \textrm{g} \textrm{g} 0 \rangle} \\
 \end{pmatrix} \, ,
    \label{Equation:Supporting:22}
\end{equation}
with dimensions $( 8 , 1 )$.

In the main text we have shown that by performing Ramsey
experiments on the control qubits Q$^{}_1$ and Q$^{}_2$ and on
the target resonator B, it is possible to obtain the quantum
phase tomography of the three-qubit XOR phase gate and of the
Toffoli-class OR phase gate (M gate). In each Ramsey experiment
one of the control qubits (or the target resonator) has to be
prepared in a $| \textrm{g} \rangle + | \textrm{e} \rangle$ (or
$| 0 \rangle + | 1 \rangle$) state, while the other control
qubit and the target resonator (or the two control qubits) are
prepared in all four possible combinations of ground and
excited state. In the case of an M gate, for example, the
twelve states for each Ramsey experiment are reported in the
first three columns of
Table~\ref{Table:Supporting:3:Matteo:Mariantoni:201107}. The
fourth column shows the ideal value of the phase difference
associated with each Ramsey
experiment~\cite{ramsey:phase:difference}. A similar Table can
easily be obtained for the XOR gate (not shown).

\begin{table}[b!]
    \centering
    \caption{\footnotesize \textbf{M gate Ramsey table.}
             The first three columns indicate the state of the control
             qubits Q$^{}_1$ and Q$^{}_2$ and the state of the target resonator B
             for the twelve Ramsey experiments needed for quantum phase tomography.
             The fourth column shows the phase difference obtained
             from each Ramsey measurement for an ideal M gate.}
    \vspace{13.0pt}
    \begin{tabular}{c|c|c|c}
    \hline \hline
        Q$^{}_1$ & Q$^{}_2$ & B & phase difference (rad) - ideal case \\
    \hline
        $| \textrm{g} \rangle + | \textrm{e} \rangle$ & $| \textrm{g} \rangle$ & $| 0 \rangle$ & 0 \\
    \hline
        $| \textrm{g} \rangle + | \textrm{e} \rangle$ & $| \textrm{g} \rangle$ & $| 1 \rangle$ & $\pi$ \\
    \hline
        $| \textrm{g} \rangle + | \textrm{e} \rangle$ & $| \textrm{e} \rangle$ & $| 0 \rangle$ & $0$ \\
    \hline
        $| \textrm{g} \rangle + | \textrm{e} \rangle$ & $| \textrm{e} \rangle$ & $| 1 \rangle$ & $0$ \\
    \hline
        $| \textrm{g} \rangle$ & $| \textrm{g} \rangle + | \textrm{e} \rangle$ & $| 0 \rangle$ & $0$ \\
    \hline
        $| \textrm{g} \rangle$ & $| \textrm{g} \rangle + | \textrm{e} \rangle$ & $| 1 \rangle$ & $\pi$ \\
    \hline
        $| \textrm{e} \rangle$ & $| \textrm{g} \rangle + | \textrm{e} \rangle$ & $| 0 \rangle$ & $0$ \\
    \hline
        $| \textrm{e} \rangle$ & $| \textrm{g} \rangle + | \textrm{e} \rangle$ & $| 1 \rangle$ & $0$ \\
    \hline
        $| \textrm{g} \rangle$ & $| \textrm{g} \rangle$ & $| 0 \rangle + | 1 \rangle$ & $0$ \\
    \hline
        $| \textrm{g} \rangle$ & $| \textrm{e} \rangle$ & $| 0 \rangle + | 1 \rangle$ & $\pi$ \\
    \hline
        $| \textrm{e} \rangle$ & $| \textrm{g} \rangle$ & $| 0 \rangle + | 1 \rangle$ & $\pi$ \\
    \hline
        $| \textrm{e} \rangle$ & $| \textrm{e} \rangle$ & $| 0 \rangle + | 1 \rangle$ & $\pi$ \\
    \hline \hline
    \end{tabular}
        \label{Table:Supporting:3:Matteo:Mariantoni:201107}
\end{table}

We note that the states of the control qubits Q$^{}_1$ and
Q$^{}_2$ and of the target resonator B displayed in the first
three columns of
Table~\ref{Table:Supporting:3:Matteo:Mariantoni:201107}
constitute a general set of states for quantum phase tomography
and, thus, can be used to characterize any type of three-qubit
controlled-phase quantum gate. The phase differences associated
with these states can be grouped in a column vector ${\bm
\varphi}^{}_{}$ defined as
\begin{equation}
 {\bm \varphi}^{}_{} {} = {}
 \begin{pmatrix}
  \phi^{}_{| \textrm{e} \textrm{g} 0 \rangle} - \phi^{}_{| \textrm{g} \textrm{g} 0 \rangle} \\
  \phi^{}_{| \textrm{e} \textrm{g} 1 \rangle} - \phi^{}_{| \textrm{g} \textrm{g} 1 \rangle} \\
  \phi^{}_{| \textrm{e} \textrm{e} 0 \rangle} - \phi^{}_{| \textrm{g} \textrm{e} 0 \rangle} \\
  \phi^{}_{| \textrm{e} \textrm{e} 1 \rangle} - \phi^{}_{| \textrm{g} \textrm{e} 1 \rangle} \\
  \phi^{}_{| \textrm{g} \textrm{e} 0 \rangle} - \phi^{}_{| \textrm{g} \textrm{g} 0 \rangle} \\
  \phi^{}_{| \textrm{g} \textrm{e} 1 \rangle} - \phi^{}_{| \textrm{g} \textrm{g} 1 \rangle} \\
  \phi^{}_{| \textrm{e} \textrm{e} 0 \rangle} - \phi^{}_{| \textrm{e} \textrm{g} 0 \rangle} \\
  \phi^{}_{| \textrm{e} \textrm{e} 1 \rangle} - \phi^{}_{| \textrm{e} \textrm{g} 1 \rangle} \\
  \phi^{}_{| \textrm{g} \textrm{g} 1 \rangle} - \phi^{}_{| \textrm{g} \textrm{g} 0 \rangle} \\
  \phi^{}_{| \textrm{g} \textrm{e} 1 \rangle} - \phi^{}_{| \textrm{g} \textrm{e} 0 \rangle} \\
  \phi^{}_{| \textrm{e} \textrm{g} 1 \rangle} - \phi^{}_{| \textrm{e} \textrm{g} 0 \rangle} \\
  \phi^{}_{| \textrm{e} \textrm{e} 1 \rangle} - \phi^{}_{| \textrm{e} \textrm{e} 0 \rangle} \\
 \end{pmatrix} \, ,
    \label{Equation:Supporting:23}
\end{equation}
with dimensions $( 12 , 1 )$.

The aim of quantum phase tomography is to obtain the seven
phase differences in vector ${\bm \tau^{}_{}}$ from the twelve
phase differences in vector ${\bm \varphi}$, which are measured
by means of Ramsey experiments. In order to facilitate the
explanation of quantum phase tomography, we now introduce a
geometric representation of the phases associated with any
three-qubit controlled-phase quantum gate.
Figure~\ref{Figure:Supporting:11:Matteo:Mariantoni:201107}A
shows a cube, hereafter termed the phase-gate cube, the
vertices of which contain information on the diagonal elements
of a three-qubit controlled-phase quantum gate. The vertices of
the phase-gate cube are enumerated according to the notation
given in the space between Fig.~4C and Fig.~4F of the main
text. Note that the phase-gate cube can directly be generalized
to $N$-qubit gates, in which case it has to be promoted to an
$N$-dimensional hypercube.

Here, we will only consider the case of three-qubit gates,
namely the XOR gate and M gate. The quantum logic circuits of
these gates are shown in Fig.~4A and Fig.~4D of the main text,
respectively. For convenience, these circuits are also
displayed in
Fig.~\ref{Figure:Supporting:11:Matteo:Mariantoni:201107}, B and
D. The sign of each element $\tau^{\textrm{XOR}}_k$, with $k {}
\in {} \{ 0 , 1 , \ldots , 7 \}$, of vector ${\bm
\tau}^{\textrm{XOR}}_{}$ and of each element
$\tau^{\textrm{M}}_k$ of vector ${\bm \tau}^{\textrm{M}}_{}$
(cf.~main text) is given on the vertices of the phase-gate cube
(cf.~Fig.~\ref{Figure:Supporting:11:Matteo:Mariantoni:201107},
C and E, respectively). As explained in the main text, a
positive sign corresponds to a $0$ phase and a negative sign to
a $\pi$ phase. As shown in
Fig.~\ref{Figure:Supporting:11:Matteo:Mariantoni:201107}, C and
E, the difference between the phases associated with each pair
of vertices connected by a segment of the cube is indicated on
the segment connecting that pair of vertices. For each gate,
this gives a total of twelve phase differences corresponding to
the elements in vector ${\bm \varphi}$ of
Eq.~\ref{Equation:Supporting:23}.

We now use the phase-gate cube to determine the transformation
matrix $T^{}_{\varphi \, \tau}$ between vector ${\bm \varphi}$
and vector ${\bm \tau^{}_{}}$. Each row of $T^{}_{\varphi \,
\tau}$ must correspond to a segment of the phase-gate cube, and
each column to a vertex, giving a matrix with dimensions $( 12
, 8 )$. The first row of $T^{}_{\varphi \, \tau}$ is associated
with the phase difference between states $| \textrm{e}
\textrm{g} 0 \rangle$ and $| \textrm{g} \textrm{g} 0 \rangle$,
$\phi^{}_{| \textrm{e} \textrm{g} 0 \rangle} - \phi^{}_{|
\textrm{g} \textrm{g} 0 \rangle}$
(cf.~Eq.~\ref{Equation:Supporting:23} and, for the case of the
M gate,
Table~\ref{Table:Supporting:3:Matteo:Mariantoni:201107}).
Adopting the enumeration in
Fig.~\ref{Figure:Supporting:11:Matteo:Mariantoni:201107}A,
state $| \textrm{e} \textrm{g} 0 \rangle$ corresponds to the
vertex ($4$) of the phase-gate cube, and state $| \textrm{g}
\textrm{g} 0 \rangle$ to the vertex ($0$). In the case of the M
gate, for example, the first raw of $T^{}_{\varphi \, \tau}$
must then be $-1 , \, 0 , \, 0 , \, 0 , \, 1 , \, 0 , \, 0 , \,
0$. Following a similar procedure for all twelve segments of
the phase-gate cube for the M gate, we readily find the entire
M gate transformation matrix $T^{}_{\varphi \, \tau}$,
\begin{equation}
 T^{}_{\varphi \, \tau} {} = {}
 \begin{pmatrix}
  \,\, -1 \quad & 0 \quad & 0 \quad & 0 \quad & 1 \quad & 0 \quad & 0 \quad & 0 \,\, \\
  \,\, 0 \quad & -1 \quad & 0 \quad & 0 \quad & 0 \quad & 1 \quad & 0 \quad & 0 \,\, \\
  \,\, 0 \quad & 0 \quad & -1 \quad & 0 \quad & 0 \quad & 0 \quad & 1 \quad & 0 \,\, \\
  \,\, 0 \quad & 0 \quad & 0 \quad & -1 \quad & 0 \quad & 0 \quad & 0 \quad & 1 \,\, \\
  \,\, -1 \quad & 0 \quad & 1 \quad & 0 \quad & 0 \quad & 0 \quad & 0 \quad & 0 \,\, \\
  \,\, 0 \quad & -1 \quad & 0 \quad & 1 \quad & 0 \quad & 0 \quad & 0 \quad & 0 \,\, \\
  \,\, 0 \quad & 0 \quad & 0 \quad & 0 \quad & -1 \quad & 0 \quad & 1 \quad & 0 \,\, \\
  \,\, 0 \quad & 0 \quad & 0 \quad & 0 \quad & 0 \quad & -1 \quad & 0 \quad & 1 \,\, \\
  \,\, -1 \quad & 1 \quad & 0 \quad & 0 \quad & 0 \quad & 0 \quad & 0 \quad & 0 \,\, \\
  \,\, 0 \quad & 0 \quad & -1 \quad & 1 \quad & 0 \quad & 0 \quad & 0 \quad & 0 \,\, \\
  \,\, 0 \quad & 0 \quad & 0 \quad & 0 \quad & -1 \quad & 1 \quad & 0 \quad & 0 \,\, \\
  \,\, 0 \quad & 0 \quad & 0 \quad & 0 \quad & 0 \quad & 0 \quad & -1 \quad & 1 \,\, \\
 \end{pmatrix} \, .
    \label{Equation:Supporting:24}
\end{equation}
Notably, the rank of the matrix $T^{}_{\varphi \, \tau}$ of
Eq.~\ref{Equation:Supporting:24} is $7$, as expected from the
number of physically independent phases of the unitary matrix
of a general three-qubit controlled-phase quantum gate,
$U^{\phi}_{}$. A similar procedure can be used to obtain the
transformation matrix associated with the XOR gate (not shown)
or any other three-qubit controlled-phase quantum gate.

Given the shape of vectors ${\bm \tau^{}_{}}$ and ${\bm
\varphi}$, and of the matrix $T^{}_{\varphi \, \tau}$, vectors
${\bm \tau^{}_{}}$ and ${\bm \varphi}$ are related by the
simple linear system
\begin{equation}
T^{}_{\varphi \, \tau} \, \cdot \, {\bm \tau^{}_{}} {} = {}
{\bm \varphi} \, .
    \label{Equation:Supporting:25}
\end{equation}
Since the phase differences in ${\bm \varphi}$ are the only
phases measured in the experiments, the system of
Eq.~\ref{Equation:Supporting:25} has to be solved in order to
find ${\bm \tau^{}_{}}$. The matrix $T^{}_{\varphi \, \tau}$ is
actually not invertible. However, the system of
Eq.~\ref{Equation:Supporting:25} is overconstrained by the
experimental data and so it can be solved in a least-squares
best fit sense, allowing us to obtain ${\bm \tau^{}_{}}$.

Figure~\ref{Figure:Supporting:12:Matteo:Mariantoni:201107}A
shows the twelve Ramsey fringes used to measure the phase
differences plotted in
Fig.~\ref{Figure:Supporting:12:Matteo:Mariantoni:201107}C in
the case of the XOR gate.
Figure~\ref{Figure:Supporting:12:Matteo:Mariantoni:201107}, B
and D, shows similar results for the M gate. The phase
differences associated with each Ramsey fringe are indicated in
the space between panels A and B and, together with the
corresponding pair of vertices of the phase-gate cube, in the
space between panels C and D. Solving the system of
Eq.~\ref{Equation:Supporting:25} for the phase differences
shown in
Fig.~\ref{Figure:Supporting:12:Matteo:Mariantoni:201107}, C and
D, finally allows us to obtain the phases shown in Fig.~4, C
and F, of the main text, thus realizing a full quantum phase
tomography of the XOR and M gate.

We note that the quantum phase tomography used here is
inherently different from that developed in
Ref.~\cite{rudner:2008:qphasetomomit}, where the time evolution
of the quantum phase of the qubit state was used to infer
information on the qubit dephasing mechanisms.

% *******************************
% *** Supplementary Figure 11 ***
% *******************************
%
\begin{figure}[p!]
    \centering
    \includegraphics[width=1.14\columnwidth]{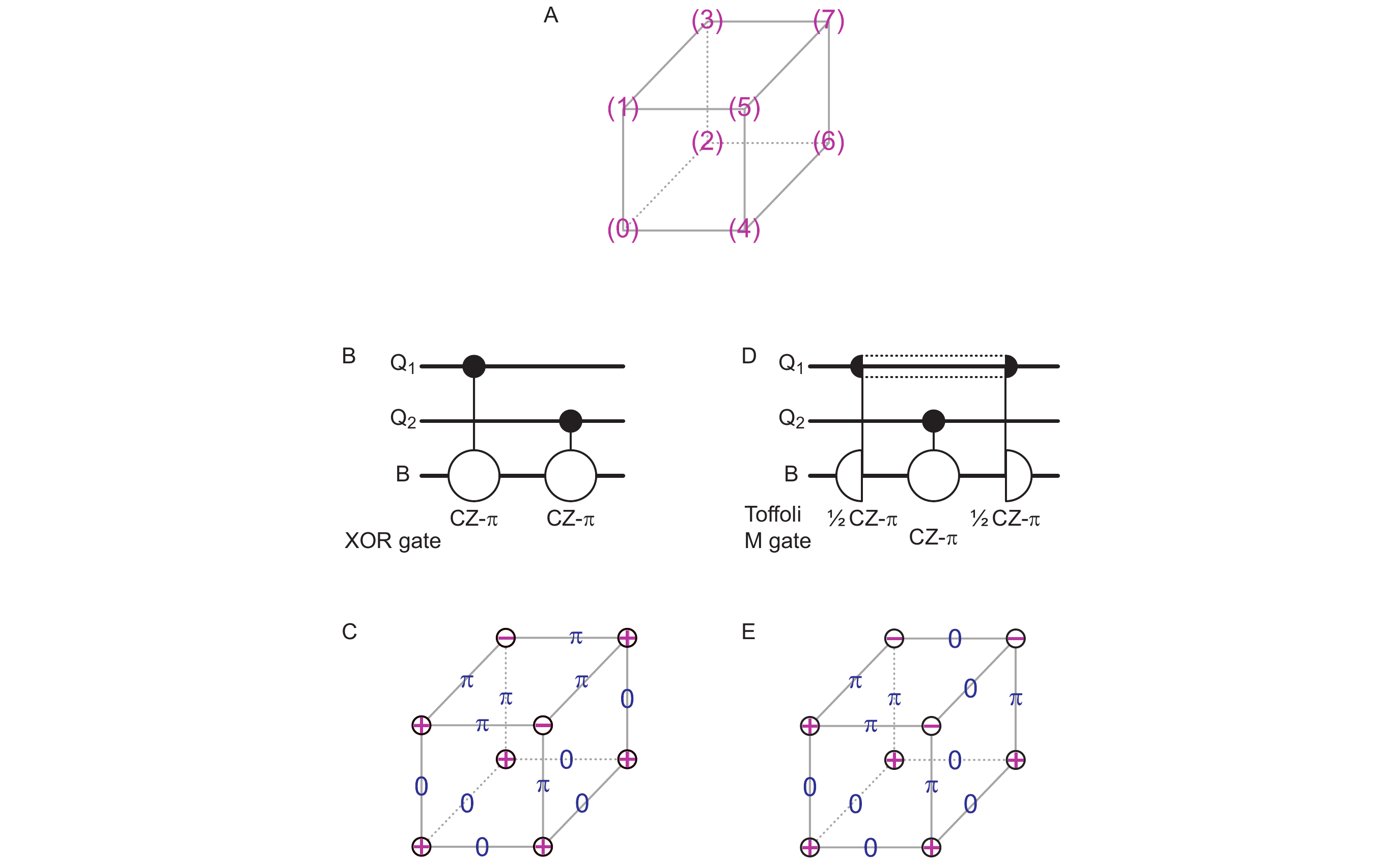}
    \caption{\footnotesize
\textbf{Geometric representation of the phases associated with
the three-qubit XOR phase gate and the three-qubit
Toffoli-class M gate.} (\textbf{A}) The phase-gate cube. The
eight vertices of the cube correspond to the diagonal elements
of the gate unitary matrix. The vertices are numbered from
($0$) to ($7$), following the same enumeration as for the
results of quantum phase tomography (cf.~space between Fig.~4C
and Fig.~4F in the main text). (\textbf{B}) Quantum logic
circuit for the XOR gate, as in Fig.~4A of the main text.
(\textbf{C}) The sign of each element $\tau^{\textrm{XOR}}_k$
of vector ${\bm \tau}^{\textrm{XOR}}_{}$ (cf.~main text) is
given on the vertices of the phase-gate cube. Each sign, $+ 1$
or $- 1$, corresponds to a phase, $0$ or $\pi$ (cf.~main text).
The difference between the phases associated with each pair of
vertices connected by a segment of the cube (a total of twelve
phase differences) is indicated on the segment connecting that
pair of vertices. (\textbf{D}) Quantum logic circuit for the M
gate, as in Fig.~4D of the main text. (\textbf{E}) As in C, but
for the M gate.
    }
    \label{Figure:Supporting:11:Matteo:Mariantoni:201107}
\end{figure}
    \clearpage

% *******************************
% *** Supplementary Figure 12 ***
% *******************************
%
\begin{figure}[p!]
    \centering
    \includegraphics[width=0.88\columnwidth]{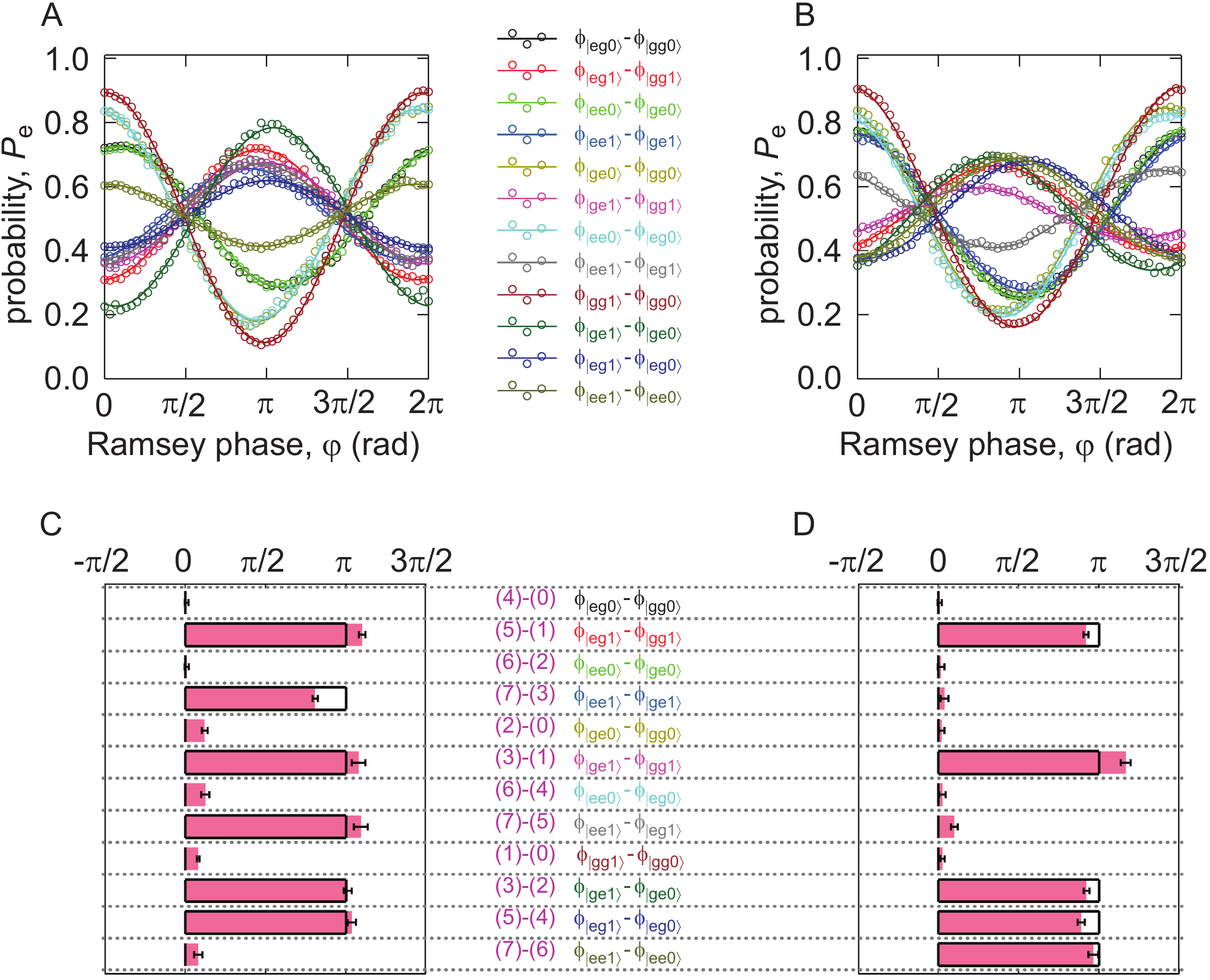}
    \caption{\footnotesize
\textbf{Quantum phase tomography for the XOR and M gate.}
(\textbf{A}) Probability $P^{}_{\textrm{e}}$ to measure either
of the control qubits Q$^{}_1$ or Q$^{}_2$ or target B in state
$| \textrm{e} \rangle$ or $| 1 \rangle$ vs. Ramsey phase
$\varphi$ for the XOR gate. (\textbf{B}) As in A, but for the M
gate. Open circles: Data. Solid lines: Least-squares fits to
the data. The legend to the phases is indicated in the space
between the panels. (\textbf{C}) The twelve phase differences
in vector ${\bm \varphi}$ obtained from the Ramsey fringes in
A. (\textbf{D}) As in C, but for the phase differences obtained
from the Ramsey fringes in B. The pair of vertices of the
phase-gate cube and the corresponding phase differences are
indicated in the space between the panels. The error bars are
due to the confidence intervals to the fits in A and B. Such
confidence intervals propagate through the quantum phase
tomography process generating the error bars in Fig.~4, C and
F, of the main text.
    }
    \label{Figure:Supporting:12:Matteo:Mariantoni:201107}
\end{figure}
    \clearpage

% ************************
% *** The bibliography ***
% ************************
%

%
    \clearpage


\begin{thebibliography}{10}
\bibitem{q:phase:tomo}
    \bibinfo{title}{A full gate characterization via quantum process tomography
                    was not possible as we could only simultaneously measure two qubits,
                    with the resonator acting as the third qubit}.
\end{thebibliography}
\begin{thebibliography}{10}
\bibitem{two:level:approximation}
    \bibinfo{title}{Note that, in general, the two-level approximation
                    is unsuitable for describing a phase qubit.
                    In fact, the phase qubit nonlinearity is typically small
                    and, thus, at least the three lowest eigenstates
                    $| \textrm{g} \rangle$, $| \textrm{e} \rangle$,
                    and $| \textrm{f} \rangle$ should be taken into account
                    for a more correct description. However, for the purposes
                    of this section we will only consider states
                    $| \textrm{g} \rangle$ and $| \textrm{e} \rangle$}.
\bibitem{simga:z:driving}
    \bibinfo{title}{A more complete description of the qubit-driving
                    Hamiltonian should also include a term $\approx {}
                    h \gamma \Omega^{}_{\textrm{D}} ( \tau ) \hat{\sigma}^{}_z
                    \sin ( 2 \pi f^{}_{\textrm{D}} \tau + \phi^{}_{\textrm{del}} )$,
                    where $\gamma$ is a small parameter ($\gamma {} \simeq {} 0$).
                    This term slightly shifts the qubit
                    eigenenergies and, for a small-amplitude driving
                    (as in our case), it can safely be neglected}.
\bibitem{ramsey:phase:difference}
    \bibinfo{title}{In the experiments, the phase difference
                    associated with the states in the first row of Table~\ref{Table:Supporting:3:Matteo:Mariantoni:201107}
                    is used as a reference phase for each phase difference
                    associated with the states in the remaining rows.
                    A similar approach is followed for the XOR gate}.
\end{thebibliography}
\end{document}